%% file: neurips_data_2024.tex
\documentclass{article}
\PassOptionsToPackage{numbers, compress}{natbib}
\usepackage[final]{neurips_data_2024}
\usepackage[pdftex]{graphicx}
\usepackage[utf8]{inputenc} 
\usepackage[T1]{fontenc}    
\usepackage{hyperref}       
\usepackage{url}            
\usepackage{booktabs}       
\usepackage{amsfonts}       
\usepackage{nicefrac}       
\usepackage{siunitx}
\usepackage{microtype}      
\usepackage{xcolor}         
\usepackage{tikz}
\usepackage{bm}
\usepackage{scalerel}
\usepackage{stackengine,wasysym}
\newcommand{\images}{./images}

\newcommand{\figref}[1]{Figure~\ref{#1}}

\newcommand\reallywidetilde[1]{\ThisStyle{%
  \setbox0=\hbox{$\SavedStyle#1$}%
  \stackengine{-.1\LMpt}{$\SavedStyle#1$}{%
    \stretchto{\scaleto{\SavedStyle\mkern.2mu\AC}{.5150\wd0}}{.6\ht0}%
  }{O}{c}{F}{T}{S}%
}}
\usepackage{subcaption}
\usepackage{bibunits}
\makeatletter
\newcommand{\maketitlepage}{%
  \begin{titlepage}
    \let\thanks\@gobble
    \let\footnote\@gobble
    \if@twocolumn
      \ifnum \col@number=\@ne
        \@maketitle
      \else
        \twocolumn[\@maketitle]%
      \fi
    \else
      \@maketitle
    \fi
    \thispagestyle{empty}
  \end{titlepage}%
}
\makeatother
\title{WindsorML: High-Fidelity Computational Fluid Dynamics Dataset For Automotive Aerodynamics}

\author{%
  Neil Ashton\thanks{Now at NVIDIA, Corresponding Author:  nashton@nvidia.com , contact@caemldatasets.org } \\
  Amazon Web Services\\
  60 Holborn Viaduct\\
  London, EC1A 2FD \\
\And
  Jordan B. Angel\\
  Volcano Platforms Inc.\\
  Stanford Research Park, 3240 Hillview Ave \\
  Palo Alto, CA 94304 \\
\And
  Aditya S. Ghate\\
  Volcano Platforms Inc.\\
  Stanford Research Park, 3240 Hillview Ave \\
  Palo Alto, CA 94304 \\
\And
  Gaetan K. W. Kenway\\
  Volcano Platforms Inc.\\
  Stanford Research Park, 3240 Hillview Ave \\
  Palo Alto, CA 94304 \\
\And
  Man Long Wong\\
  Volcano Platforms Inc.\\
  Stanford Research Park, 3240 Hillview Ave \\
  Palo Alto, CA 94304 \\
\And
  Cetin Kiris\\
  Volcano Platforms Inc.\\
  Stanford Research Park, 3240 Hillview Ave \\
  Palo Alto, CA 94304 \\
\And
  Astrid Walle\\
  Siemens Energy\\
  Huttenstraße 12, 10553 \\
  Berlin, Germany \\ 
\And
  Danielle C. Maddix\\
  AWS AI Labs\\
  2795 Augustine Dr., \\
  Santa Clara, CA 95054, United States \\
\And
  Gary Page\\
  Loughborough University\\
  Epinal Way, Loughborough  \\
  LE11 3TU, United Kingdom \\
}

\begin{document}

\maketitle

\begin{abstract}
This paper presents a new open-source high-fidelity dataset for Machine Learning (ML) containing 355 geometric variants of the Windsor body, to help the development and testing of ML surrogate models for external automotive aerodynamics. Each Computational Fluid Dynamics (CFD) simulation was run with a GPU-native high-fidelity Wall-Modeled Large-Eddy Simulations (WMLES) using a Cartesian immersed-boundary method using more than 280M cells to ensure the greatest possible accuracy. The dataset contains geometry variants that exhibits a wide range of flow characteristics that are representative of those observed on road-cars. 
The dataset itself contains the 3D time-averaged volume \& boundary data as well as the geometry and force \& moment coefficients. This paper discusses the validation of the underlying CFD methods as well as contents and structure of the dataset. To the authors knowledge, this represents the first, large-scale high-fidelity CFD dataset for the Windsor body with a permissive open-source license (CC-BY-SA). 

\end{abstract}
\begin{bibunit}
\input{sections/introduction}   
\input{sections/test-case} 
\input{sections/validation}
\input{sections/dataset}

\input{sections/conclusions}
\clearpage
\begin{ack}
Thank you to members of the AutoCFD4 AI/ML TFG for early feedback on the dataset. Thank you to Nate Chadwick, Peter Yu, Mariano Lizarraga, Pablo Hermoso Moreno, Shreyas Subramanian and Vidyasagar Ananthan from Amazon Web Services for the useful contributions and feedback as well as debugging of the dataset in preparation for release. Also, thank you to Bernie Wang for providing additional feedback on the paper itself. 
\end{ack}
\putbib[references,manualRefs,applicationI,applicationIII,numerics_thesis, references_add]
\end{bibunit}

\setcounter{section}{0}
\title{WindsorML: Supplementary Information}
\maketitlepage
\tableofcontents
\newpage
\appendix
\begin{bibunit}
\input{sections/appendix-methodology}
\input{sections/appendix-dataset-windsor}

\input{sections/appendix-validation}
\input{sections/appendix-mlevaluation}

\newpage
\input{sections/appendix-datasheet}

\newpage
\putbib[references,manualRefs,applicationI,applicationIII,numerics_thesis, references_add]
\end{bibunit}

\end{document}

%% file: sections/introduction.tex
\section*{Introduction}

The use of Machine Learning (ML) to augment Computational Fluid Dynamics (CFD) is gaining the attention of academia and industry because of its potential to offer an additional tool to explore new designs in near real-time compared to traditional CFD simulations \cite{ananthan2024,vinuesa2022,haber2023,lino2023,li2023}. Whilst these methods are still in their infancy, there have been promising results that show their ability to predict surface, volume and/or forces and moments on unseen geometries or boundary conditions \cite{bonnet2022,jacob2022,ananthan2024,Elrefaie2024,baque2018}. However the majority of these examples have been for 2D cases \cite{bonnet2022} with limited examples on more complex 3D cases \cite{ananthan2024,baque2018,jacob2022,Elrefaie2024}. One of the reasons for the limited dissemination of 3D examples is the lack of publicly available 3D training data. Unlike Large-Language Models (LLMs) where the training data is readily available, public training data for CFD, based upon 3D realistic geometries is limited and where such examples exist they are often using lower-fidelity CFD methods \cite{Elrefaie2024,bonnet2022} that are not representative of industry state of the art. Generating such datasets also requires a combination of large-scale High-Performance Computing (HPC) resources, human time to build an efficient workflow and in-depth knowledge of the underlying test-case to ensure rigour and correlation to the corresponding experimental data.

To help to address this lack of 3D training data, Ashton et al. \cite{ashton2024ahmed} recently created the AhmedML dataset, that contains 500 different geometry variants of the Ahmed car body, simulated using a high-fidelity hybrid RANS-LES approach using OpenFOAM on meshes of approximately 20M cells. This open-source dataset contains time-averaged volume \& boundary flow-field variables as well as forces and moment data to provide a rich dataset to aid ML model development. Whilst this dataset enables the development and testing of ML methods, there are a number of limitations to this dataset that have motivated the creation of the WindsorML dataset. 

Firstly, in order to thoroughly assess an ML method, more than one dataset is required. Just as traditional CFD approaches are validated on a number of cases to ensure general applicability. Secondly, the AhmedML dataset is based upon mesh sizes of approximately 20M cells which is below the industry standard of $>100$M cells for scale-resolving simulations of road-cars \cite{hupertz2022} and limits the ability to test out the scaling of ML methods using realistic mesh sizes. This is an important point as some ML methods that work well for cases at 20M cells may become inefficient as you scale beyond 300M points \cite{ananthan2024} and/or require the use of downsampling and data reduction techniques. Moreover from a flow physics point of view, the Windsor body \cite{Varney2020a,page2022} is closer to modern road-cars compared to the Ahmed car body and contains more detailed and up to date experimental data e.g tomographic 3D PIV data.

Finally, the choice of the Windsor body is also motivated by its use within the Automotive CFD Prediction Workshops (AutoCFD) \footnote{https://autocfd.org} that aim to bring the fluid dynamics and automotive community together to improve the state of the art for CFD prediction of road vehicles. The dataset described in this paper was used as a test-case during the 4th AutoCFD workshop, for which future papers will discuss the resulting comparisons from a range of different ML methods. Concurrently the DrivAerML dataset \cite{ashton2024drivaer} based upon the open-source Drivaer has been created, that is described in it's own paper. Used together, the AhmedML \cite{ashton2024ahmed}, WindsorML and DrivAerML \cite{ashton2024drivaer} datasets, provide ML developers with a broad set of data to use that has been formatted in a consistent fashion and is openly available to download and use with a permissive license (CC-BY-SA). A dedicated website \footnote{https://caemldatasets.org} for these datasets has been created to provide the community with up to date information.  

\subsection*{Main Contributions}
\label{sec:introduction-novelty}

This paper's novel contributions are summarized as follows:
\begin{itemize} 
    \item 355 variations in the Windsor body geometry that cover a broad range of pressure and geometry induced flow separation; 
    \item the use of high-fidelity wall-modeled Large-Eddy Simulation (WMLES) CFD method which ensures the best possible correlation to the ground truth;
    \item first ever, freely available, open-source (CC-BY-SA) large-scale dataset based upon the Windsor body that can be used to train ML methods for automotive aerodynamics use cases.
\end{itemize}

The paper is organised as follows: first, the Windsor test-case is described together with the available experimental data. Next, the validation of the baseline geometry and test-case are described. The results section focuses on both global force coefficients and also selected off-body quantities such as planes in the wake of the car. Next, the dataset itself is detailed, including the choice of the geometry variations and the specific outputs that are included in the dataset. Finally, conclusions are drawn and future work is described. The Supplementary Information (SI) contains additional details on the dataset and validation.

%% file: sections/test-case.tex
\section*{Test-Case Description}
\label{gen_inst}
The Windsor body was created by Steve Windsor and Jeff Howell of Rover and later Jaguar Land Rover (JLR) \cite{Varney2020a,page2022} to improve upon the well studied Ahmed car body \cite{Lienhart2003,Ahmed1984,meile2011,Ashton2015a} by creating a geometry whose flow features more closely match a road vehicle. Whilst the Ahmed car body has been used in many experimental and CFD studies, the long flat surface between the front and rear of the body decouples the aerodynamic behaviour of these two regions and makes it more relevant to commercial vehicles. Two variants of the Windsor body were experimentally tested at the Loughborough University wind-tunnel \cite{Varney2020a} and have been the subject of numerous CFD studies - that most recently have taken place within the 2nd, 3rd and 4th Automotive CFD Prediction Workshops \cite{page2022}. The Windsor body was tested in numerous configurations but for this study we follow the 4th AutoCFD4 setup, where the vehicle has no wheels (which simplifies a common source of uncertainty) and the squareback rear shape is used, as the baseline geometry. As discussed later, this baseline geometry is then adapted according to 7 parameters to generate 355 different geometries. 

\begin{figure}[h]
  \centering
\includegraphics[width=0.9\linewidth]{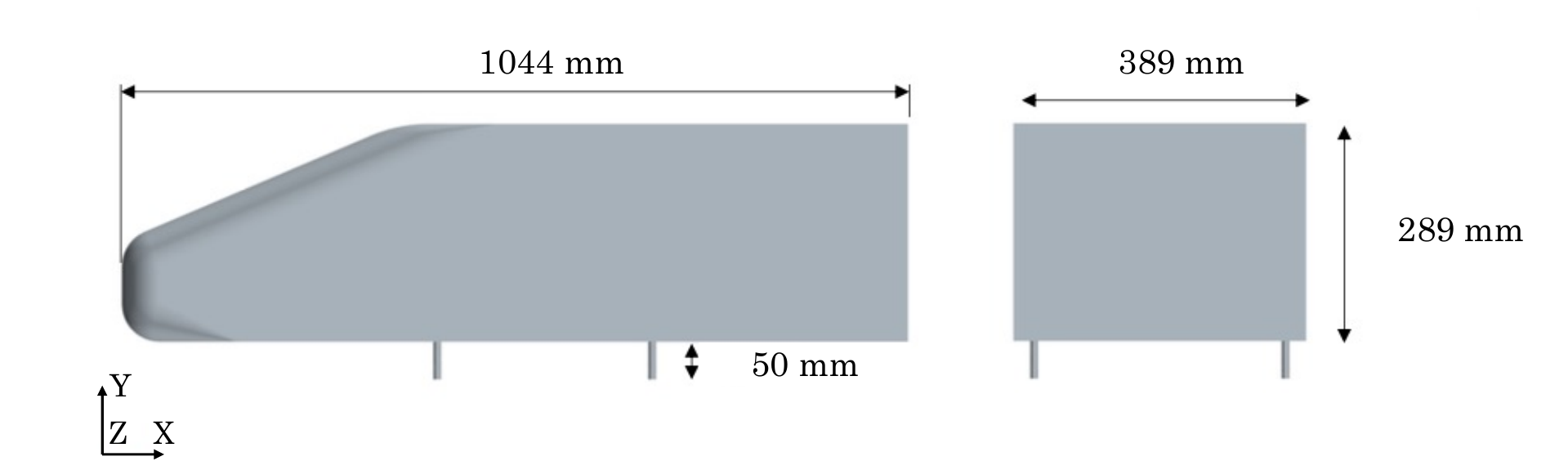}
  \caption{Windsor Geometry}
  \label{fig:windsor-geo}
\end{figure}

\subsection*{Geometry }
The model geometry is shown in \ref{fig:windsor-geo}. The reference frontal area is defined by the vehicle height and width and for the baseline geometry is $\qty{0.112}{\m^2}$. 
The reference length used for pitching moment is the wheelbase $\qty{0.6375}{\m}$.
The CAD geometry of the model has its origin on the ground plane, in the symmetry plane midway between the wheels (which are not included in this particular geometry) i.e $x,y,z=(0,0,0)$. This is also the moment reference center for the calculating of the moments. 

The model is yawed by $-\qty{2.5}{\degree}$ around the $y$-axis (please note that for all the simulations in this work, $y$ is upwards, which differs to the original Windsor geometry where $z$ is upwards) , so generating a positive side force consistent with the experiment. The experimental forces and moments are in the coordinate system of the yawed model, whilst the PIV measurements are in the wind tunnel coordinate system. The choice of $-\qty{2.5}{\degree}$ yaw was to avoid the well documented bi-stability that can often occur on these bluff bodies \cite{Volpe2015} that would cause unnecessary uncertainties (but nevertheless deserves future investigation).

The model is mounted in the wind tunnel with four pins at a ground clearance of
$\qty{50}{\mm}$ and zero pitch. In order to more closely match the underlying
experimental data, the simulations themselves include a simplified wind tunnel domain
i.e. $\qty{3.2}{\m}$ long working section with a $\qty{1.92}{\m} \times
\qty{1.32}{\m}$ (width by height) cross section. There is no moving ground
plane so a boundary layer grows along the groundplane. Experimental
measurements~\cite{Varney2020a} at the centre of the working section quote a
boundary layer thickness of $\qty{60}{\mm}$, displacement thickness of
$\qty{9.4}{\mm}$ and momentum thickness of $\qty{5.5}{\mm}$. The maximum turbulence
intensity was measured to be approximately 3\% at the edge of the boundary
layer.

\subsection*{Solver Setup}
All simulations were run using the Volcano ScaLES code by Volcano Platforms, which solves the explicit compressible Navier--Stokes equations using a
nominally 4th order spatially accurate finite difference discretization with
favorable Kinetic Energy and Entropy Consistency properties making it suitable
for Large Eddy Simulations (LES) over a  range of flow regimes. The viscous
flux discretization utilizes a mix of 4th order and 2nd order discretizations
with high spectral bandwidth thereby allowing further robustness for high
Reynolds number flows without additional numerical dissipation via the inviscid
flux numerical operator in turbulence resolving regions. Physics-based numerical
sensors that are functions of local velocity gradients as well as pressure and
density fluctuations allow for spatio-temporally localized use of limiters
needed to capture flows with discontinuities such as shocks and steep density
gradients.  
The
Strong Stability Preserving (SSP) 3rd order
Runge--Kutta scheme developed by Gottlieb \& Shu~\cite{gottlieb1998total} is
utilized for all the work presented in this paper.

Geometries are represented in the numerical formulation using an
immersed boundary algorithm capable of enforcing inviscid no-penetration
boundary conditions as well as viscous skin friction with appropriate Reynolds
number asymptotic properties using an equilibrium wall-model. The wall-model
that provides a shear-stress constraint at the wall uses solution information
interpolated via probing at a fixed distance of $1.5\Delta$ into the fluid away
from walls, where $\Delta$ is the local grid spacing at the wall. Additional methods to damp generation of spurious
numerical noise at walls as well as at coarse-fine interfaces are also utilized
in the present work. All simulations presented in this work utilize the constant
coefficient Vreman subgrid scale closure~\cite{vreman2004eddy} to model the
subgrid-scale stresses. 

\subsubsection*{Computational Mesh} The Volcano ScaLES solver uses Cartesian grids
generated using a recursive octree approach. Grids around engineering-quality
configurations consist of an unstructured tree-of-cubes with millions of leaf
cubes each containing a structured grid of $4^3$ or $8^3$ cells.

\begin{figure}
    \centering
    \begin{tikzpicture}
        \draw[use as bounding box, draw=none] (0,0) rectangle (\textwidth,5);
        \draw (1.5cm,0) node[anchor=south west] {\includegraphics[width=0.85\textwidth]{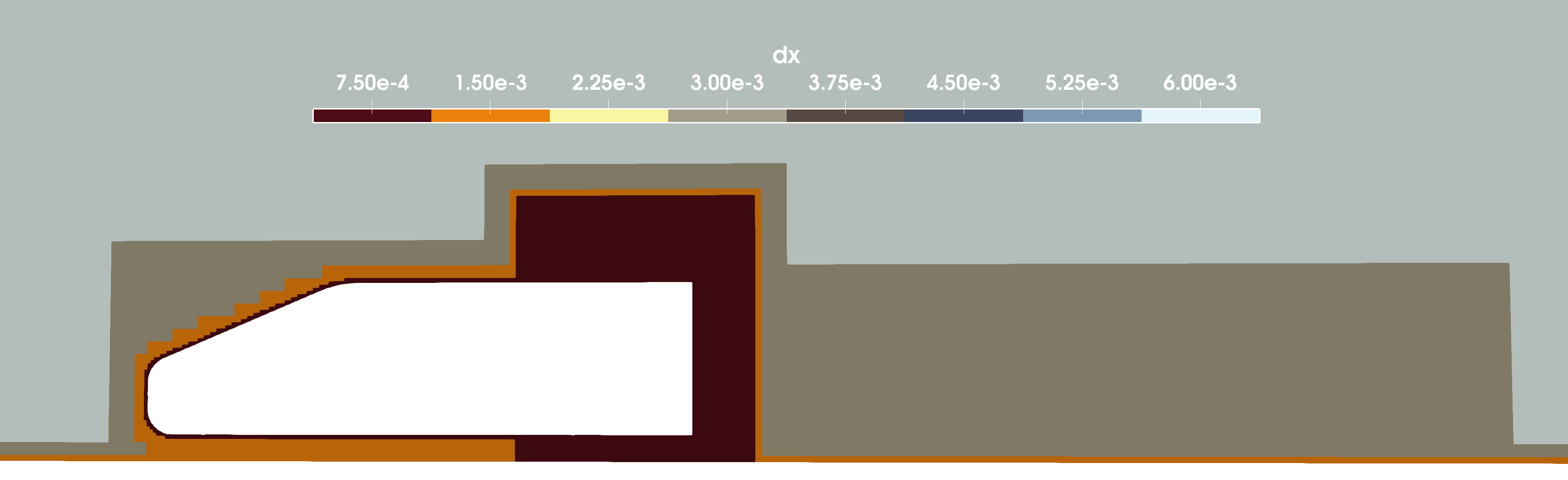}};
    \end{tikzpicture}
    \caption{Grid spacing (dx) for a slice through $y=\qty{0}{\m}$ for the Windsor body.}
    \label{fig:windsor_mesh_wake}
\end{figure}

\figref{fig:windsor_mesh_wake} depicts the nested refinements regions for the
optimized mesh. Refinement is emphasized for the boundary layer development
from the nose of the car and for the wake and off-body vorticity directly behind
the body. Section 3 of this paper discusses in detail the validation of this code against experimental data for the baseline windsor model and the rationale behind the choice of computational mesh.

\subsection*{Boundary Conditions}
\begin{figure}[hbt]
    \centering
    \begin{tikzpicture}
        \draw[use as bounding box, draw=none] (0, .00cm) rectangle (0.8\linewidth,8.5cm);
        \draw ( 0.80cm, 0.00cm) node[anchor=south west] {\includegraphics[width=0.75\textwidth, trim = {0cm, 0cm, 0cm, 0cm}, clip]{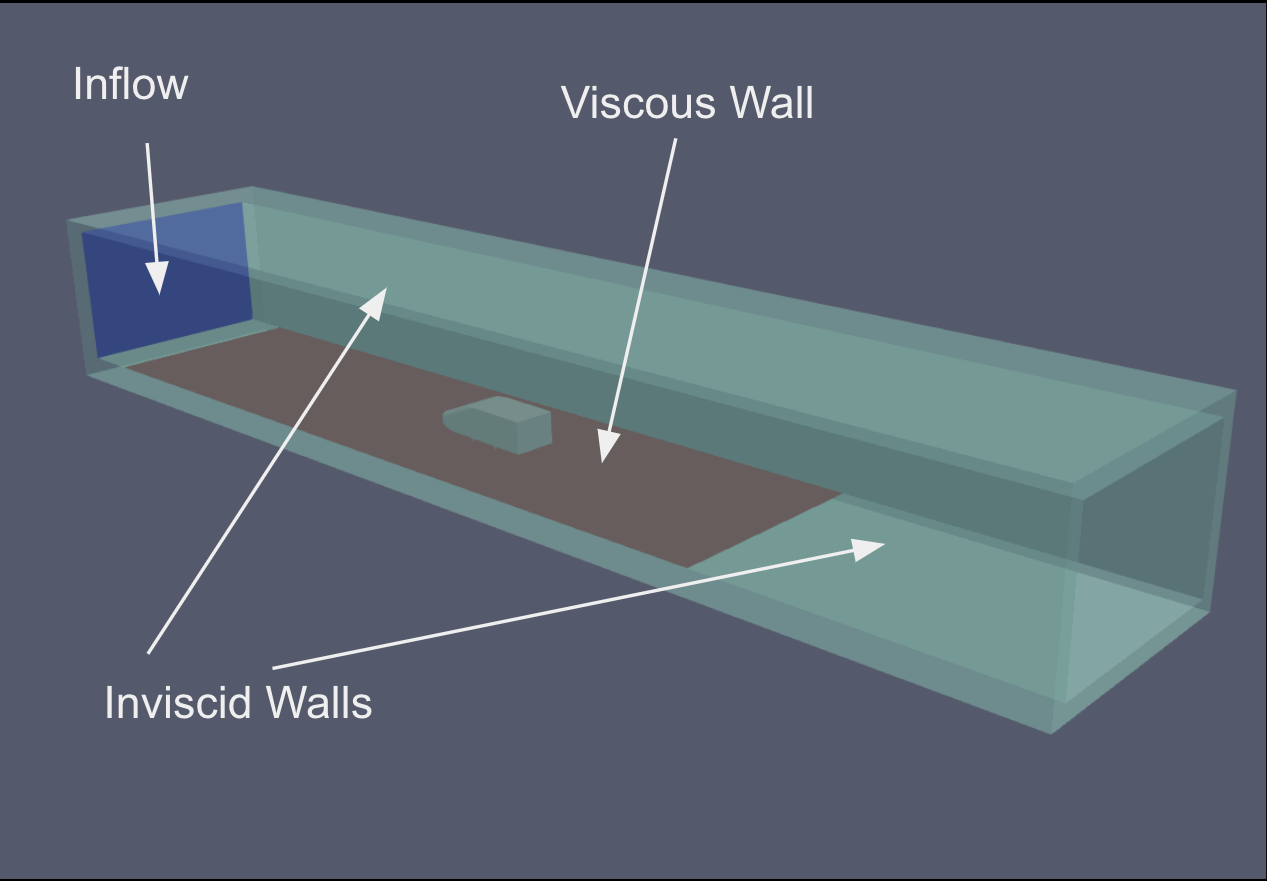}};
    \end{tikzpicture}
    \caption{Depiction of the computational domain for the Windsor body simulations. }
    \label{fig:windsor_domain}
\end{figure}

To follow the setup of the AutoCFD 3 and 4 workshops, the Windsor body is yawed
2.5 degrees and the Reynolds number is $Re=2.9\times 10^6$ based on the body
length and the freestream velocity. A nominal inlet velocity of
$\qty{40}{\m\per\s}$ is given upstream of the car and a sensor is placed at
$(x,y,z) = (\qty{-2}{\m}, \qty{1.3}{\m}, \qty{0}{\m})$ to measure local static
pressure and velocity. For the validation of the method against experimental data for the baseline geometry, these are used to normalize the simulation outputs,   however for the dataset itself, the reference inlet velocity is used to normalize all outputs (see SI for further details),. The remaining boundary conditions are shown in Figure \ref{fig:windsor_domain}. 

The domain required for this case (see Figure~\ref{fig:windsor_domain}) represents the wind tunnel confinement but with the following modifications:
\begin{enumerate}
  \item The Windsor body sidewalls are aligned with the $x$-direction, the tunnel is yawed.
  \item Only the ground plane has a no slip condition and hence has boundary layer growth; 
the top and side walls should be treated as a slip or ``inviscid'' wall. 
  \item  A long parallel inlet run is used in order to grow a boundary layer on the ground plane
of approximately the correct thickness. 
  \item A parallel exit run is added downstream to avoid interactions with the wake.
The domain extends upstream to $x=-\qty{5}{\m}$ and downstream to $x=+\qty{6}{\m}$ (the model is $x=-\qty{0.56}{\m}$ nose to $x=+\qty{0.48}{\m}$ base). The width and height of the CFD domain matches the wind tunnel.
\item The ``outlet'' for the tunnel is not a domain boundary for these simulations even though the volume data does not extend further. 
 To minimize numerical reflections from the tunnel exit, the tunnel was allowed to vent into a large reservoir which is initialized with atmospheric conditions and given
 simple extrapolation boundary conditions far from the tunnel. This means there is no fixed pressure condition or any other boundary condition enforced at the tunnel exit.
\end{enumerate}

\subsection*{HPC setup}
All simulations were run on Amazon Web Services, using a dynamic HPC cluster provisioned by AWS ParallelCluster v3.9. Amazon EC2 g5.48xlarge nodes were used for each case, which contain 8 Nvidia A10g GPU cards and a second generation AMD EPYC processor with 96 physical cores and 768GB of CPU memory. These are connected using a 100Gbit/s Elastic Fabric Adapter (EFA) interconnect \cite{Shalev2020}. A 300TB Amazon Fsx for Lustre parallel file-system was used as the main data location during the runs, which were later transferred to object storage in Amazon S3. Each simulation was run on a single g5.48xlarge node i.e 8 A10g GPUs, and each case took approximately \qty{2}{\min} to mesh and \qty{28}{\hour} to solve.

%% file: sections/validation.tex
\section*{Validation}
\begin{table}
    \centering
    \caption{Computational cost for the Windsor body.}
    \begin{tabular}{@{}cccccc@{}} \toprule
        Label      & Min. grid spacing      & No. of cells  & Wall time / CTU  & $C_d$\\ 
                   & [\si{\mm}]  &   [$10^6$]     &  [min]          &       \\ \midrule
        $G_1$      & 1.00                 & 151            & 6               & 0.31 \\
        $G_2$      & 0.75                 & 275            & 15              & 0.34 \\
        $G_3$      & 0.625                & 468            & 27              & 0.33  \\ 
        Exp.       & -                    & -              & -               & 0.33  \\ \bottomrule
    \end{tabular}
    \label{tab:cost_windsor}
\end{table}
For the design case study, we choose to keep the mesh refinement regions fixed for
all design cases. Using the fixed refinement topology, we consider a consistent
grid refinement study for the Windsor body. An example of a grid in this family
is depicted in Figure~~\ref{fig:windsor_mesh_wake}.
The surface of the body is refined to the finest level with a large wake block immediately behind the body. The
block is taller than necessary for the Windsor body but since we are choosing to keep
the refinement regions fixed we define a region to accommodate all designs ride height.
Table~\ref{tab:cost_windsor} shows results of a consistent grid refinement
study using a sequence of grids $\{G_j\}$ for $j=1,2$ and $3$. The columns of the table show the minimum grid spacing used,
the number of computational cells in the grid, the wall-clock time per
convective time unit\footnote{Nondimensional time is defined to be $t^* = tU_\infty/L$ where $L$ is the length of the Windsor body, $U_\infty$ is the free-stream velocity, and $t$ is the time.} 
(CTU) in minutes and the drag coefficient $C_d$. An
additional row is included for the experimental drag coefficient value.
Based on the results, $G_2$ is chosen as a representative baseline case
 and the remainder of the experimental comparisons are carried out for the $G_2$ grid. This was to achieve a balance between accuracy, computational cost and reach a mesh size that is typically used by the automotive industry for external aerodynamics.

\begin{figure}
    \centering
    \begin{tikzpicture}
        \draw (0.1cm,8.75cm) node[anchor=south west,draw, inner sep=0] {\includegraphics[width=0.30\textwidth, trim={5.cm 5.cm 15.0cm 5.0cm}, clip]{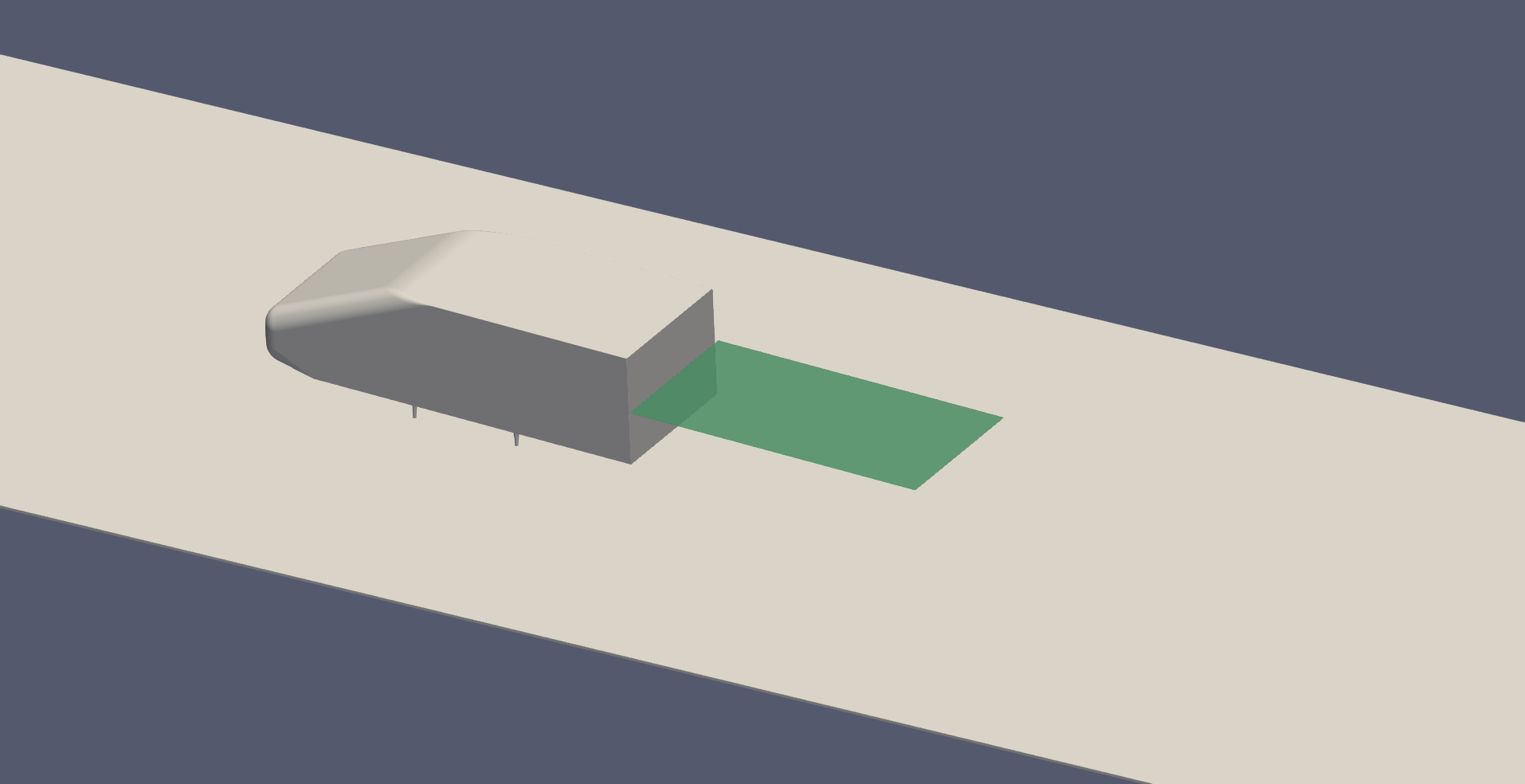}};
        \draw (0.1cm,5.20cm) node[anchor=south west,draw, inner sep=0] {\includegraphics[width=0.45\textwidth, trim={19cm 4.cm 19.0cm 9.0cm}, clip]{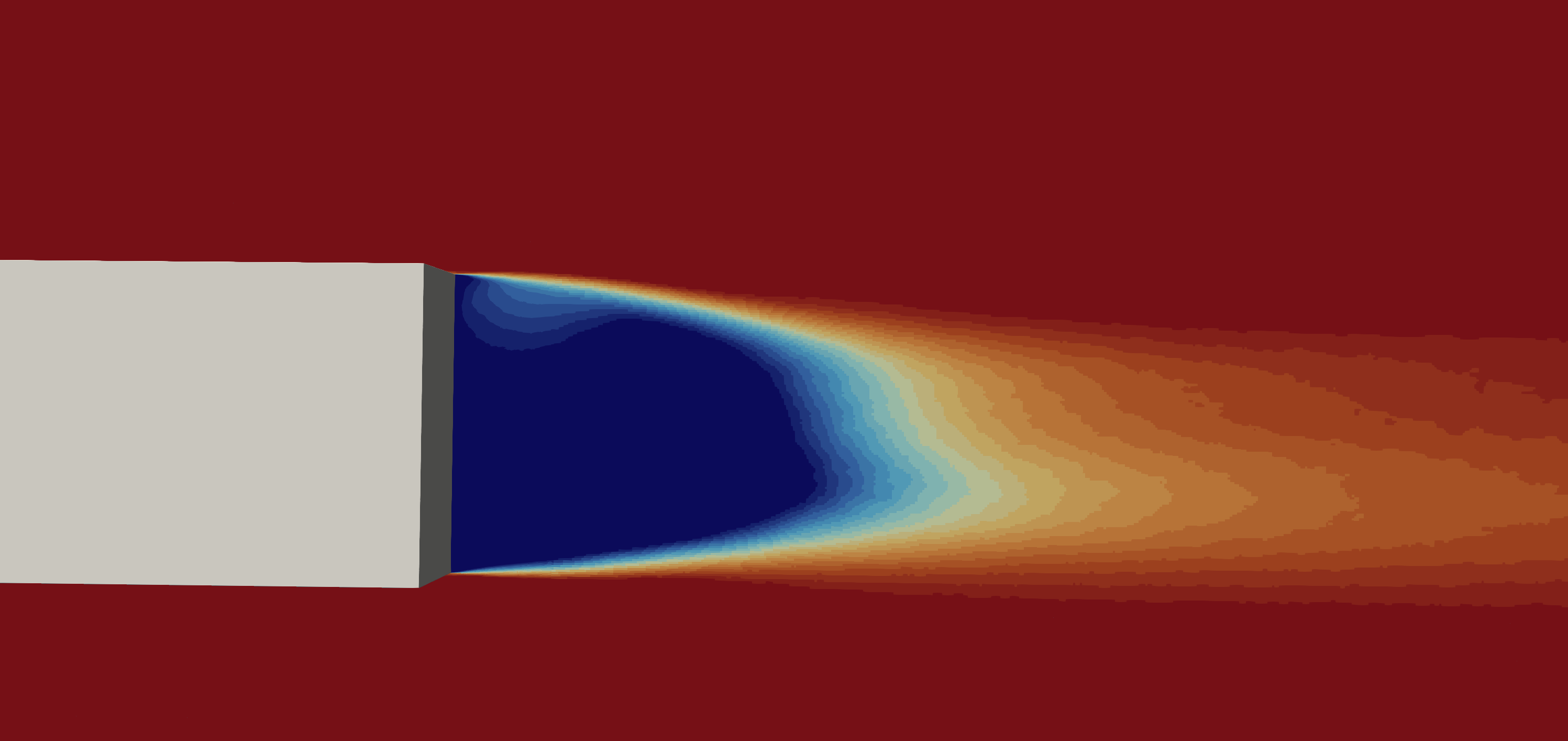}};
        \draw (7.0cm,5.20cm) node[anchor=south west,draw, inner sep=0] {\includegraphics[width=0.45\textwidth, trim={19cm 4.cm 19.0cm 9.0cm}, clip]{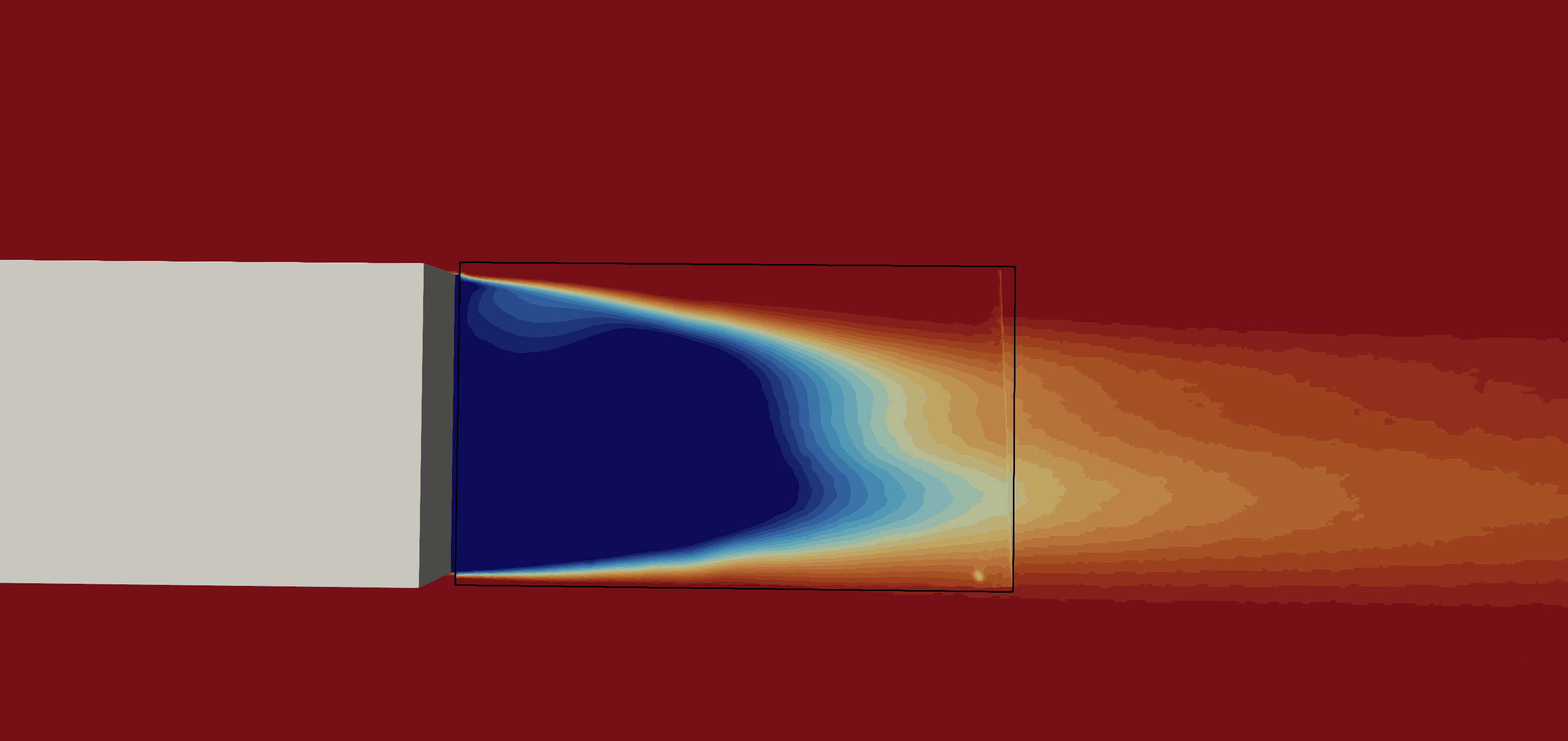}};
        \draw (2.0cm, 7.9cm) node[anchor=south west,draw,fill=white,rounded corners] {Averaged Solution};
        \draw (9.0cm, 7.9cm) node[anchor=south west,draw,fill=white,rounded corners] {With PIV Data};
        \draw (4.6cm,9.1cm) node[anchor=south west,draw]   {\includegraphics[width=0.6\textwidth]{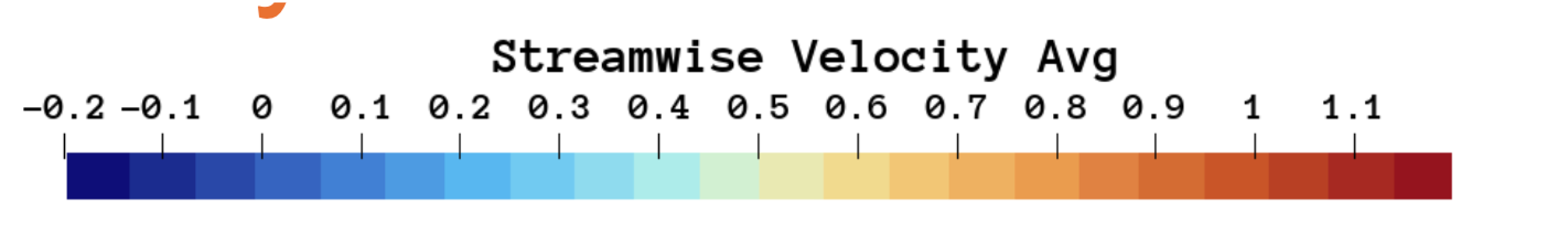}};
    \end{tikzpicture}
    \caption{Time-averaged streamwise velocity at $y=\qty{0.195}{\m}$ with PIV data}
    \label{fig:windsor_piv_z0p195}
\end{figure}

\begin{figure}
    \centering
    \begin{tikzpicture}
        \draw (0.1cm,5.00cm) node[anchor=south west,draw, inner sep=0] {\includegraphics[width=0.30\textwidth, trim={5.cm 5.cm 15.0cm 5.0cm}, clip]{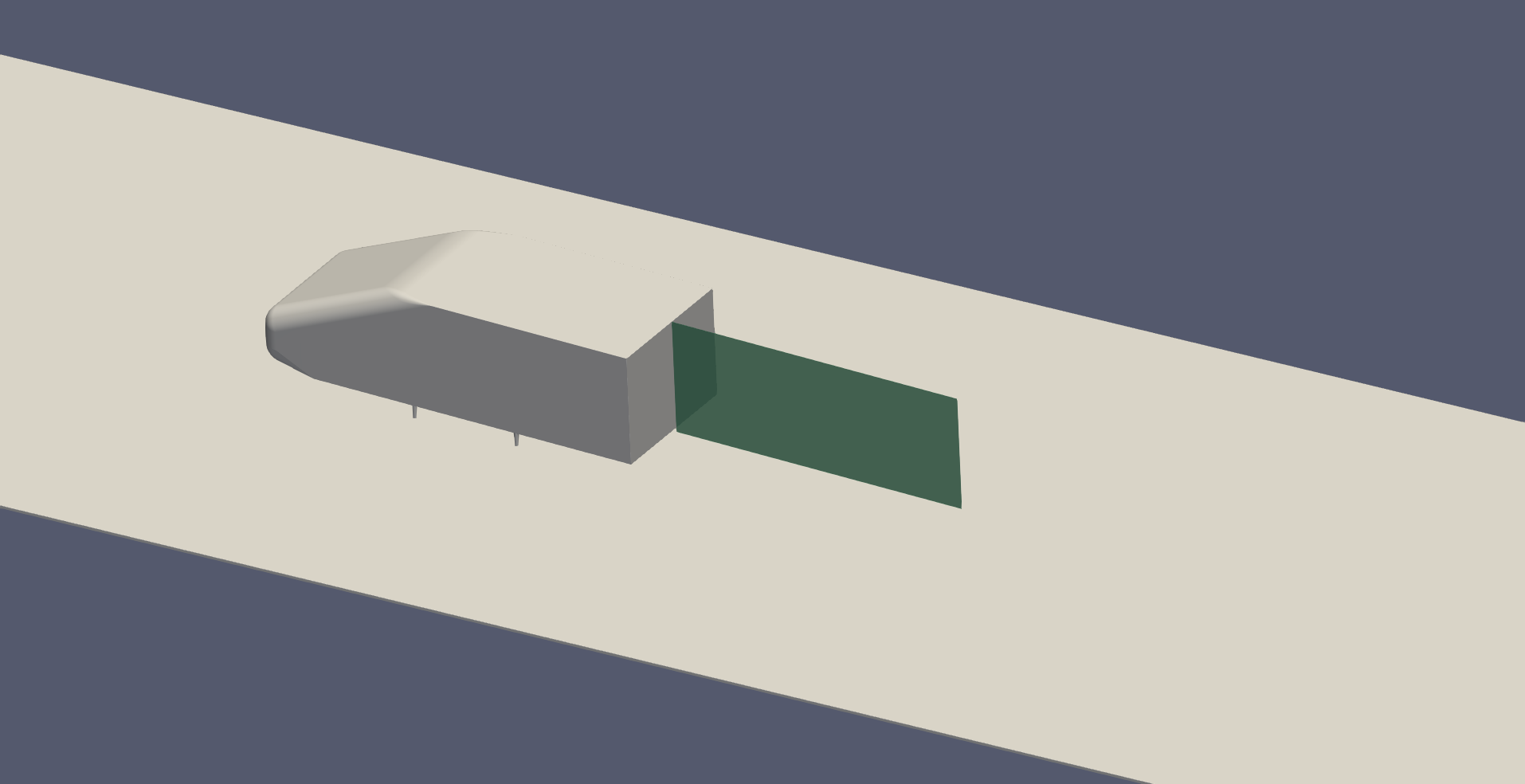}};
        \draw (0.1cm,0.25cm) node[anchor=south west,draw, inner sep=0] {\includegraphics[width=0.45\textwidth, trim={4.5cm 5.cm 15.0cm 5.0cm}, clip]{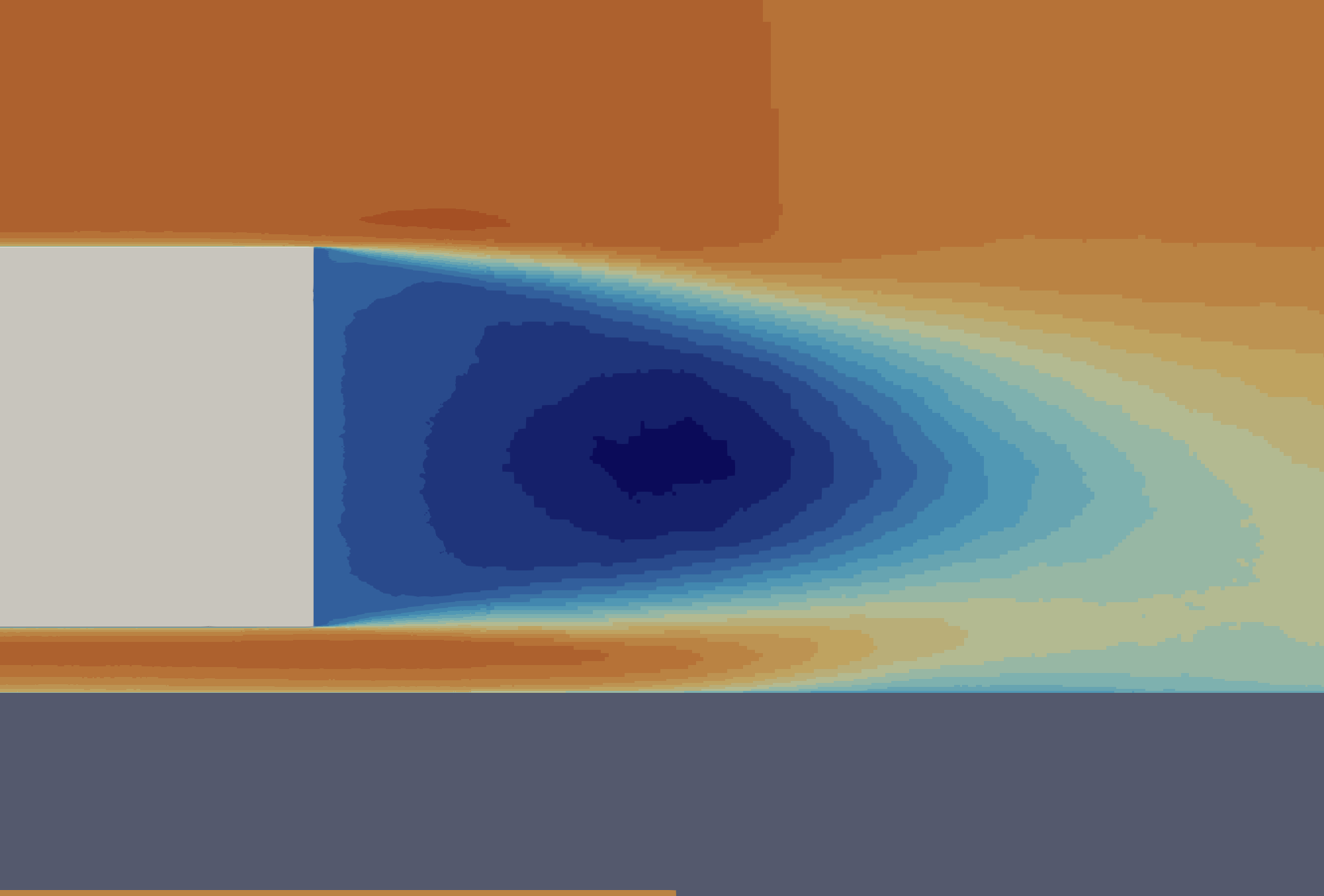}};
        \draw (7.0cm,0.25cm) node[anchor=south west,draw, inner sep=0] {\includegraphics[width=0.45\textwidth, trim={4.5cm 5.cm 15.0cm 5.0cm}, clip]{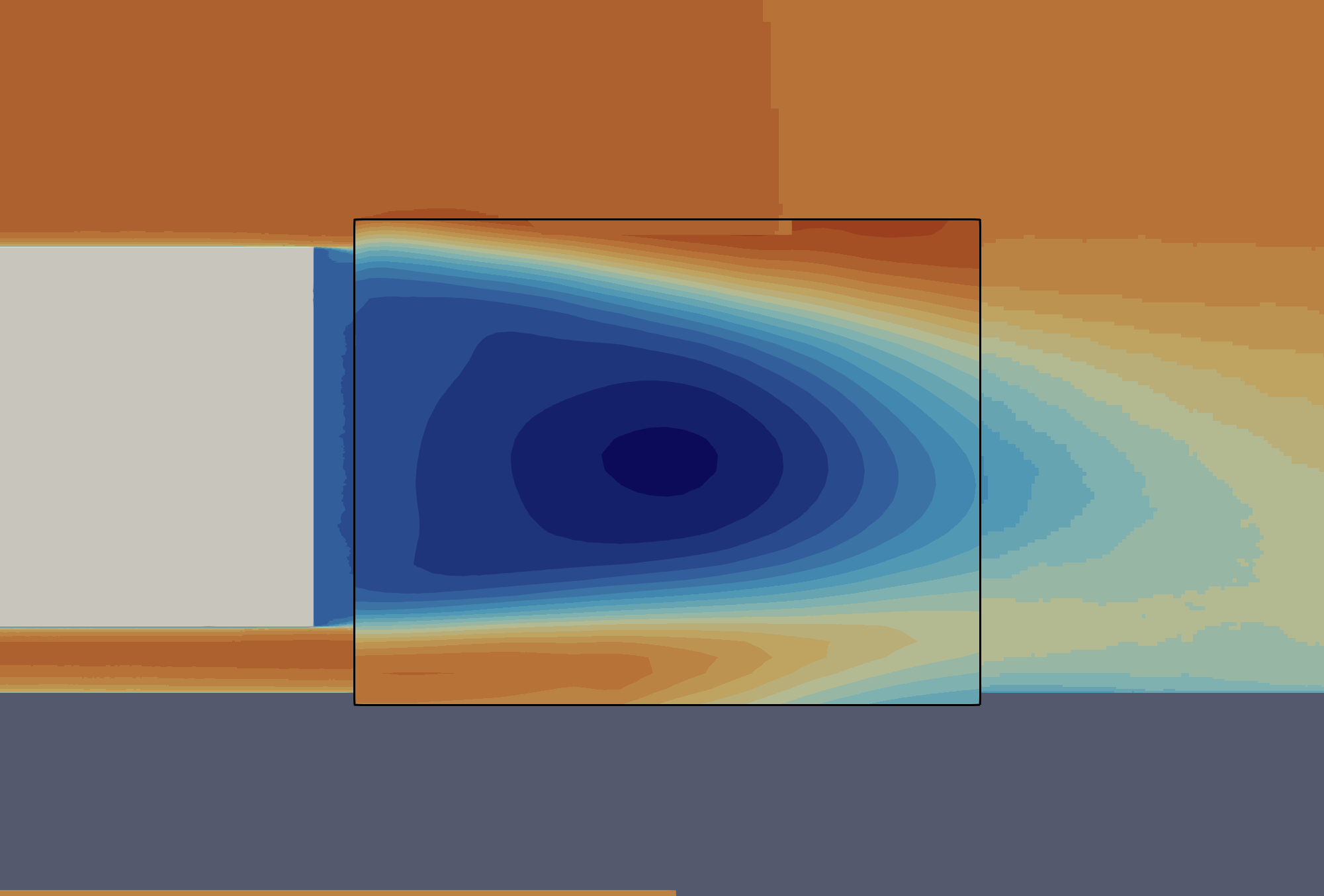}};
        
        \draw (2.0cm, 4.0cm) node[anchor=south west,draw,fill=white,rounded corners] {Averaged Solution};
        \draw (9.0cm, 4.0cm) node[anchor=south west,draw,fill=white,rounded corners] {With Tomo. PIV Data};
        \draw (4.6cm,5.35cm) node[anchor=south west,draw]   {\includegraphics[width=0.6\textwidth]{\images/vel-legend.png}};
    \end{tikzpicture}
    \caption{Time-averaged streamwise velocity at $z=\qty{0}{\m}$ with tomographic PIV data.}
    \label{fig:windsor_piv_y0}
\end{figure}

  {\bf Velocity:}
Typical of bluff-bodies, the Windsor body shows a large recirculation region at the base.
The data for the
$y=\qty{0.195}{\m}$ slice was taken with a freestream velocity of $\qty{40}{\m\per\s}$ which
is the same as our simulation but, in addition to this plane, tomographic PIV
was taken for a 3D region behind the car. This data was recorded for a flow with
a freestream velocity of $\qty{30}{\m\per\s}$. To make comparisons with the tomographic PIV
data, we normalize the velocities by their respective freestream velocity. With that
caveat, Figures ~\ref{fig:windsor_piv_z0p195} \& \ref{fig:windsor_piv_y0} shows a comparison of the
normalized time-averaged streamwise velocity for $y=\qty{0.195}{\m}$ compared with the PIV measurements and 
the same solution at $z=\qty{0}{\m}$ compared to the tomographic PIV data. 
The overall shape of the wake region is well
captured as well as is its interaction with the floor boundary layer which can be
seen in the velocity contours. See SI for a detailed discussion on the validation of the underlying CFD method. 

%% file: sections/dataset.tex
\section*{Dataset description} 

\subsection*{Geometry variations}
\begin{figure}
  \centering
\includegraphics[width=1\textwidth]{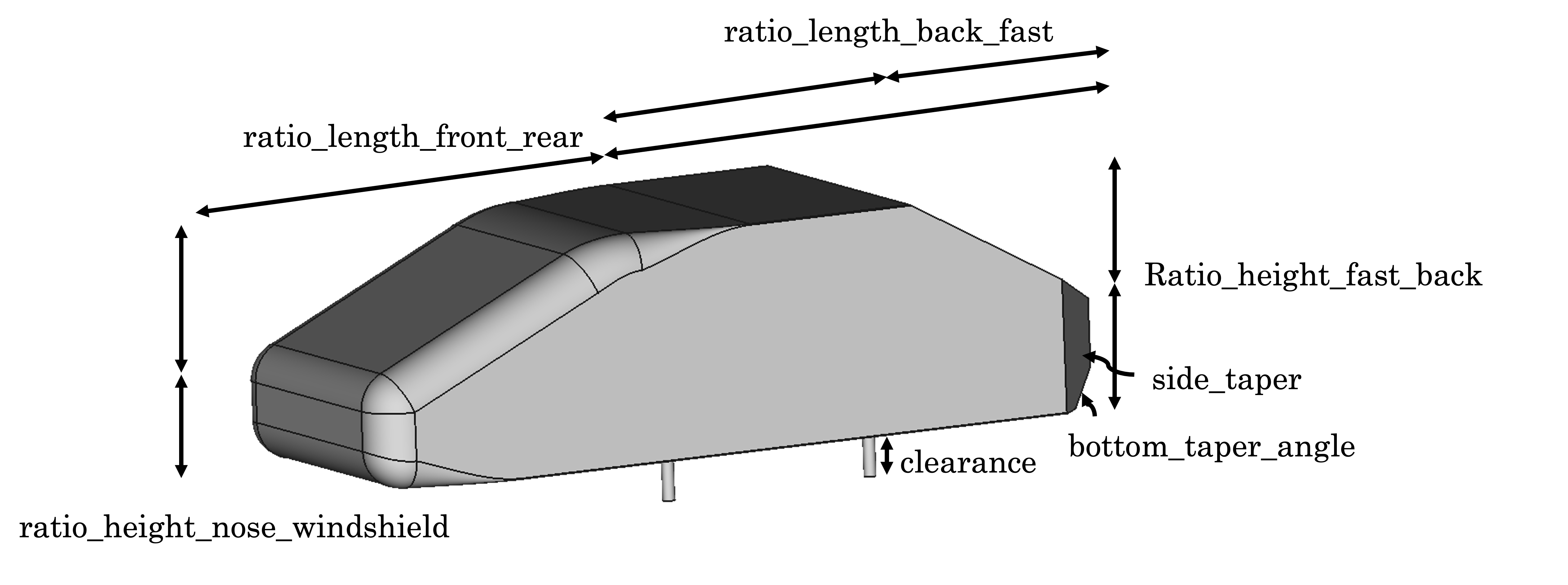}
  \caption{Parameterized CAD model}
  \label{fig:windsor-cad}
\end{figure}

The baseline geometry is transferred into a parameterized CAD model to allow for automated sampling in the design space. The chosen sampling algorithm is a Halton sequence, which results in a quasi-random low discrepancy sequence with the advantages of being deterministic and extendable. The parameters and respective ranges are shown in Figure \ref{fig:windsor-cad} and Table \ref{geo-table}. The parameters were chosen to provide a suitable range of geometries that would exhibit different flow patterns e.g pressure vs geometry induced separation. The min/max values of each value were based upon engineering judgement and to avoid completely unrealistic shapes as well as avoid invalid shapes during geometry creation. The choice of 355 geometries  were also based upon a mixture of computational budget as well as matching what may be possible within industry i.e. how many geometries would realistically be generated within an engineering company. Future work could be to expand this dataset further if required.

\begin{figure}[htb]
    \begin{subfigure}[b]{0.48\textwidth}
        \centering
        \includegraphics[width=1\textwidth]{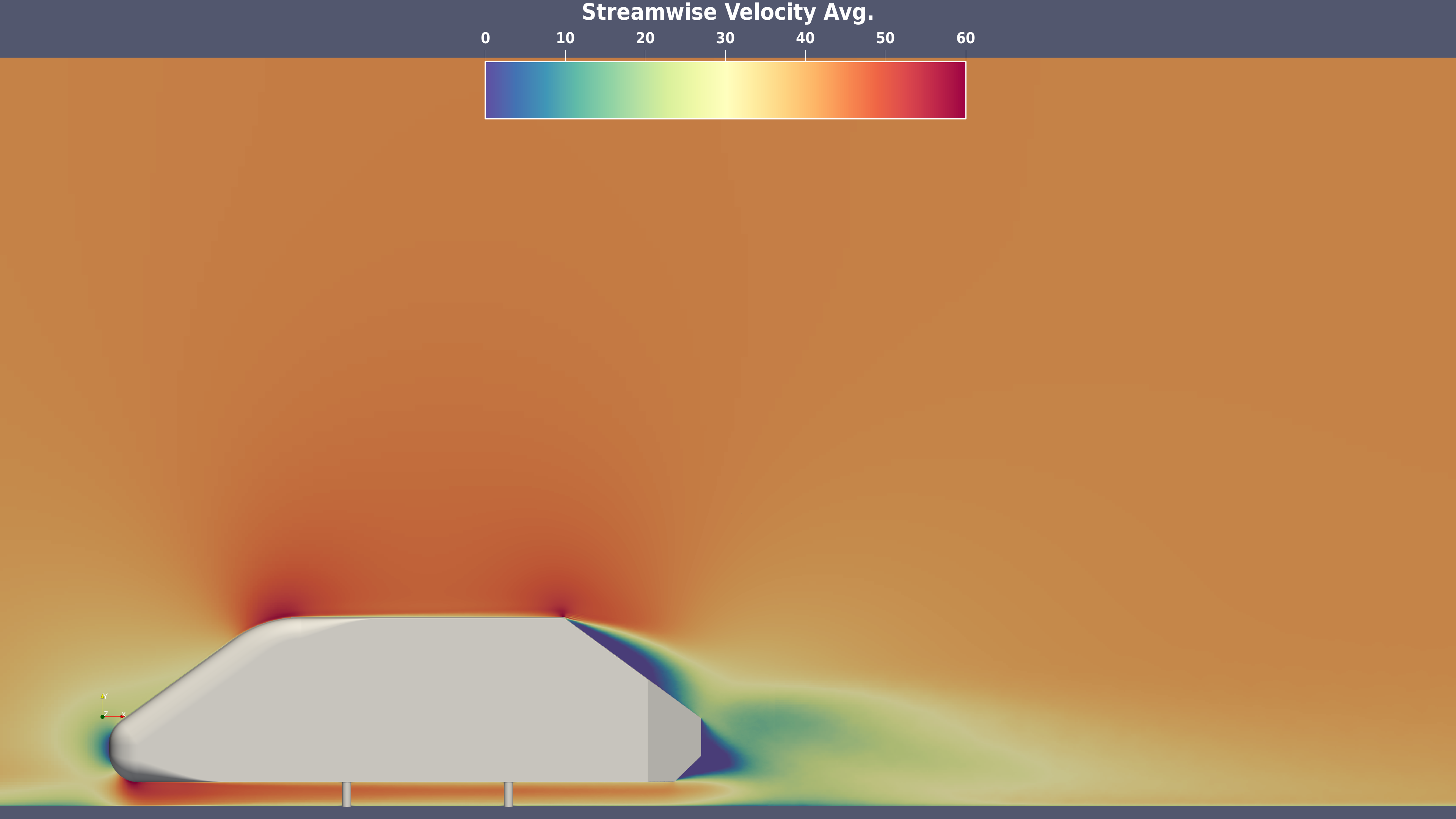}
        \caption{Mean streamwise velocity for high drag geometry variant example}
        \label{fig:run340}
    \end{subfigure}
    \hfill
    \begin{subfigure}[b]{0.48\textwidth}
        \centering
        \includegraphics[width=1\textwidth]{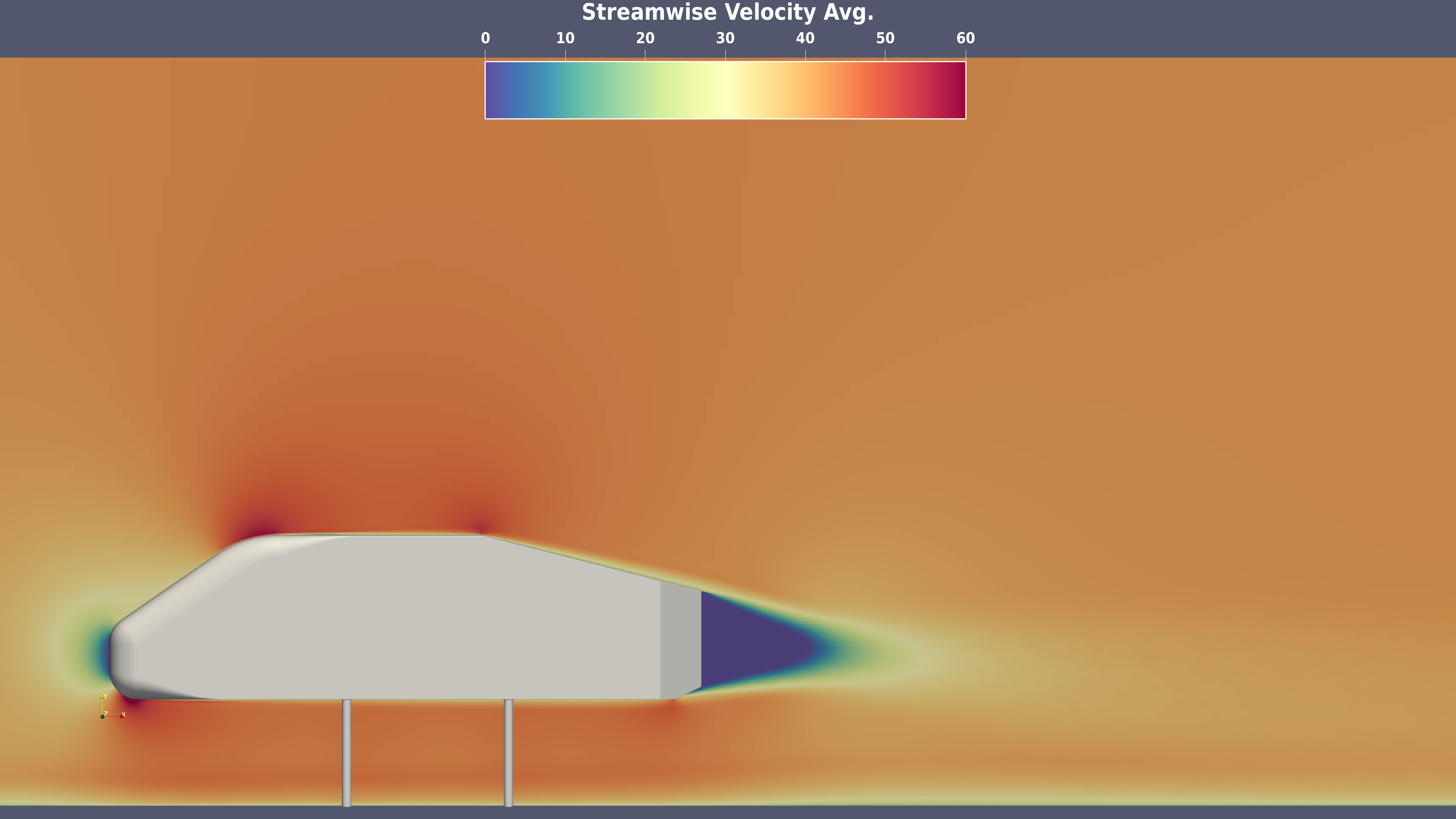}
        \caption{Mean streamwise velocity for low drag geometry variant example}
        \label{fig:run98}
    \end{subfigure}
    \hfill
    \begin{subfigure}[b]{0.48\textwidth}
        \centering
        \includegraphics[width=\textwidth]{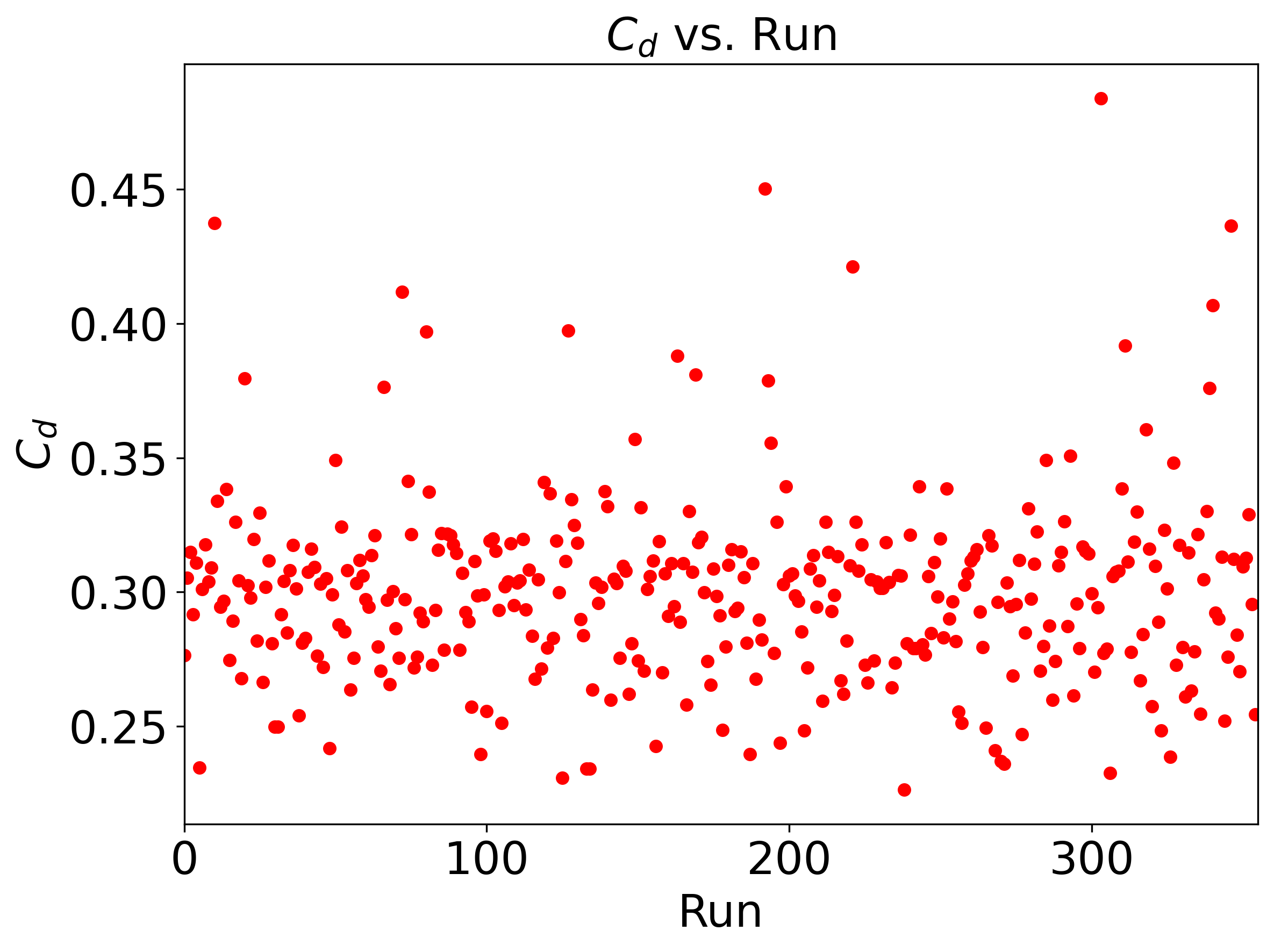}
        \caption{Variation of drag coefficient against run number}
        \label{fig:cdvar}
    \end{subfigure}
    \hfill
    \begin{subfigure}[b]{0.48\textwidth}
        \centering
        \includegraphics[width=\textwidth]{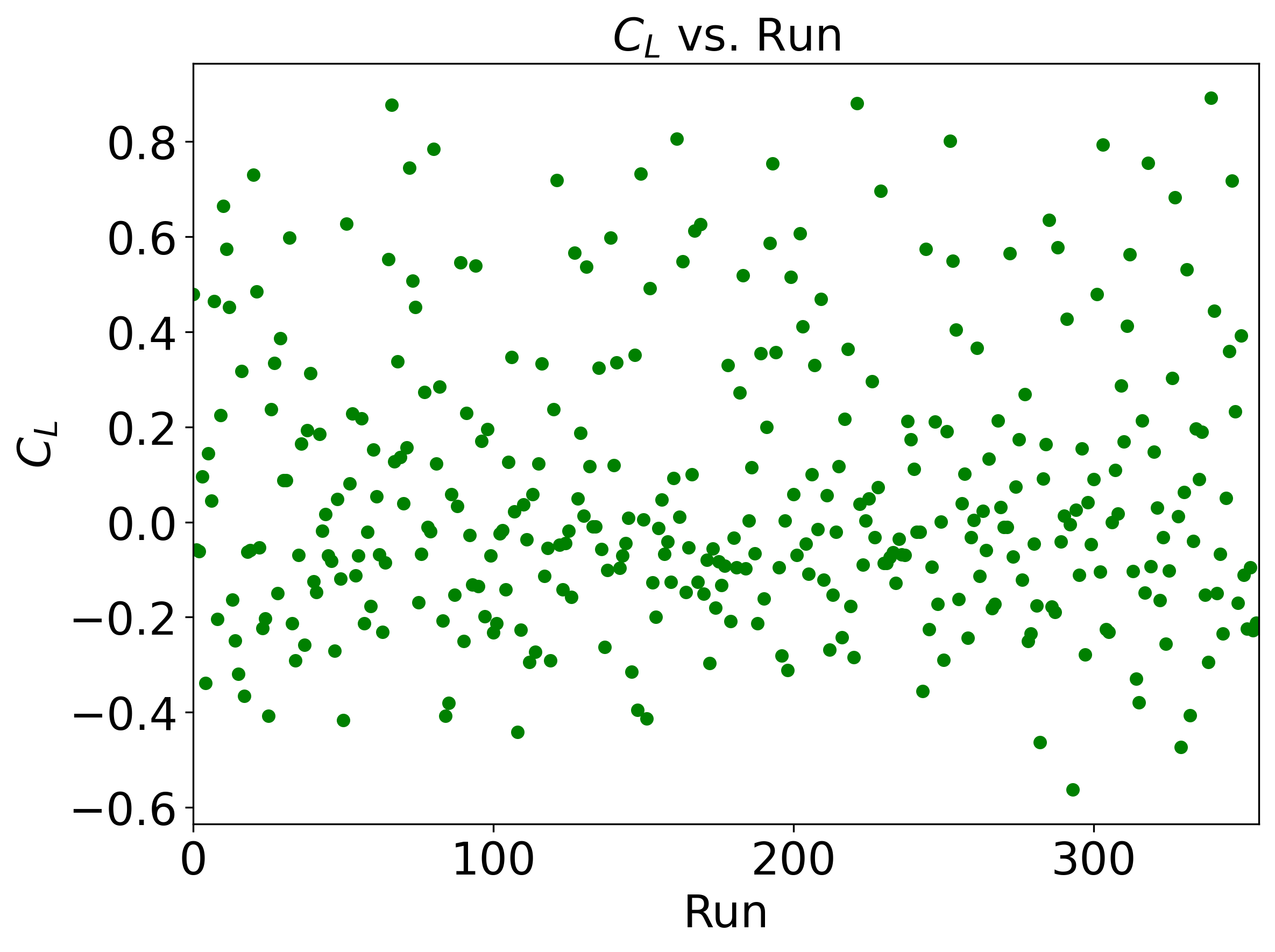}
        \caption{Variation of lift coefficient against run number}
        \label{fig:clvar}
    \end{subfigure}
    \caption{Variation of mean streamwise velocity and force coefficients across a sample of the dataset}
    \label{fig:datasetexample} 
    \end{figure}

The geometry choice within the dataset results in a broad range of flow physics - that is partly illustrated in Figures \ref{fig:cdvar} and \ref{fig:clvar} which show the large range of drag and lift coefficients across the dataset. Figures \ref{fig:run340} and \ref{fig:run98} take two examples from the dataset that shows the resulting flow-field (mean streamwise velocity) for a geometry producing a high drag coefficient and another one for a low drag coefficient (see SI for more detailed discussion of the dataset outputs).

\begin{table}[h]
  \caption{Geometry variants of the Windsor body}
  \label{geo-table}
  \centering
  \begin{tabular}{@{}lll@{}} \toprule
    Variable                              & Min    & Max  \\ \midrule
    {\tt ratio\_length\_front\_rear}      & 0     & 0.8     \\
    {\tt ratio\_length\_back\_fast}       & 0.08    & 0.5      \\
    {\tt ratio\_height\_nose\_windshield} & 0.3    & 0.7  \\
    {\tt ratio\_height\_fast\_back}       & 0    & 0.9  \\
    {\tt side\_taper [mm]}                     & 50    & 100  \\
    {\tt clearance [mm]}                       & 10     & 200  \\ 
    {\tt bottom\_taper\_angle [°]}            & 1     & 50  \\  \bottomrule
  \end{tabular}
\end{table}

\subsection*{How to access the dataset}
The dataset is open-source under the CC-BY-SA license \footnote{https://creativecommons.org/licenses/by-sa/4.0/)} and available to download via Amazon S3. In order to download, no AWS account is required and the full details are provided in the SI, the dataset README \footnote{https://caemldatasets.s3.us-east-1.amazonaws.com/windsor/dataset/README.txt} and the website \footnote{https://caemldatasets.org}. Further mirroring sites are being explored and details will be on the website once available. 

\subsection*{Dataset description}

The dataset follows the same structure as two concurrently developed datasets; AhmedML \cite{ashton2024ahmed} and DrivAerML \cite{ashton2024drivaer}. A summary of the dataset is provided in Table \ref{output-table2} (see SI for full details). The purpose of the dataset is to provide a rich dataset to support the development of a broad range of potential ML approaches. For this reason, we provide all possible outputs such as the full volume flow-field, the boundary surface, images of the flow-field as well as the time-averaged force and moment coefficients. All the outputs are provided in commonly used open-source formats (vtp, vtu, etc.) which  allow for the the broadest compatibility and ease of use for developers.   

\subsection*{Limitations of the dataset}
Whilst this high-fidelity, large-scale dataset has numerous benefits over prior work that was based upon lower-fidelity methods, there are a number of limitations:

\begin{itemize}
\item The dataset has purely geometrical differences with no variation in boundary conditions. Extending the dataset to include boundary condition changes, such as the inflow velocity, would help ML developers to use this dataset for more than just geometry prediction.
\item Whilst the Windsor body improves over the Ahmed car body it lacks the complexity of a real-life vehicle (addressed by the upcoming related DrivAerML dataset \cite{ashton2024drivaer}) 
\item The dataset only includes time-averaged data rather than time-series. Future work could be to extend the dataset to also include a limited number of transient outputs i.e volume/boundary outputs at each time-step.
\end{itemize}

\begin{table}[htp]
  \caption{Summary of the dataset contents}
  \label{output-table2}
  \centering
\begin{tabular}{p{0.25\linewidth} | p{0.5\linewidth}}
    \toprule
    \cmidrule(r){1-2}
    Output     & Description  \\
    \toprule
    \multicolumn{2}{c}{Per run (inside each run\_i folder)}                   \\
    \cmidrule(r){1-2}
    windsor\_i.stl & surface mesh of the windsor body geometry  \\
    \midrule
    windsor\_i.stp & surface CAD of the windsor body geometry  \\
    \midrule
    boundary\_i.vtu     &  time-averaged flow quantities (Pressure Coefficient, Skin-Friction Coefficient, $y^{+}$) on the Windsor car body surface    \\
    \midrule
    volume\_i.vtu     &  time-averaged flow quantities (pressure, velocity, Reynolds Stresses, Turbulent Kinetic Energy) within the domain volume \\
        \midrule
    force\_mom\_i.csv & time-averaged drag, \& lift, side-force and pitching moment coefficients using constant $A_{ref}$ \& $L_{ref}$ \\           \midrule
    force\_mom\_varref\_i.csv & time-averaged drag, \& lift, side-force and pitching moment coefficients using case-dependant $A_{ref}$ \\           \midrule
    geo\_parameters\_i.csv  & parameters that define the shape (in mm)  \\
\midrule
    images  & folder containing .png images of Reynolds stresses, streamwise velocity and pressure in $X$, $Y$, $Z$ slices through the volume  \\
    \toprule
    \multicolumn{2}{c}{Other}                   \\
    \cmidrule(r){1-2}
    force\_mom\_all.csv  & time-averaged drag \& lift for all runs  \\
    \midrule
    force\_mom\_varref\_all.csv  & time-averaged drag \& lift using case-dependant $A_{ref}$ for all runs  \\
    \midrule
    geo\_parameters\_all.csv     & parameters that define the shape for all runs  \\   
    \bottomrule
  \end{tabular}
\end{table}

\section*{ML evaluation}

\begin{figure}[htb]
     \centering
     \begin{subfigure}[b]{0.49\textwidth}
         \centering
         \includegraphics[width=\textwidth]{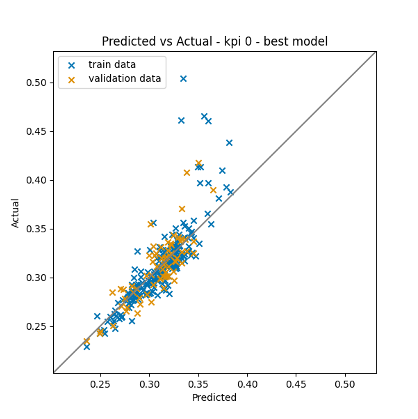}
         \caption{Predicted drag coefficient on training/validation split}
         \label{fig:mlcd1}
     \end{subfigure}
     \hfill
     \begin{subfigure}[b]{0.49\textwidth}
         \centering
         \includegraphics[width=\textwidth]{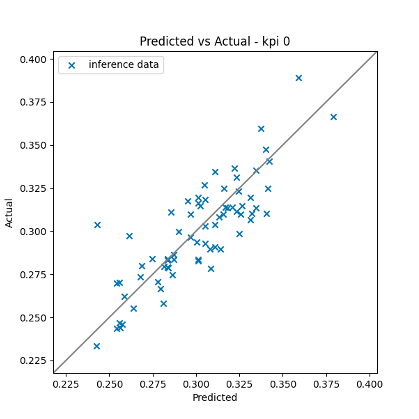}
         \caption{Predicted drag coefficient on test split}
         \label{fig:mlcd2}
     \end{subfigure}
              \caption{Prediction of the drag coefficient using the direct KPI method}
    \label{fig:mlpredictions0}
\end{figure}

We conducted an example ML evaluation using a Graph Neural Network (GNN) approach, based upon a modified version of MeshGraphNets \cite{pfaff2021learning} (more details in the SI) to demonstrate how this dataset could be used to train a ML model to predict unseen cases. We assess two approaches; firstly directly predicting the Key Performance Indicators (KPIs); in this case the drag and lift coefficient values, for each geometry using the lower resolution STL surface mesh (89k nodes) as an input, and secondly predicting the lift and drag coefficient through integration of the surface wall-shear stress and pressure from the higher resolution VTP surface mesh (4.4M nodes). Using the first method we find that using a 60/20/20 split of train, validation and test data, it is possible to obtain a MSE of less than 0.00028 for the drag coefficient and a MSE less than 0.0175 for the lift coefficient. An example of the prediction accuracy is shown in Figure \ref{fig:mlpredictions0} for the training, validation and test data for the drag coefficient. Using x8 Nvidia L40s GPUs (Amazon EC2 g6e.48xlarge instances via Amazon Web Services), the training completes in approximately 2hrs and the inference time for each new predicted geometry is 0.15 seconds. Additional results and further details of the ML setup are provided in the SI. 
Please note that these ML evaluations are preliminary and purely serve to illustrate how this dataset can be used for ML evaluation. We hope other groups will use this dataset to do a more thorough evaluation of different ML methodologies. 

%% file: sections/conclusions.tex
\section*{Conclusions}

In this paper, we present a new large-scale, open-source Computational Fluid Dynamics dataset based upon 355 geometry variants of the automotive Windsor body. 
The flow conditions are fixed and seven geometric parameters of interest to automotive designers are varied. The CFD simulations use a high-fidelity, time-accurate scale-resolving approach to ensure accurate prediction of the underlying flow structures. The dataset uses widely used open-source formats and contains the vehicle geometries, time averaged integrated forces, time averaged vehicle surface data and time average volumetric flowfield data. It is hoped that this open-source dataset will allow for faster development of data-driven and physics-driven ML approaches for vehicle aerodynamics prediction.

%% file: sections/appendix-methodology.tex
\section{Numerical methodology}
Volcano ScaLES, a GPU-based Large Eddy Simulation (LES) solver developed by Volcano Platforms Inc. is utilized to generate this database. Volcano ScaLES solves the compressible Navier-Stokes equations on Cartesian Octree grids with solid geometries discretized using an immersed boundary method suitable for high Reynolds number flows. In this section, we cover the basic formulation along with brief details about the closure modeling being employed.  
\subsection{Large Eddy Simulations}
Since a compressible Navier-Stokes formulation is utilized, Favre-averaged (density weighted) flow variables are solved
\begin{equation}
    \tilde{f} = \frac{\overline{\rho f}}{\overline{\rho}}
\end{equation}
where the $\overline{\cdot}$ denotes the standard LES low-pass filtering operator (with a length scale proportional to the grid spacing, $\Delta$), and $\rho$ denotes the density field. Using this combination of Favre-averaging and low-pass filtering, the conservation of mass equation can be written as
\begin{equation}
    \frac{\partial \overline{\rho}}{\partial t} + \frac{\partial}{\partial x_i}\left(\overline{\rho} \tilde{u}_i \right) = 0
\end{equation}
where $\tilde{u}_i$ is the Favre-filtered velocity field. 
Similarly, the momentum equation is given by 
\begin{equation}
    \frac{\partial \overline{\rho}\tilde{u}_i}{\partial t} + \frac{\partial}{\partial x_j}\left(\overline{\rho} \tilde{u}_i \tilde{u}_j \right) + \frac{\partial \overline{p}}{\partial x_i} - \frac{\partial{\hat{\sigma}_{ij}}}{\partial x_j} = \frac{\partial R_{ij}}{\partial x_j}
    \label{eq:mom}
\end{equation}
where $\overline{p}$ is the low-pass filtered static pressure and the resolved viscous stress tensor is given by: 
\begin{equation}
    \hat{\sigma}_{ij} = 2\mu(\tilde{T}) \left(\frac{1}{2}\left(\frac{\partial \tilde{u}_i}{\partial x_j} + \frac{\partial \tilde{u}_j}{\partial x_i} \right) - \frac{1}{3}\delta_{ij}\frac{\partial \tilde{u_k}}{\partial x_k} \right)
\end{equation}
Note that the dynamic viscosity, $\mu$ can be expressed as a function of Favre-filtered static temperature using an appropriate law (such as the Sutherland's law). However, in the present work, this dynamic viscosity has be assumed to a spatio-temporal constant with a value determined to achieve the required Reynolds number. Static temperature, $\tilde{T}$ is a function of the local pressure and density in accordance to the ideal gas law using a gas constant, $R$: 
\begin{equation}
    \tilde{T} = \frac{\overline{\rho  T}}{\overline{\rho}} = \frac{\overline{P}}{R \overline{\rho}}
\end{equation}
The term on the right-hand side of Equation \ref{eq:mom} is the un-closed LES term that needs to be modelled,
\begin{equation}
    R_{ij} = \overline{\rho}\left(\tilde{u}_i \tilde{u}_j - \reallywidetilde{u_i u_j}\right)
\end{equation}
Note that the residual stress given above assumes that $\hat{\sigma}_{ij} - \tilde{\sigma}_{ij}$ is negligible; this is true in high-Reynolds number flows and is exact for constant viscosity flows. Vreman's constant coefficient algebraic model\cite{vreman2004eddy} is used to close residual stress tensor:
\begin{equation}
    R_{ij} = f_\text{vreman}\left(c_\text{vreman}, \Delta, \overline{\rho}, \frac{\partial \tilde{u}_i}{\partial x_j}\right) 
\end{equation}

The governing equation for conservation of total energy is given as 
\begin{equation}
    \frac{\partial \hat{e}}{\partial t} + \frac{\partial}{\partial x_j}\left((\hat{e} + \overline{p})\tilde{u}_j)\right) + \frac{\partial \hat{q}_j}{\partial x_j} - \frac{\partial}{\partial x_j}\left(\hat{\sigma}_{ij} u_i\right) = R_e
\end{equation}
where the unclosed terms on the right hand side represent interactions between subgrid-scale energy and the resolved scale energy. Since these terms are only relevant at high Mach numbers, the residual terms in the energy equation are modelled via just a turbulent Prandtl number and the Vreman's eddy viscosity model used for the momentum equations. The positive definiteness of the SGS eddy viscosity ensures that the numerical model for the filtered energy equation is purely dissipative, i.e. it assumes that the energy is exclusively transferred from the large scales to the small scales of motion.  
\begin{equation}
    R_e = f^e_\text{vreman}\left(c_\text{vreman}, \text{Pr}_\text{turb},\Delta,\overline{\rho}, \frac{\partial \tilde{u}_i}{\partial x_j}, \frac{\partial \tilde{T}}{\partial x_j}\right)
\end{equation}

The resolved total energy, $\hat{e}$ can be expressed exclusively in terms of other resolved field variables as 
\begin{equation}
    \hat{e} = \frac{\overline{p}}{\gamma - 1} + \frac{1}{2}\overline{\rho}\tilde{u}_i \tilde{u}_i
\end{equation}

\subsection{Discretization}
Volcano ScaLES solves the governing LES equations using a nominally 4th order spatially accurate
finite difference discretization with favorable Kinetic Energy and Entropy Consistency properties making it suitable for
Large Eddy Simulations (LES) over a vast range of flow regimes. The viscous flux discretization utilizes a mix of 4th
order and 2nd order discretizations with high spectral bandwidth thereby allowing further robustness for high Reynolds number flows without additional numerical dissipation via the inviscid flux numerical operator in turbulence resolving
regions. Physics-based numerical sensors that are functions of local velocity gradients as well as pressure and density
fluctuations allow for spatio-temporally localized use of limiters needed to capture flows with discontinuities such as
shocks and steep density gradients. 

A key advantage of Cartesian grids with unit aspect ratio is
the use of explicit time-stepping scheme even in flows involving highly complex geometries; the perfectly isotropic
grid cells near the geometry do not introduce any numerical stiffness associated with complex grid topologies. There
are other significant advantages of Cartesian meshes including significantly reduced memory requirements for storing
the grid as well as efficient and high accuracy in the underlying numerical formulation. The popular Strong Stability
Preserving (SSP) variant of the classical 3rd order Runge–Kutta scheme developed by Gottlieb \& Shu\cite{gottlieb1998total} is utilized for
all the work presented in this paper.
Geometries are represented in the numerical formulation using a ghost-cell immersed boundary algorithm capable
of enforcing inviscid no-penetration boundary conditions as well as viscous skin friction with appropriate Reynolds
number asymptotic properties using an equilibrium wall-model. The wall-model that provides a shear-stress constraint
at the wall uses solution information interpolated via probing at a fixed distance of $3/2\Delta$ into the fluid away from
walls. 

%% file: sections/appendix-dataset-windsor.tex
\clearpage
\section{Dataset}
\label{app:dataset}
\subsection{Licensing terms}
\label{app:dataset-license}

The dataset is provided with the Creative Commons CC-BY-SA v4.0 license\footnote{\url{https://creativecommons.org/licenses/by-sa/4.0/deed.en}}. A full description of the license terms is provided under the following URL:

\url{https://caemldatasets.s3.us-east-1.amazonaws.com/windsor/dataset/LICENSE.txt}
\subsection{Access to dataset}
\label{app:dataset-access}

The dataset is hosted on Amazon Web Services (AWS) via an Amazon S3 bucket 

\url{s3://caemldatasets/windsor/dataset}

The dataset README.txt will be kept up to date for any changes to the dataset and can be found at the following URL:

\url{https://caemldatasets.s3.us-east-1.amazonaws.com/windsor/dataset/README.txt}

The dataset itself can be downloaded via the AWS Command Line Interface (CLI) tool, which is free-of-charge. An instruction about how to install the AWS CLI tool is given here: \url{https://docs.aws.amazon.com/cli/latest/userguide/getting-started-install.html}. After installing AWS CLI, please follow the example instructions in the README.txt on ways to download the data from Amazon S3 e.g to download all the files you can run the following (however note the dataset size is more than 20TB in total) : 

\begin{verbatim}
  aws s3 cp --recursive s3://caemldatasets/windsor/dataset .
\end{verbatim}

Note 1 : If you don't have an AWS account you will need to add --no-sign-request within your AWS command i.e aws s3 cp --no-sign-request --recursive etc... \\
Note 2 : If you have an AWS account, please note the bucket is in us-east-1, so you will have the fastest download if you have your AWS service or EC2 instance running in us-east-1.

\subsection{Long-term hosting/maintenance plan}
The data is hosted on Amazon S3 as it provides high durability, fast connectivity and is accessible without first requesting an account or credentials (only AWS CLI tools, described above, which are free to download and use). A dedicated website has been created for this dataset and the two associated datasets; AhmedML \cite{ashton2024ahmed} and DrivAerML datasets \cite{ashton2024drivaer} at https://caemldatasets.org to help further clarify where the data is hosted and to communicate any additional mirroring sites.   

\subsection{Intended use}
The dataset was created with the following intended uses:

\begin{itemize}
\item Development and testing of data-driven or physics-driven ML surrogate models for the prediction of surface, volume and/or force coefficients
\item a stepping stone dataset between the simplier AhmedML \cite{ashton2024ahmed} and more complex DrivAerML dataset \cite{ashton2024drivaer} 
\item Large-scale dataset to study bluff-body flow physics
\end{itemize}

\subsection{DOI}

At present there is no specific DOI for the dataset (only for this associated paper) - however the authors are investigating ways of assigning DOI to this Amazon S3 hosted dataset.  

\subsection{Dataset contents}
\label{app:dataset-dataDesc}

For each geometry there is a separate folder (e.g \verb|run_1|, \verb|run_2|, ..., \verb|run_i|, etc.) , where "i" is the run number that ranges from 0 to 354. All run folders contain the same time-averaged data which is summarized in Table \ref{output-table3}.

\begin{table}[htp]
  \caption{Summary of the dataset contents}
  \label{output-table3}
  \centering
\begin{tabular}{p{0.25\linewidth} | p{0.5\linewidth}}
    \toprule
    \cmidrule(r){1-2}
    Output     & Description  \\
    \toprule
    \multicolumn{2}{c}{Per run (inside each run\_i folder)}                   \\
    \cmidrule(r){1-2}
    windsor\_i.stl & surface mesh of the windsor body geometry  \\
    \midrule
    windsor\_i.stp & surface CAD of the windsor body geometry  \\
    \midrule
    boundary\_i.vtu     &  time-averaged flow quantities (Pressure Coefficient, Skin-Friction Coefficient, $y^{+}$) on the Windsor car body surface    \\
    \midrule
    volume\_i.vtu     &  time-averaged flow quantities (pressure, velocity, Reynolds Stresses, Turbulent Kinetic Energy) within the domain volume \\
        \midrule
    force\_mom\_i.csv & time-averaged drag, \& lift, side-force and pitching moment coefficients using constant $A_{ref}$ \& $L_{ref}$ \\           \midrule
    force\_mom\_varref\_i.csv & time-averaged drag, \& lift, side-force and pitching moment coefficients using case-dependant $A_{ref}$  \\           \midrule
    geo\_parameters\_i.csv  & parameters that define the shape (in mm)  \\
\midrule
    images  & folder containing .png images of Reynolds stresses, streamwise velocity and pressure in $X$, $Y$, $Z$ slices through the volume  \\
    \toprule
    \multicolumn{2}{c}{Other}                   \\
    \cmidrule(r){1-2}
    force\_mom\_all.csv  & time-averaged drag \& lift for all runs  \\
    \midrule
    force\_mom\_varref\_all.csv  & time-averaged drag \& lift using case-dependant $A_{ref}$ for all runs  \\
    \midrule
    geo\_parameters\_all.csv     & parameters that define the shape for all runs  \\   
    \bottomrule
  \end{tabular}
\end{table}

\subsubsection{Cartesian to Unstructured output}
In order to simplify the reading of the dataset into existing tools, the Cartesian data was converted into a fully unstructured VTK mesh made of hexahedra. Solution data at cell centers of the hexahedra are averaged to cell corners accounting for block neighbor's data. This change makes the data easier to consume since VTK is widely used in the scientific community and tools to read and process these files are readily available (e.g. ParaView). 

\subsubsection{Time-Averaging}
The dataset only includes time-averaged data rather than time-series due to the extremely large number of time-steps that would be required to output i.e $> 100,000$. However future work could be to include this for a limited number of runs to help develop models to capture the time-history.
\subsubsection{Force Coefficients}
The drag, lift and side-force coefficients are defined as follows (please note that for all the simulations in this work $y$ is upwards, which differs to the original Windsor work where $z$ is upwards):
\begin{equation}
    C_\mathrm{D} = \frac{F_{x}}{0.5 \, \rho_\infty \, |U_\infty|^2 \, A_{ref}} \, , \quad
        C_\mathrm{L} = \frac{F_{y}}{0.5 \, \rho_\infty \, |U_\infty|^2 \, A_{ref}} \, , \quad
        C_\mathrm{S} = \frac{F_{z}}{0.5 \, \rho_\infty \, |U_\infty|^2 \, A_{ref}} \, , \quad
    \label{eq:forceCoeffs}
\end{equation}
Where $F$ is the integrated force, $A$ is the frontal area of the geometry, and  $\rho_\infty$ is the reference density (see Table \ref{tab:dataset-refCond} for reference values). 

The pitching moment is defined as:
\begin{equation}
    C_\mathrm{My} = \frac{M_{y}}{0.5 \, \rho_\infty \, |U_\infty|^2 \, A_{ref} \, L_{ref}} \, , \quad
    \label{eq:forceCoeffs2}
\end{equation}

Where $M$ is the integrated moment, $A_{ref}$ is the frontal area of the geometry, $L_{ref}$ is the reference length, and  $\rho_\infty$ is the reference density (see Table \ref{tab:dataset-refCond} for reference values). 

Two outputs are provided for the forces; one in which the reference area is kept constant across all details (force\_mom\_i.csv), and secondly one where it based upon the frontal-area of each different geometry variant (force\_mom\_varref\_i.csv).

\subsubsection{Boundary data}
\newcommand{\pref}{p_\text{ref}}
\newcommand{\qref}{q_\text{ref}}
\newcommand{\uref}{U_\text{ref}}
\newcommand{\dref}{\rho_\text{ref}}

\newcommand{\ptref}{\tilde{p}_\text{ref}}
\newcommand{\qtref}{\tilde{q}_\text{ref}}
\newcommand{\utref}{\tilde{U}_\text{ref}}
\newcommand{\dtref}{\rho_\text{ref}}

Time-averaged surface data for nondimensional force coefficients are provided.
The surface pressure coefficient $C_p = (p - \pref)/\qref$ where the
"freestream" dynamic pressure is $\qref = \frac{1}{2}\dref |\uref|^2$ and $\pref$ is
the "freestream" pressure. Similarly, $cf_x$, $cf_y$ and $cf_z$ correspond to
the wall-strear-stress tensor (units Pascal) projections in the three coordinate
directions normalized by $\qref$.  Lift and drag
coefficients can be computed via surface integration of the local
$C_p$, $cf_x$, etc. using a simple surface quadrature rule (rectangle-rule in the
present case). Note that lift and drag use the "reference area" to normalize the
differential area that shows up for surface integration.

As previously mentioned, the time-averaged quantities from the numerical sensor (measured at
$(x,y,z) = (\qty{-2}{\m}, \qty{1.3}{\m}, \qty{0}{\m})$) are used to normalize the force and pressure
coefficients for the baseline experimental setup, but not for this particular dataset (given there is no wind-tunnel data for the geometries). 

\begin{table}[htb]
    \caption{Nominal freestream conditions and reference dimensions.}
    \label{tab:dataset-refCond}
    \centering
    \begin{tabular}{@{}lr@{}} \toprule
        $\uref$                 & $40\;\si{m/s}$ \\
        $\dref$                 & $1.31\;\si{kg/m^3}$ \\
        $\pref$                 & $37330.4\;\si{Pa}$ \\
        $A_\mathrm{ref}$        &  $0.112\;\si{m^2}$\\
        $L_\mathrm{ref}$        &  $0.6375\;\si{m}$\\ \bottomrule
    \end{tabular}
\end{table}

\subsubsection{Volume data}

The ``outlet'' for the tunnel is not a domain boundary for these simulations even though the volume data does not extend further. 
 To minimize numerical reflections from the tunnel exit, the tunnel was allowed to vent into a large reservoir which is initialized with atmospheric conditions and given
 simple extrapolation boundary conditions far from the tunnel. This means there is no fixed pressure condition or any other boundary condition enforced at the tunnel exit.

 \subsubsection{Images}

 Table \ref{output-table-variables-images} describes the .png images that are created in $x$, $y$ and $z$ directions to either enable ML methods to directly predict them and/or provide a quick way to inspect the flow fields. 

\subsubsection{Data formats}
All provided data is in the open source format VTK (i.e. *.vtp and *.vtu) to ensure the broadest compatibility.

\clearpage
\begin{table}[htb]
    \caption{List of output quantities in the provided dataset files, all quantities are time-averaged.}
    \label{tab:dataset-output-quantities}
    \centering
    \begin{tabular}{p{0.08\linewidth} p{0.08\linewidth} p{0.35\linewidth} p{0.4\linewidth}}
        \toprule
        \multicolumn{4}{c}{\textbf{volume\_i.vtu}}                   \\
        Symbol                          & Units                 & Field name                                      & Description  \\
        \toprule
        $\overline{p^*}$                & $[\si{\m^2/\s^2}]$    & \texttt{pressureavg}                              & relative kinematic pressure \\
        \midrule
        $\overline{U_x}$                & $[\si{\m/\s}]$        & \texttt{velocityxavg}                              & velocity component in $x$ \\
        \midrule
        $\overline{U_y}$                & $[\si{\m/\s}]$        & \texttt{velocityyavg}                              & velocity component in $y$ \\
        \midrule
        $\overline{U_z}$                & $[\si{\m/\s}]$        & \texttt{velocityzavg}                              & velocity component in $z$ \\
            \midrule
        $\overline{u'_xu'_x}$           & $[\si{\m^2/\s^2}]$    & \texttt{reynoldsstressxx}                        & resolved Reynolds stress xx \\  
            \midrule
        $\overline{u'_yu'_y}$           & $[\si{\m^2/\s^2}]$    & \texttt{reynoldsstressyy}                        & resolved Reynolds stress yy \\ 
            \midrule
        $\overline{u'_zu'_z}$           & $[\si{\m^2/\s^2}]$    & \texttt{reynoldsstresszz}                        & resolved Reynolds stress zz \\ 
            \midrule
        $\overline{u'_xu'_y}$           & $[\si{\m^2/\s^2}]$    & \texttt{reynoldsstressxy}                        & resolved Reynolds stress xy \\ 
            \midrule
        $\overline{u'_xu'_z}$           & $[\si{\m^2/\s^2}]$    & \texttt{reynoldsstressxz}                        & resolved Reynolds stress xz \\
            \midrule
        $\overline{u'_yu'_z}$           & $[\si{\m^2/\s^2}]$    & \texttt{reynoldsstressyz}                        & resolved Reynolds stress yz \\
            \midrule
        $\overline{k}$              & $[\si{\m^2/\s^3}]$      & \texttt{tke}                            & turbulent kinetic energy \\  
        \toprule
        \multicolumn{4}{c}{\textbf{boundary\_i.vtu}}                   \\
        Symbol                          & Unit                  & Field name                                      & Description  \\
        \toprule
        $\overline{y^{+}}$    & $[ - ]$    & \texttt{yplusavg}                        & $y^{+}$ \\
            \midrule
        $\overline{C_fx}$             & $[ - ]$    & \texttt{cfxavg}                & skin-friction coefficient in $x$ \\  
            \midrule
        $\overline{C_fy}$             & $[ - ]$    & \texttt{cfyavg}                & skin-friction coefficient in $y$ \\ 
            \midrule
        $\overline{C_fz}$             & $[ - ]$    & \texttt{cfzavg}                & skin-friction coefficient in $z$ \\ 
            \midrule
        $\overline{C_p}$                & $[ - ]$               & \texttt{cpavg}                             & static pressure coefficient \\  
            \midrule
        $\overline{{C_p}^{2}}$                & $[ - ]$               & \texttt{cpvar}                             & static pressure variance coefficient \\ 
        \bottomrule
    \end{tabular}
\end{table}

\begin{table}[hbt]
  \caption{Description of files within the images folder }
  \label{output-table-variables-images}
  \centering
  \begin{tabular}{@{}ll@{}} \toprule
     Name            &  Description  \\ \midrule
    pressureavg (folder)      & time-averaged pressure \\   
    velocityxavg (folder)     & time-averaged $x$-component of the velocity  \\
    rstress\_xx (folder) & time-averaged $xx$-component of the Reynolds stress tensor \\ 
    rstress\_yy (folder) & time-averaged $yy$-component of the Reynolds stress tensor \\ 
    rstress\_zz (folder) & time-averaged $zz$-component of the Reynolds stress tensor \\
    windsor\_i.png              & picture of the geometry itself \\ \bottomrule
     view1\_constz\_scan\_0000-0009.png              & 10 slices of the above variables in z from z= -0.4 to 0.4  \\
     view2\_constx\_scan\_0000-0079.png              & 80 slices of the above variables in x from x=-0.5 to 1.0  \\
     view3\_consty\_scan\_0000-0019.png              & 20 slices of the above variables in y from y=0.03 to 0.55  \\
     \bottomrule   
  \end{tabular}
\end{table}

\clearpage
\subsection{Geometry variants}
\label{app:dataset-geomVars}

\begin{figure}[h]
  \centering
\includegraphics[width=1\textwidth]{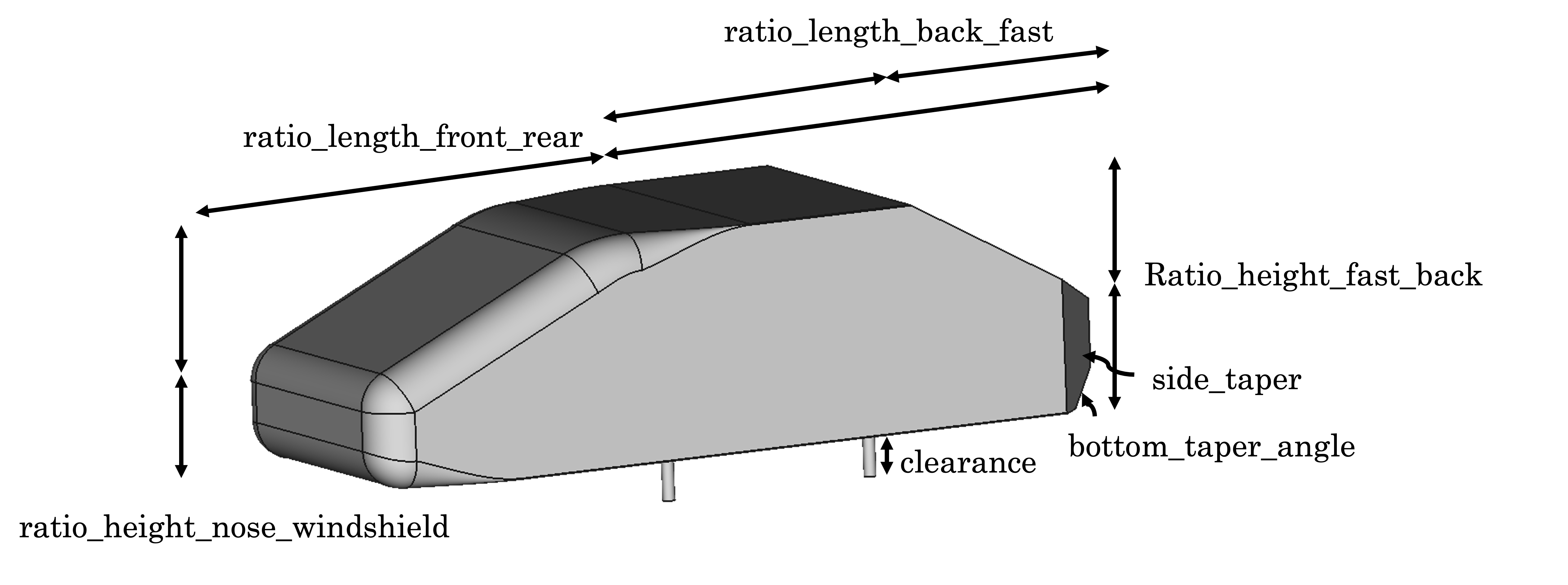}
  \caption{Parameterized CAD model}
  \label{fig:windsor-cad2}
\end{figure}

The baseline geometry is transferred into a parameterized CAD model to allow for automated sampling in the design space. The chosen sampling algorithm is a Halton sequence, which results in a quasi-random low discrepancy sequence with the advantages of being deterministic and extendable. The parameters and respective ranges are shown in Figure \ref{fig:windsor-cad2} and Table \ref{geo-table2}. The parameters were chosen to provide a suitable range of geometries that would exhibit different flow patterns e.g pressure vs geometry induced separation. The min/max values of each value were based upon engineering judgement and to avoid completely unrealistic shapes as well as avoid invalid shapes during geometry creation. The choice of 355 geometries  were also based upon a mixture of computational budget as well as matching what may be possible within industry i.e how many geometries would realistically be generated within an engineering company. Future work could be to expand this dataset further if required.

\begin{figure}
    \centering
    \includegraphics[width=0.8\textwidth]{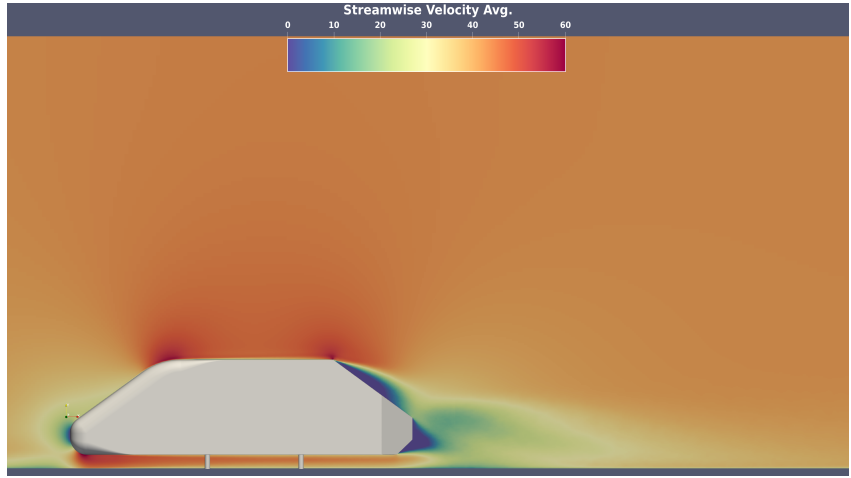}
\caption{Mean streamwise velocity for high drag geometry variant example}
    \label{fig:run3402}
\end{figure}

\begin{figure}
    \centering
    \includegraphics[width=0.8\textwidth]{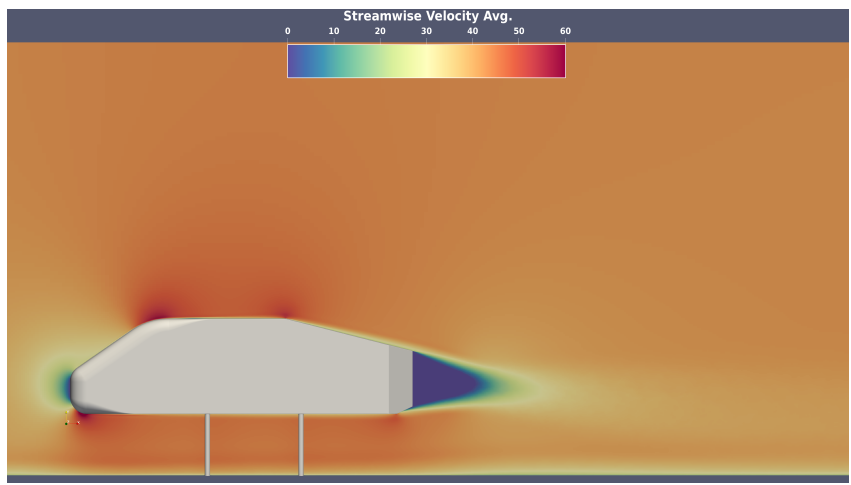}
\caption{Mean streamwise velocity for low drag geometry variant example}
    \label{fig:run982}
\end{figure}

\begin{figure}
    \centering
    \includegraphics[width=0.8\textwidth]{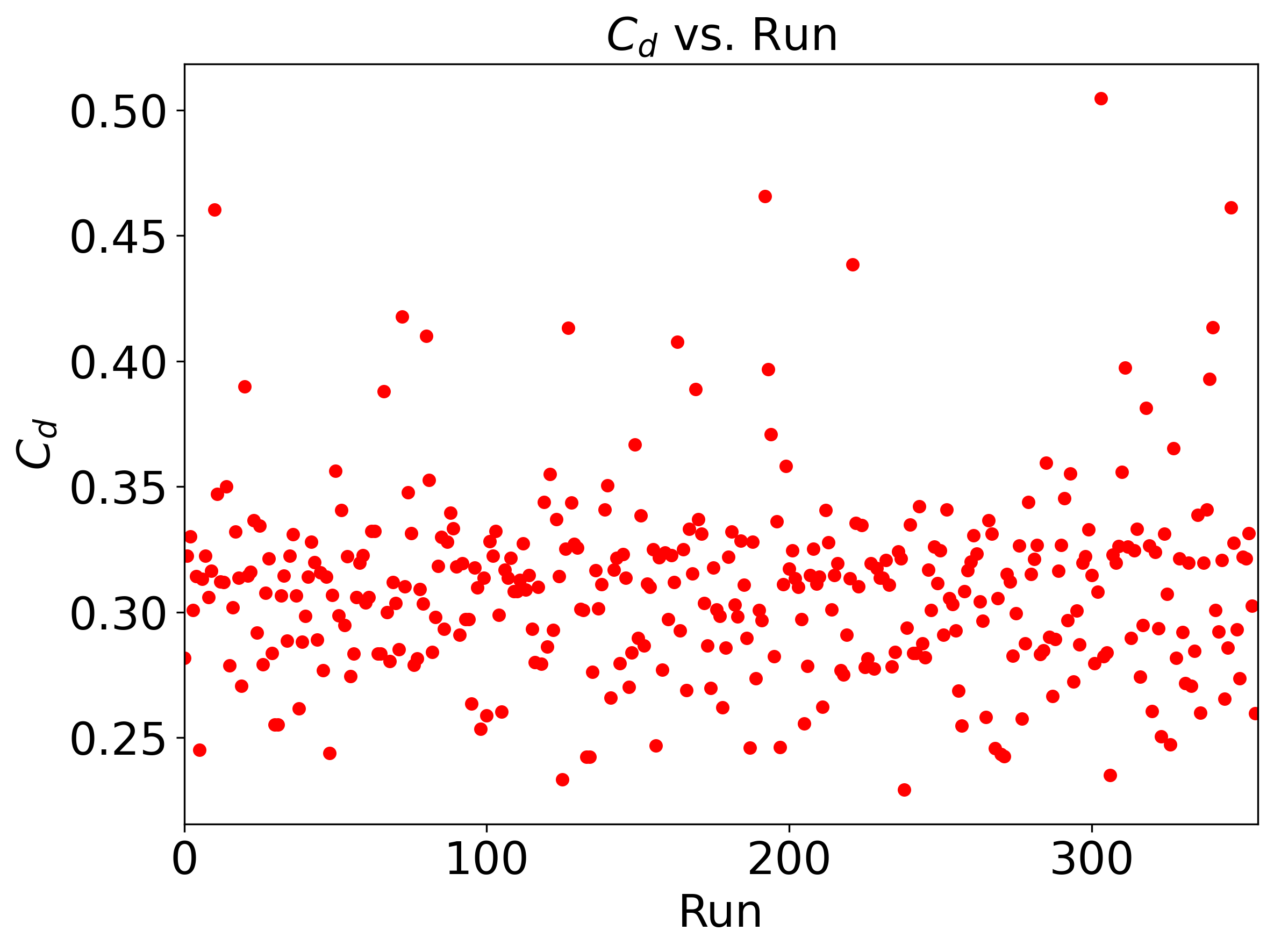}
\caption{Variation of drag coefficient against run number}
    \label{fig:cdvar2}
\end{figure}

\begin{figure}
    \centering
    \includegraphics[width=0.8\textwidth]{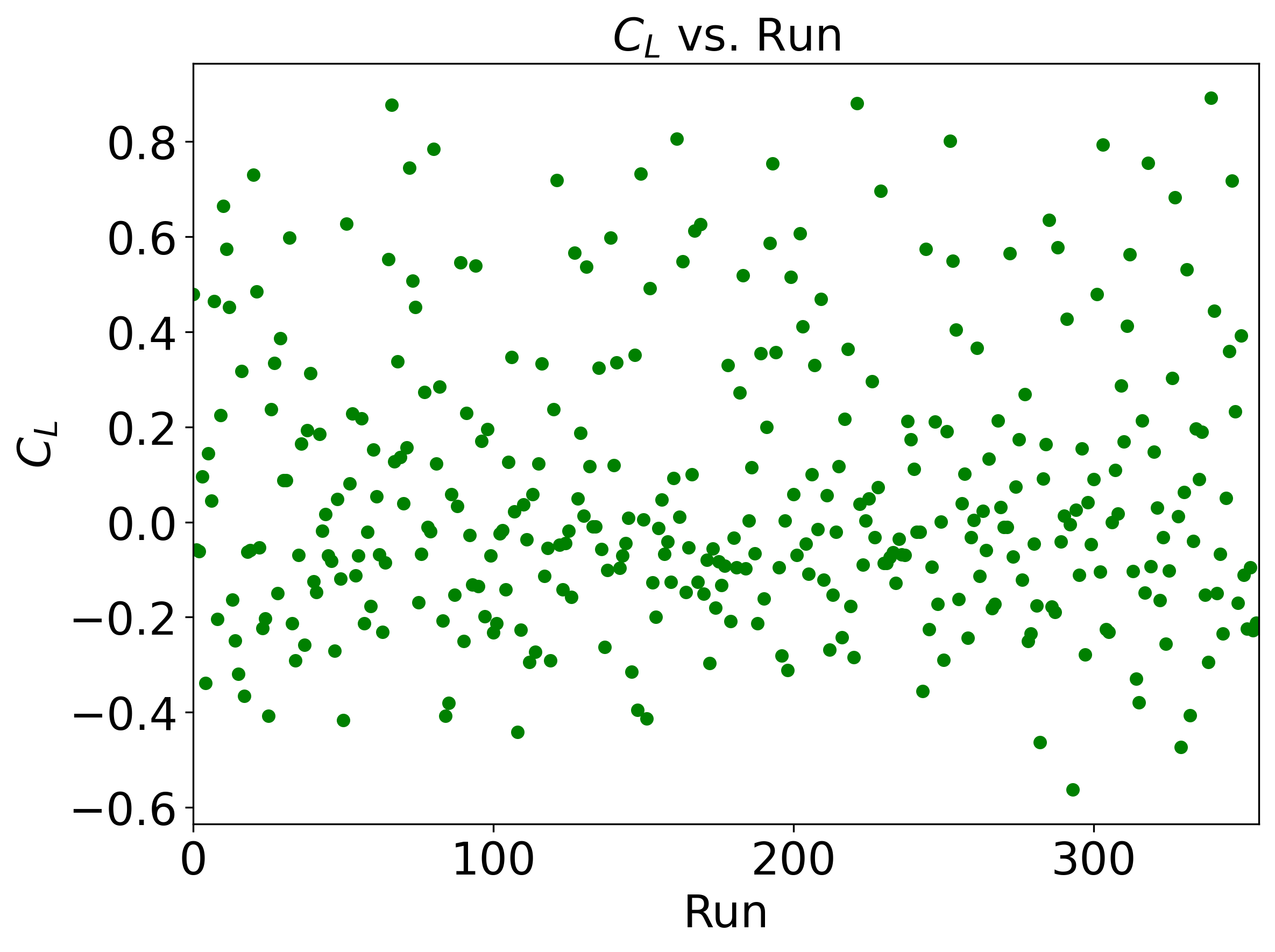}
\caption{Variation of lift coefficient against run number}
    \label{fig:clvar2}
\end{figure}

The geometry choice within the dataset results in a broad range of flow physics - that is partly illustrated in Figures \ref{fig:cdvar2} and \ref{fig:clvar2} which show the large range of drag and lift coefficients across the dataset. Figures \ref{fig:run3402} and \ref{fig:run982} take two examples from the dataset that shows the resulting flow-field (mean streamwise velocity) for a geometry producing a high drag coefficient and another one for a low drag coefficient. 

\begin{table}[h]
  \caption{Geometry variants of the Windsor body}
  \label{geo-table2}
  \centering
  \begin{tabular}{@{}lll@{}} \toprule
    Variable                              & Min    & Max  \\ \midrule
    {\tt ratio\_length\_front\_rear}      & 0     & 0.8     \\
    {\tt ratio\_length\_back\_fast}       & 0.08    & 0.5      \\
    {\tt ratio\_height\_nose\_windshield} & 0.3    & 0.7  \\
    {\tt ratio\_height\_fast\_back}       & 0    & 0.9  \\
    {\tt side\_taper [mm]}                     & 50    & 100  \\
    {\tt clearance [mm]}                       & 10     & 200  \\ 
    {\tt bottom\_taper\_angle [°]}            & 1     & 50  \\  \bottomrule
  \end{tabular}
\end{table}

\begin{figure}[hbt!]
    \centering
    \includegraphics[width=\textwidth]{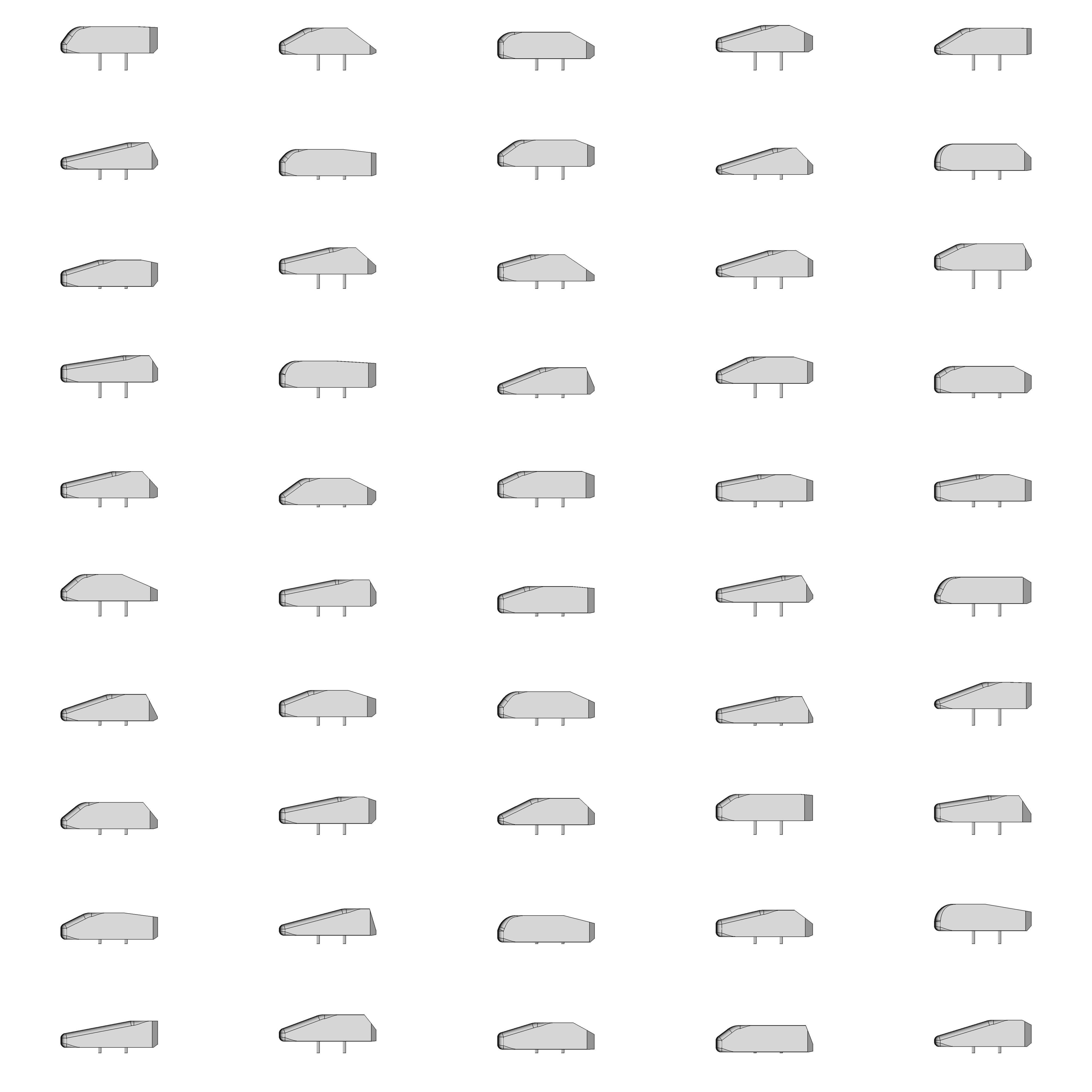}
\caption{Image of the geometry for runs 1 to 50}
    \label{fig:geo50}
\end{figure}

\begin{figure}[hbt!]
    \centering
    \includegraphics[width=\textwidth]{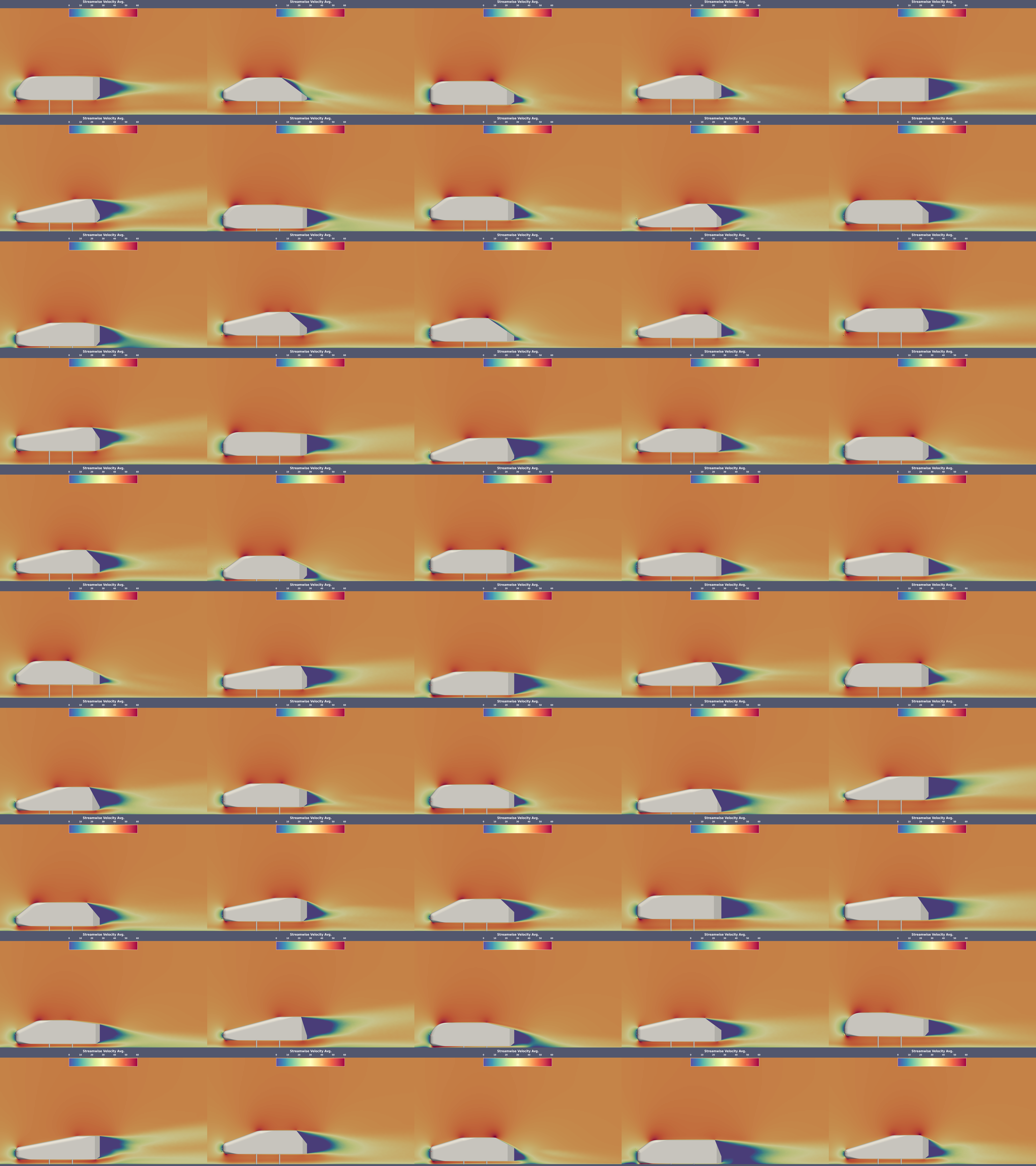}
\caption{Mean Streamwise velocity for runs 1 to 50 at $z=0$}
    \label{fig:velxz50}
\end{figure}

\begin{figure}[hbt!]
    \centering
    \includegraphics[width=\textwidth]{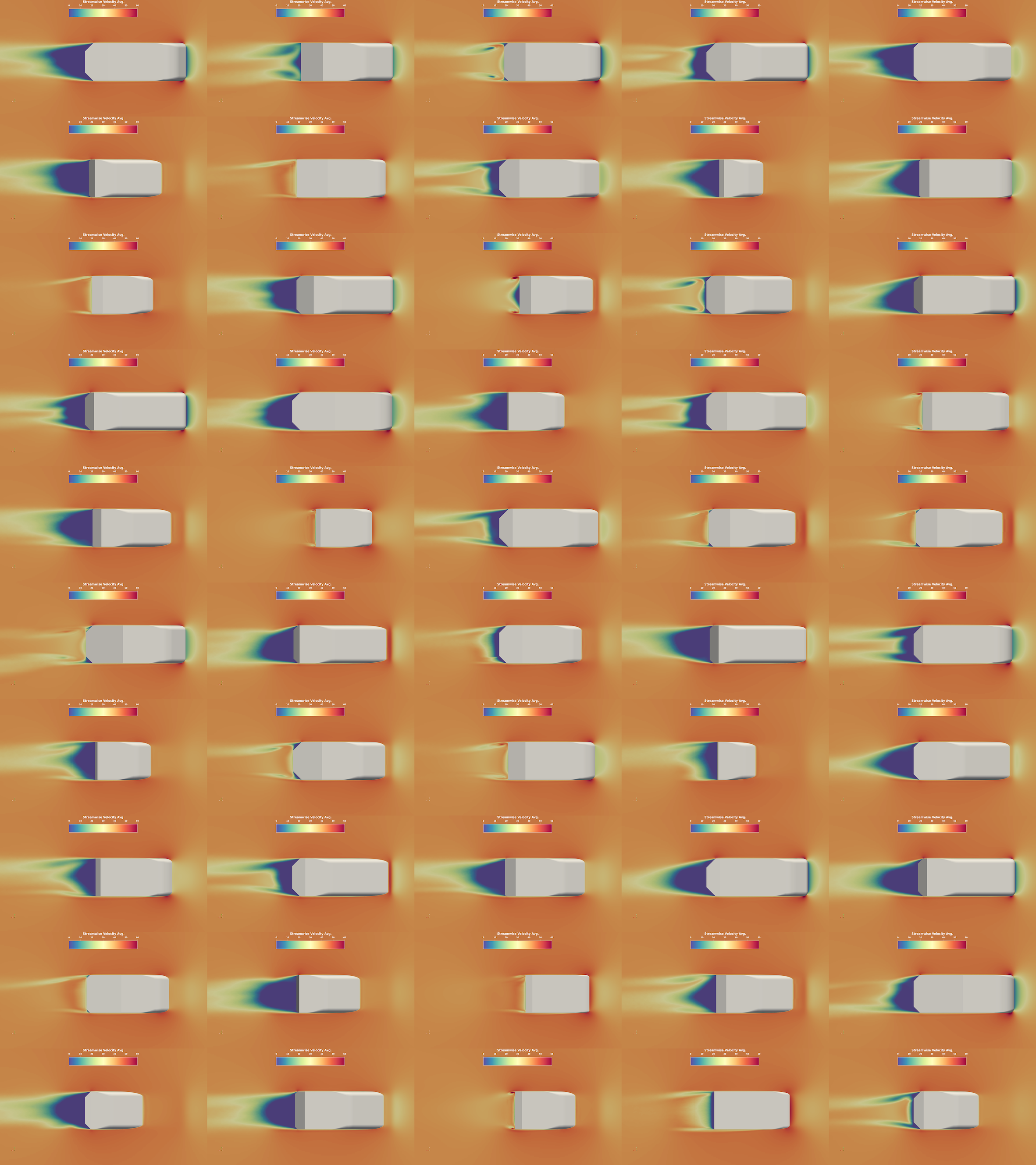}
\caption{Mean Streamwise velocity for runs 1 to 50 at $y=0$}
    \label{fig:velxy50}
\end{figure}

\begin{figure}[hbt!]
    \centering
    \includegraphics[width=\textwidth]{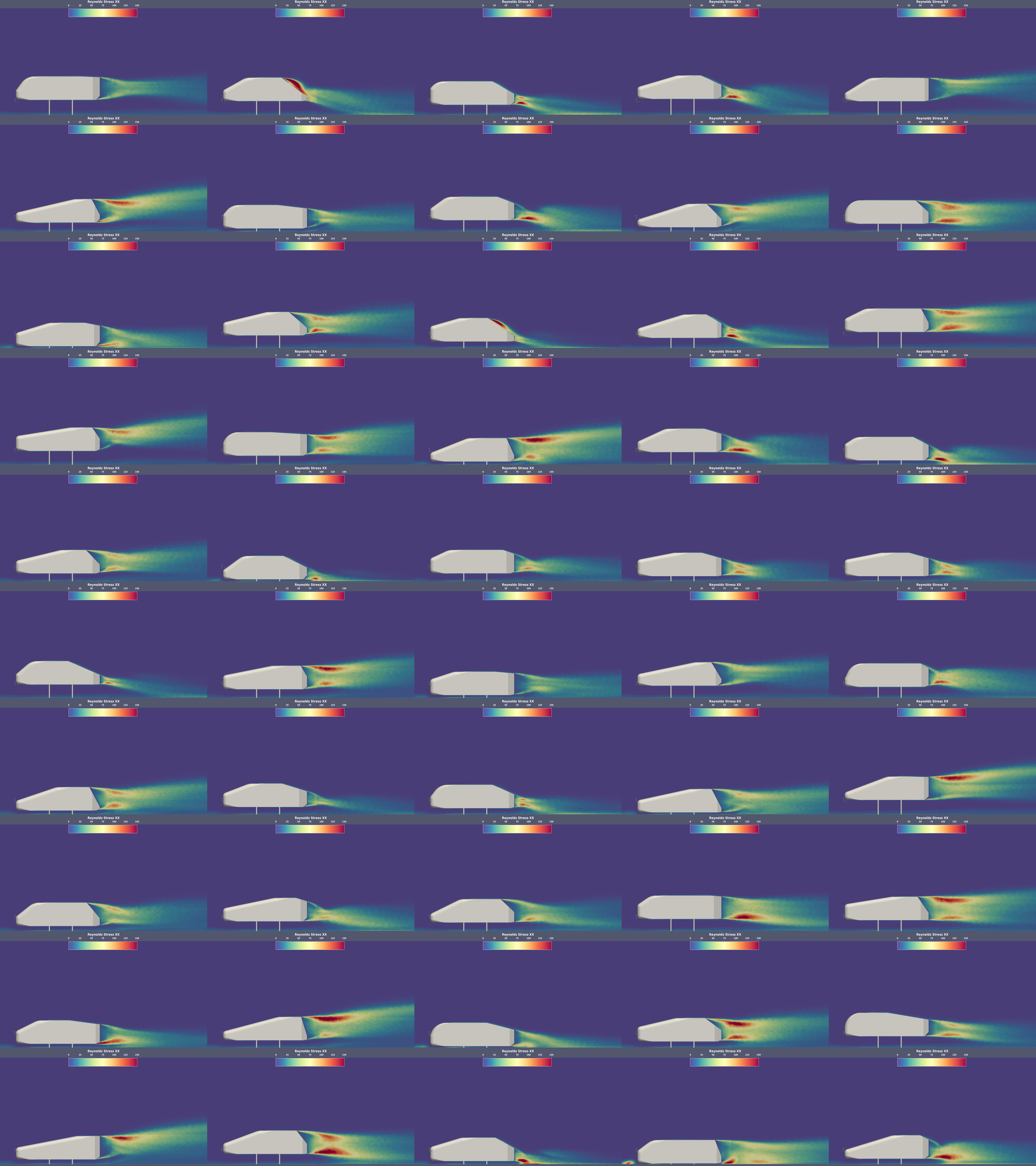}
\caption{Mean Reynolds Stress (xx) for runs 1 to 50 at $z=0$}
    \label{fig:rstressxx50}
\end{figure}

Figures \ref{fig:geo50}, \ref{fig:velxz50}, \ref{fig:velxy50}, \& \ref{fig:rstressxx50} are provided to give an illustration of the range of flow features and post-processing that makes this a rich dataset for the development and testing of ML methods.

\newpage

%% file: sections/appendix-validation.tex
\clearpage

\section{Additional Validation details}
This sections provide some additional validation details for the Windsor body
case left out of the main paper. The main paper showed grid convergence results for
the drag coefficient which motivated our choice of the $G_2$ mesh as our baseline grid.
Here we consider local experimental data in the form of surface pressure measurements and
velocity PIV data. 

\begin{figure}[hbt!]
    \centering
    \begin{tikzpicture}
        \draw[use as bounding box, draw=none] (0,0) rectangle (\textwidth,5.2cm);
        \draw (0.0cm,0.25cm) node[anchor=south west,draw, inner sep=0] {\includegraphics[width=0.49\textwidth, trim={0.0cm 0.0cm 0.0cm 0.0cm}, clip]{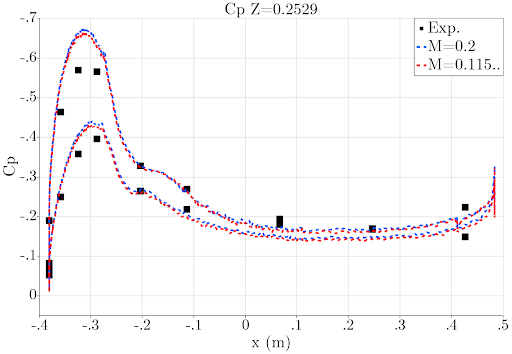}};
        \draw (7.05cm,0.25cm) node[anchor=south west,draw, inner sep=0] {\includegraphics[width=0.49\textwidth, trim={0.0cm 0.0cm 0.0cm 0.0cm}, clip]{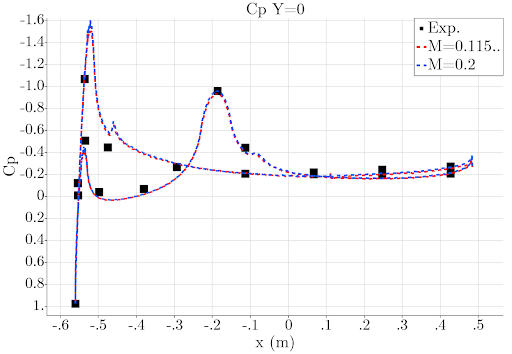}};
    \end{tikzpicture}
    \caption{Comparison of pressure coefficient cuts on the Windsor body for two simulations on the same grid with different Mach numbers but the same Reynolds number. The solution is not degraded and the maximum stable time step is increased by 55\% for the higher Mach number case.}
    \label{fig:windsor_mach_trick}
\end{figure}

We consider an additional modification to the problem to
improve efficiency. The Mach number of the problem is not actually specified but
is implied for some temperature that must be assumed. This degree of
freedom can be chosen to artificially increase the Mach number while keeping the Reynolds
number fixed. Assuming standard atmospheric conditions implies a freestream Mach
number $M\approx 0.11$. By instead taking the ambient temperature to be $T\approx
\qty{100}{\K}$, the Mach number increases to $M=0.2$ which increases the
maximum stable timestep by $55\%$. Figure~\ref{fig:windsor_mach_trick} shows the
pressure coefficient comparisons for $y=\qty{0.2595}{\m}$ and $z=\qty{0}{\m}$
with the original Mach number, the increased Mach number case and the
experimental data. Increasing the Mach number did not degrade the
quality of the solution but significantly reduces the turn around time for each
simulation considered in the dataset. Comparisons of base pressure and PIV data are
shown below for the $G_2$ grid at $M=0.2$. As defined in the AutoCFD3 Case 1 problem description, a numerical sensor is 
placed in the tunnel and average flow quantities from this probe are used to normalize the pressure and velocities
from the simulation.

\begin{figure}[hbt!]
    \centering
    \begin{tikzpicture}
        \draw[use as bounding box, draw=none] (0,0) rectangle (\textwidth,10.2);
        \draw (0.cm,0cm) node[anchor=south west] {\includegraphics[width=0.90\textwidth]{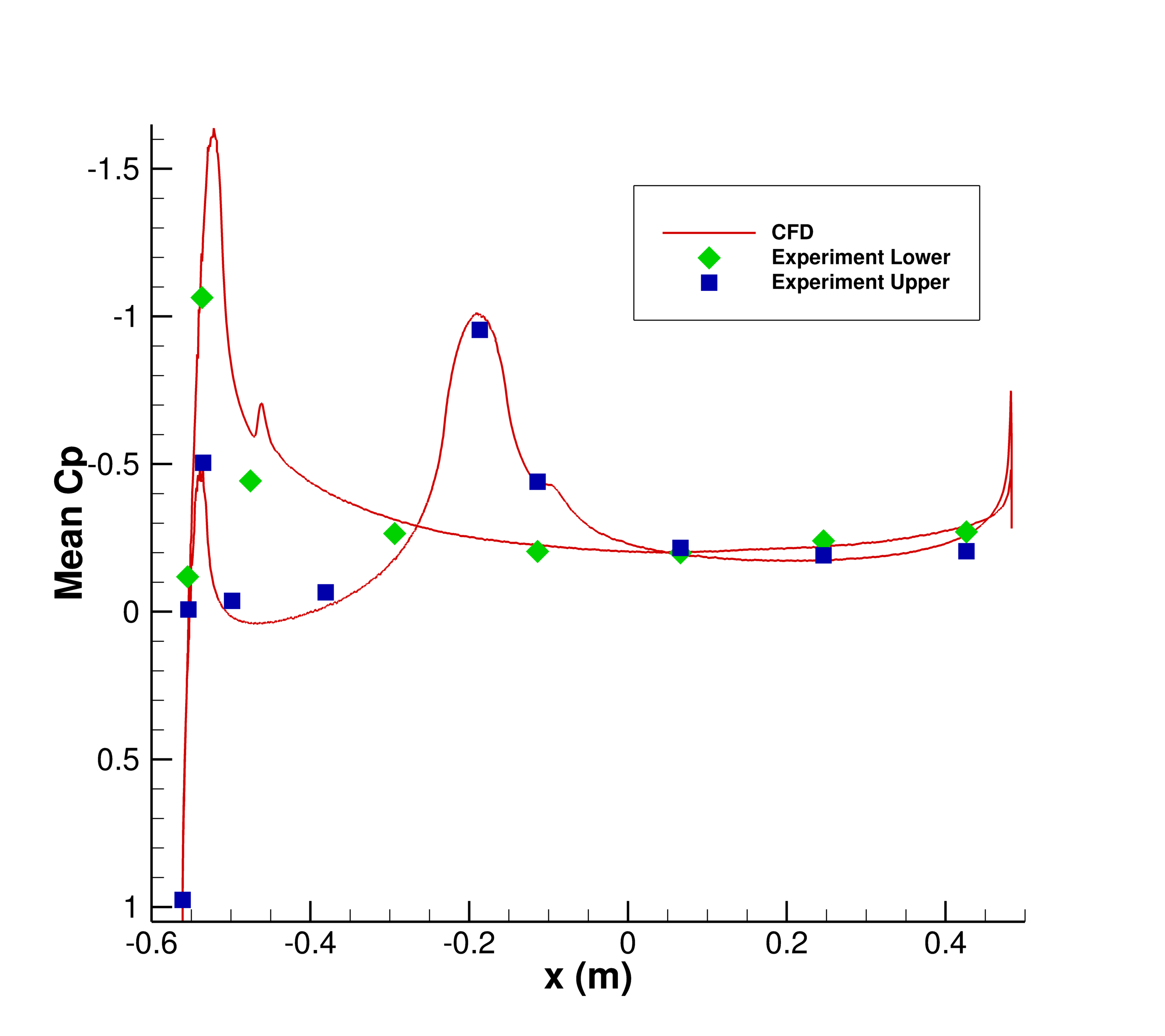}};
        \draw (6cm,2cm) node[anchor=south west,draw=black,very thick, inner sep=0mm] {\includegraphics[width=0.35\textwidth]{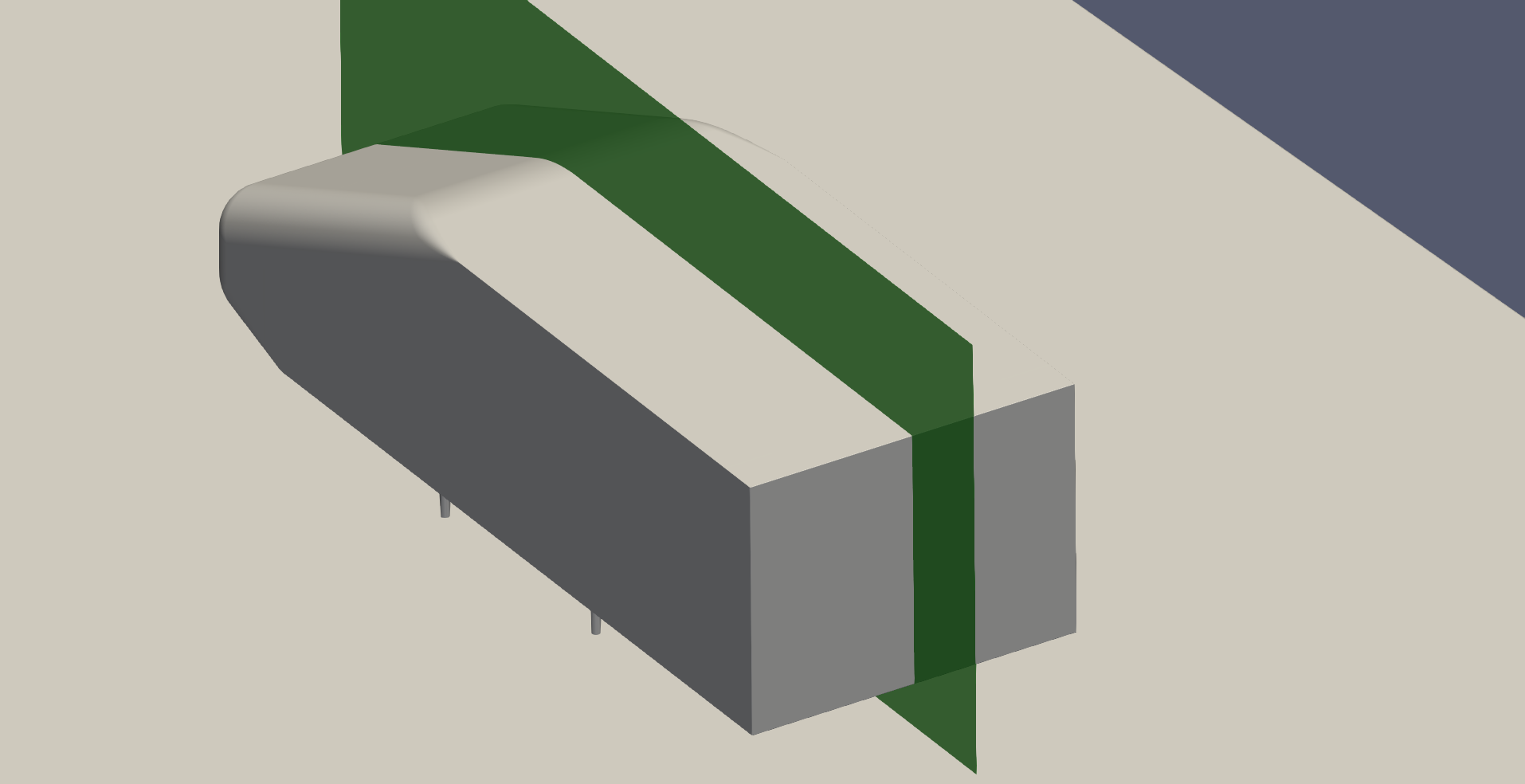}};
    \end{tikzpicture}
    \caption{Comparison of mean pressure coefficient at $z=\qty{0}{\m}$ for the Windsor body. }
    \label{fig:windsor_cpy}
\end{figure}

\begin{figure}[hbt!]
    \centering
    \begin{tikzpicture}
        \draw[use as bounding box, draw=none] (0,0) rectangle (\textwidth,10.5);
        \draw (0cm,0cm) node[anchor=south west] {\includegraphics[width=0.90\textwidth, trim = {1cm 1cm 1cm 8cm}, clip]{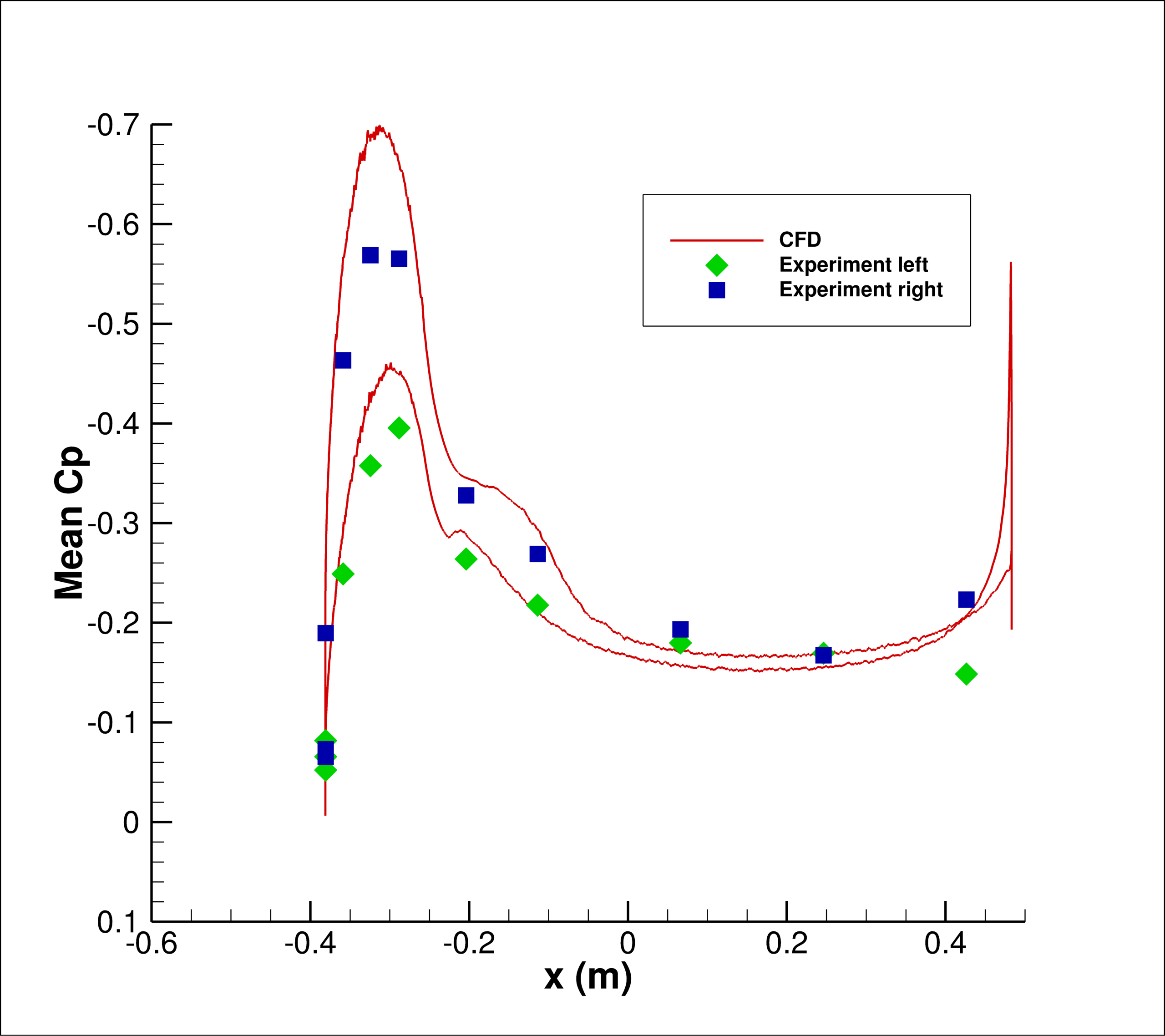}};
        \draw (6.0cm,1.6cm) node[anchor=south west,draw=black,very thick,inner sep=0mm] {\includegraphics[width=0.30\textwidth]{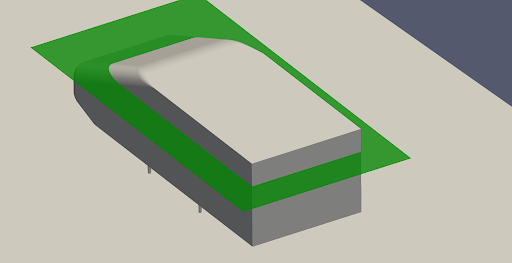}};

    \end{tikzpicture}
    \caption{Comparison of pressure coefficient at $y=\qty{0.2595}{\m}$ for the Windsor body.}
    \label{fig:windsor_cpz}
\end{figure}

\begin{figure}[hbt!]
    \centering
    \begin{tikzpicture}
        \draw[use as bounding box, draw=none] (0,0) rectangle (\textwidth,6.2cm);
        \draw (0.0cm,0.25cm) node[anchor=south west] {\includegraphics[width=0.49\textwidth, trim={0.0cm 0.0cm 0.0cm 0.0cm}, clip]{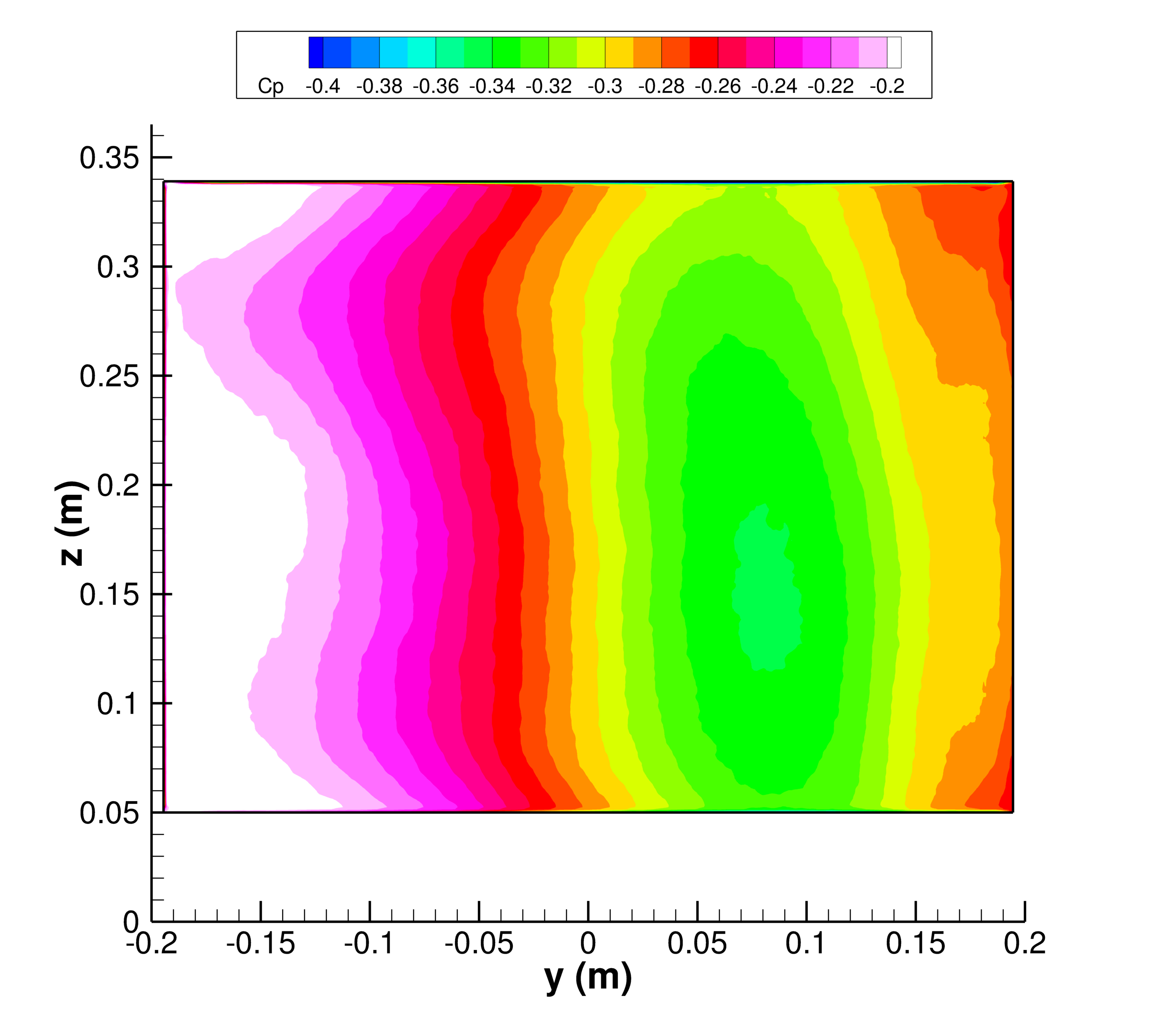}};
        \draw (7.2cm,0.25cm) node[anchor=south west] {\includegraphics[width=0.49\textwidth, trim={0.0cm 0.0cm 0.0cm 0.0cm}, clip]{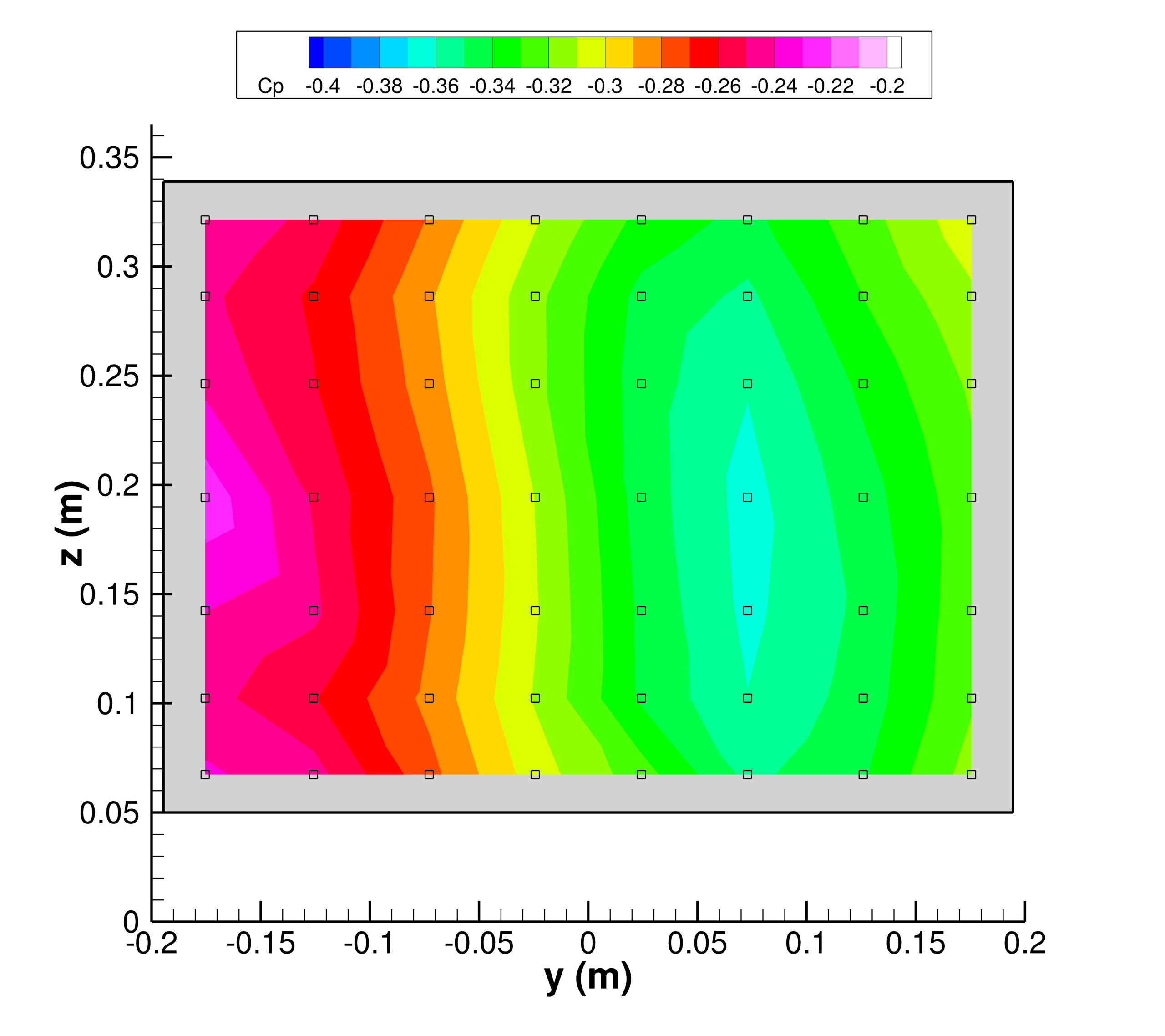}};


    \end{tikzpicture}
    \caption{Left: Time-averaged pressure coefficient at the base of the Windsor body. Right: Comparison of time-averaged pressure coefficient at the base of the Windsor body.}
    \label{fig:windsor_basepressure}
\end{figure}

{\bf Pressure:}
In Figure~\ref{fig:windsor_cpy} and Figure~\ref{fig:windsor_cpz}, we show experimental and computed 
surface pressure coefficient values, $C_p$, for the cut planes $z=\qty{0}{\m}$ and $y=\qty{0.2595}{\m}$
respectively. For the $z=\qty{0}{\m}$ plane we observe stronger suction in the CFD compared to the experiment 
near $x=\qty{-0.5}{\m}$ where the flow transitions to turbulence, which is typical in other
CFD simulations for this problem. Otherwise, there is good agreement for the
upper and lower sections.
For the $y=\qty{0.2595}{\m}$ cut,  the CFD again shows a stronger suction peak
this time in the A-pillar area. Small variations in the data suggest a longer
averaging could be done but it would not be expected to change the agreement.
Overall, the agreement is reasonably good and similar to other results seen in the AutoCFD3 results.
A comparative view of the base pressure is shown in
Figure~\ref{fig:windsor_basepressure} where similar pressure distributions are seen 
with a low pressure region on the leeward side of the vehicle
associated with the recirculation region. The simulation results show a higher pressure than the experiment 
which is most evident on the $-z$ side, however, the spatial distribution of the pressure and thus the shape of the 
contours match well.

\begin{figure}[hbt!]
    \centering
    \begin{tikzpicture}
        \draw[use as bounding box, draw=none] (0,0) rectangle (\textwidth,8.5);
        \draw (0.0cm,0.05cm) node[anchor=south west] {\includegraphics[width=0.49\textwidth, trim={0.0cm 0.0cm 0.0cm 0.0cm}, clip]{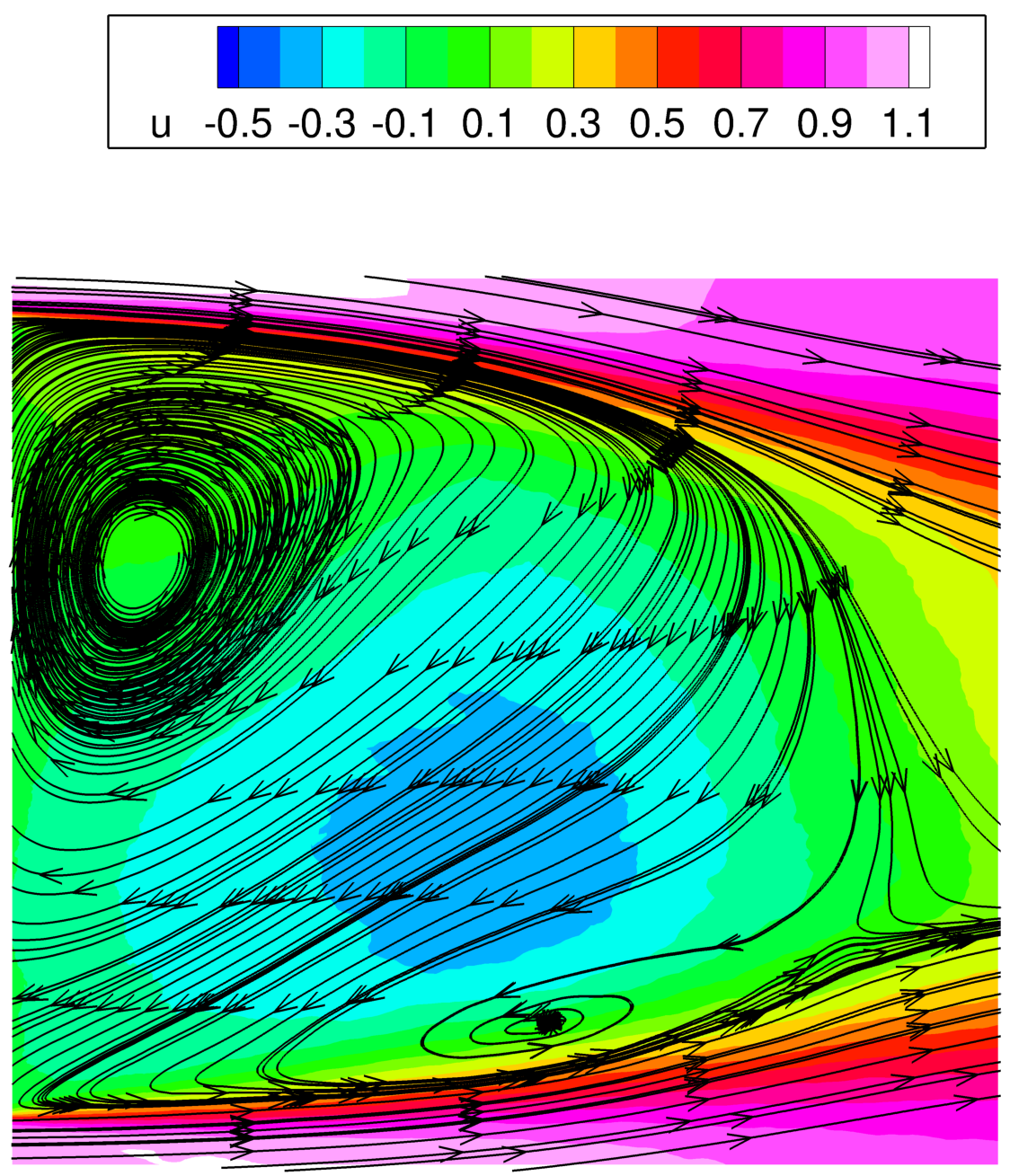}};
        \draw (7.0cm,0.0cm) node[anchor=south west] {\includegraphics[width=0.49\textwidth, trim={0.0cm 0.0cm 0.0cm 0.0cm}, clip]{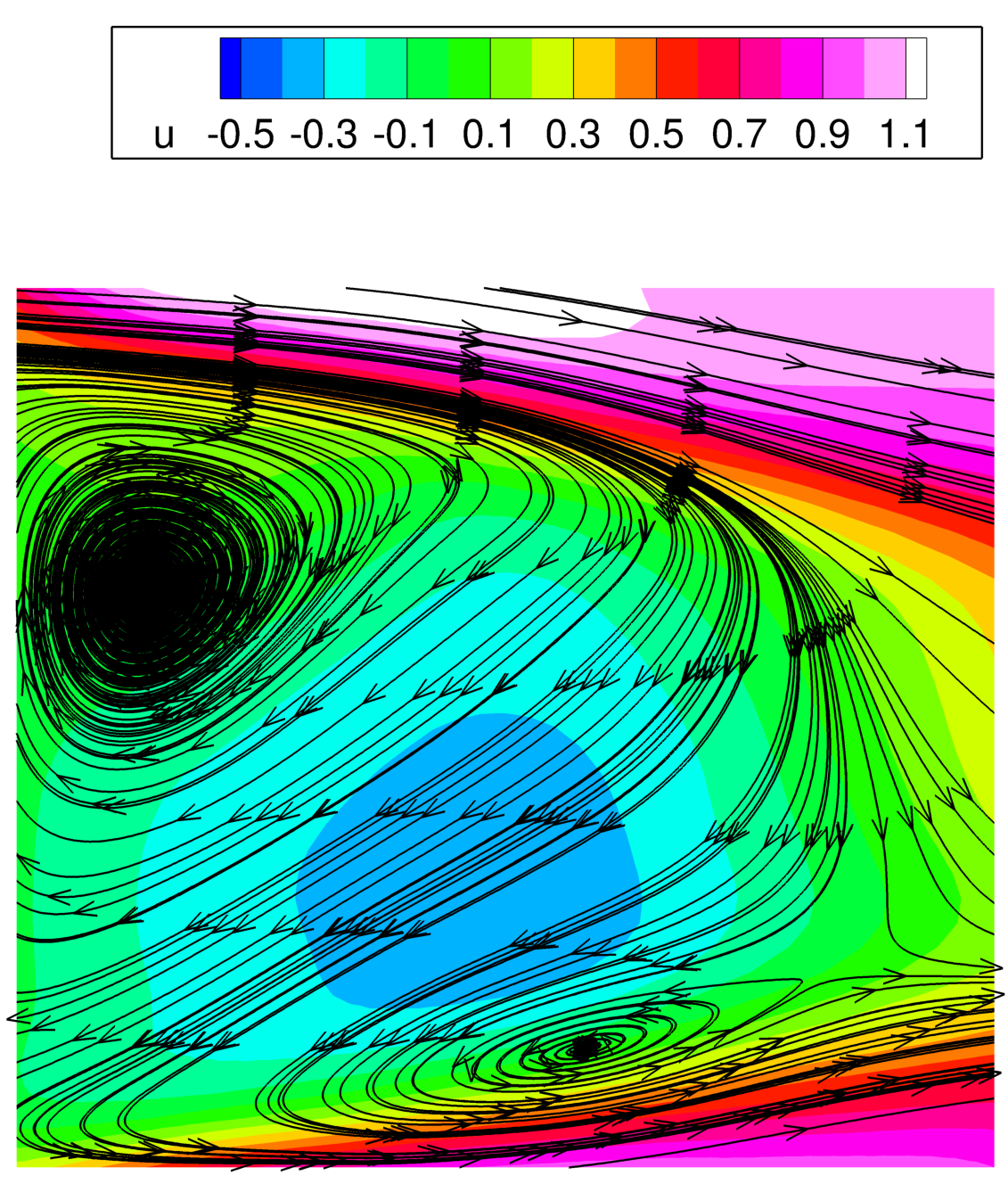}};
    \end{tikzpicture}
    \caption{Comparison of mean velocity streamlines colored by normalized streamwise velocity at $y=\qty{0.195}{\m}$ for the Windsor body.
        CFD results are shown on the left and the experiment data is plotted on the right. }.
    \label{fig:windsor_piv_z0p195_lic}
\end{figure}

\begin{figure}[hbt!]
     \centering
     \begin{tikzpicture}
         \draw[use as bounding box, draw=none] (0,0) rectangle (\textwidth,13.5cm);
         
         \draw (0.2cm,6.90cm) node[anchor=south west,draw, inner sep=0] {\includegraphics[width=0.45\textwidth, trim={12.5cm 4.3cm 12.5cm 5.0cm}, clip]{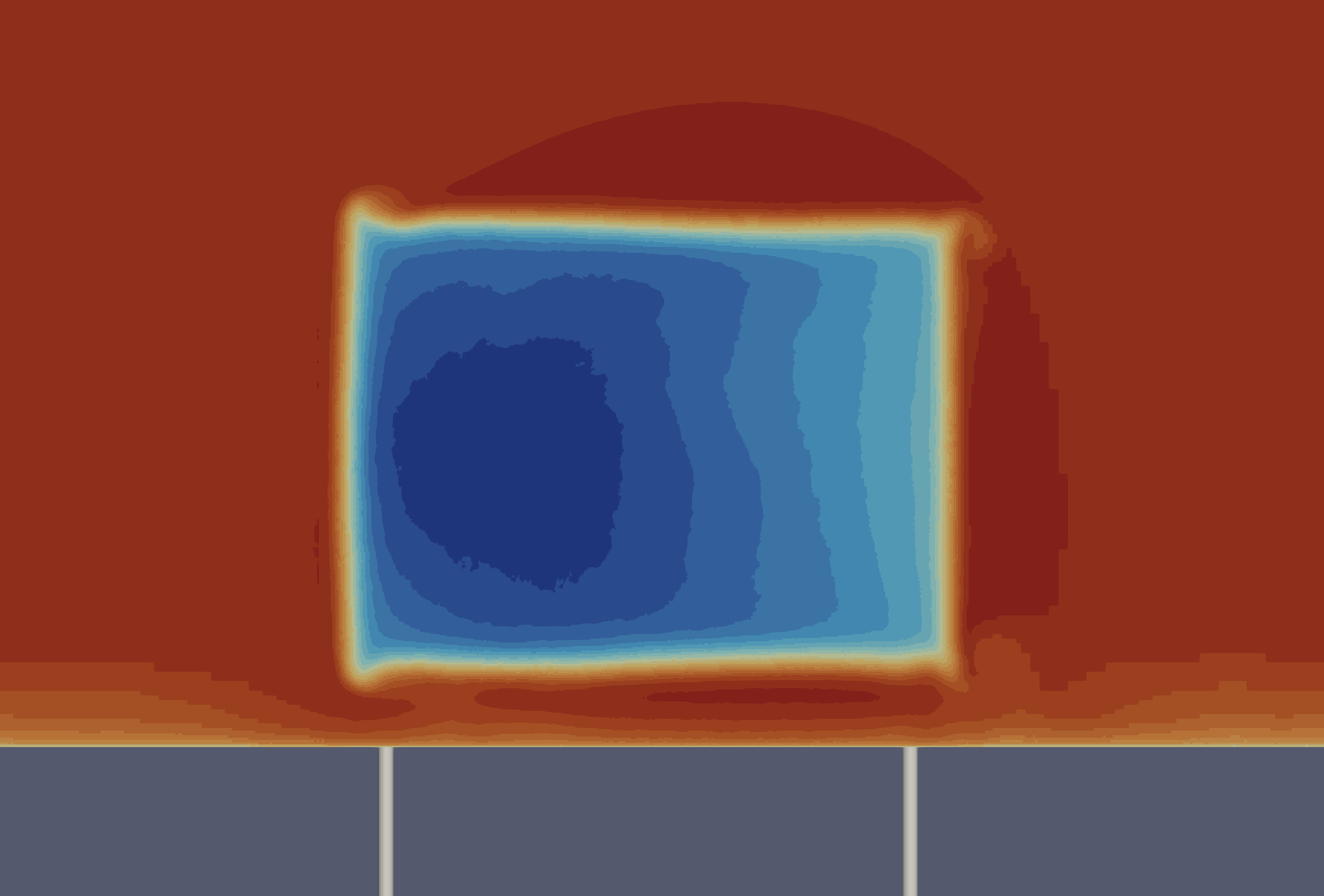}};
         \draw (7.0cm,6.90cm) node[anchor=south west,draw, inner sep=0] {\includegraphics[width=0.45\textwidth, trim={12.5cm 4.3cm 12.5cm 5.0cm}, clip]{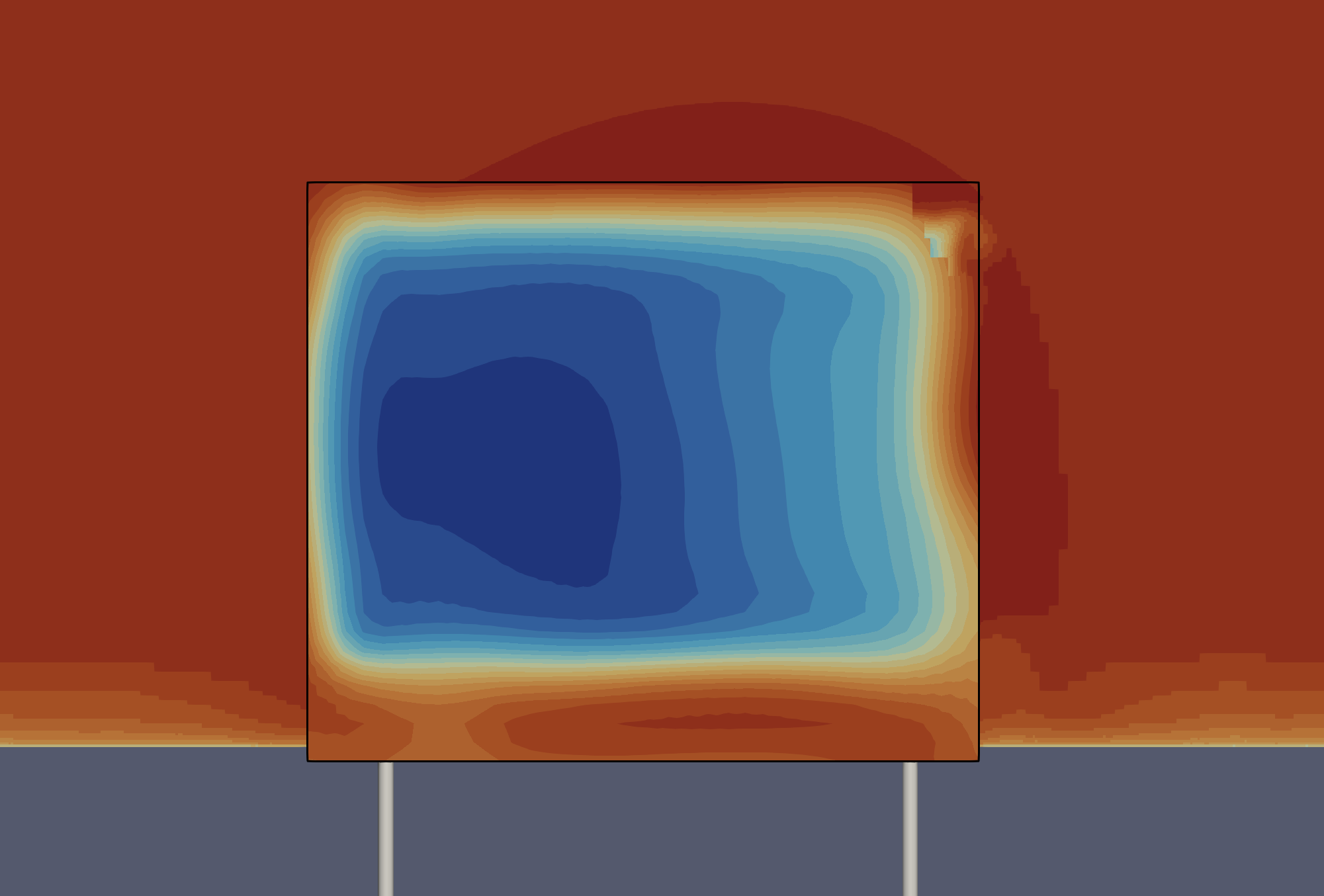}};
 
         \draw (2.0cm,  11.4cm) node[anchor=south west,draw,fill=white,rounded corners] {Averaged Solution};
         \draw (9.2cm, 11.4cm) node[anchor=south west,draw,fill=white,rounded corners] {Tomo. PIV Data};
 
         \draw (0.2cm,1.25cm) node[anchor=south west,draw, inner sep=0] {\includegraphics[width=0.45\textwidth, trim={12.5cm 4.3cm 12.5cm 5.0cm}, clip]{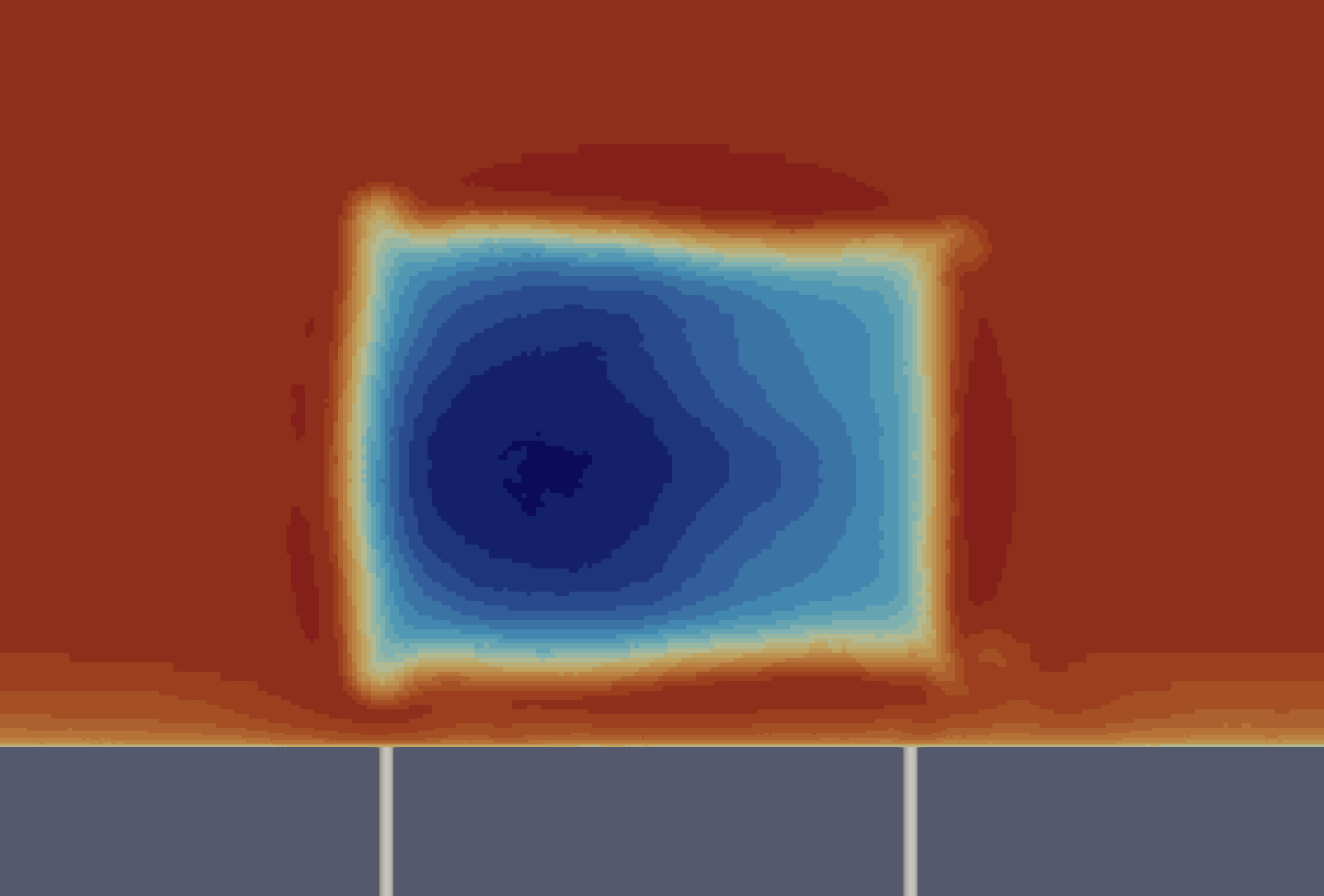}};
         \draw (7.0cm,1.25cm) node[anchor=south west,draw, inner sep=0] {\includegraphics[width=0.45\textwidth, trim={12.5cm 4.3cm 12.5cm 5.0cm}, clip]{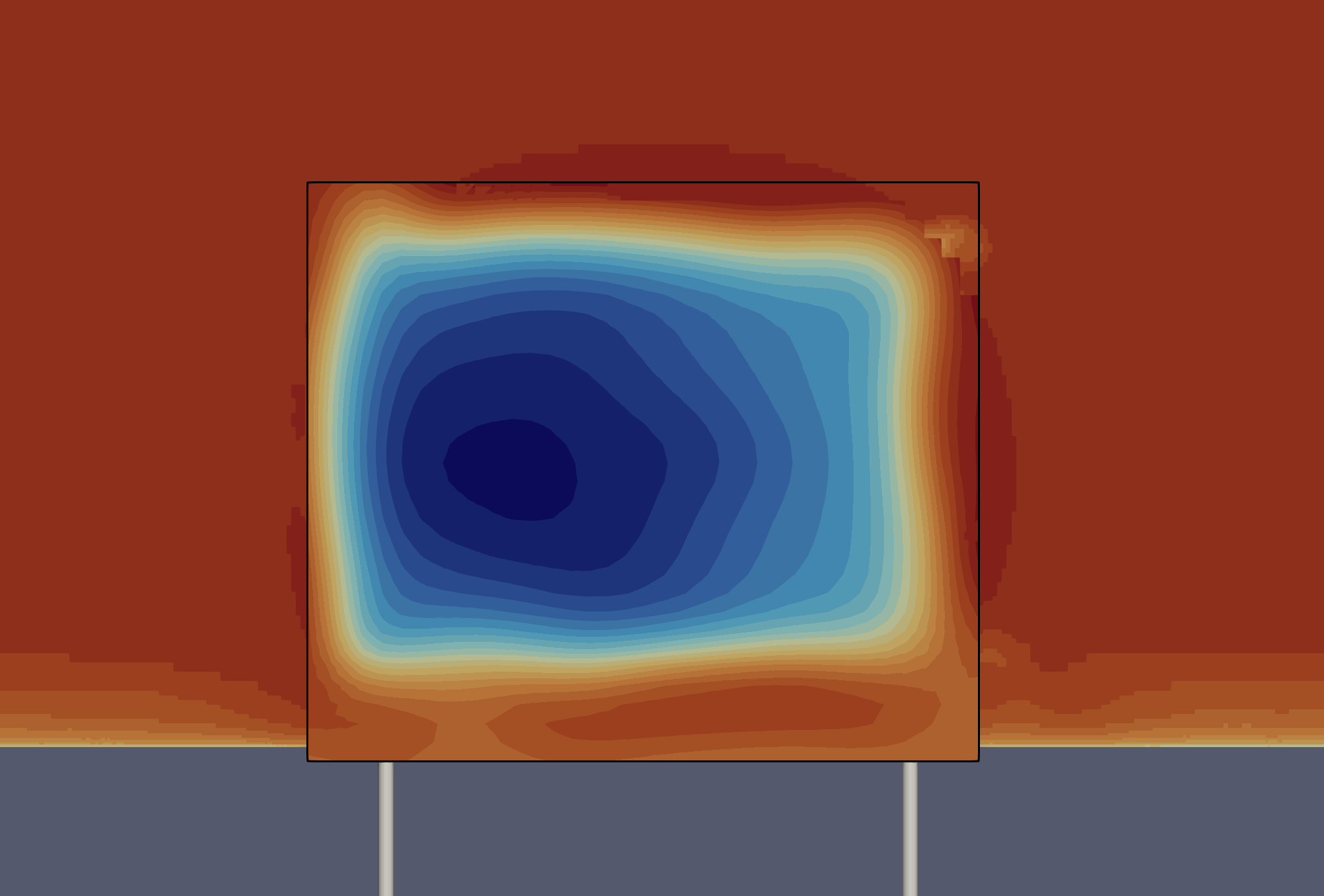}};
 
         \draw (5.0cm,11.25cm) node[anchor=south west,draw, inner sep=0] {\includegraphics[width=0.25\textwidth, trim={5.cm 9.cm 19.0cm 5.0cm}, clip]{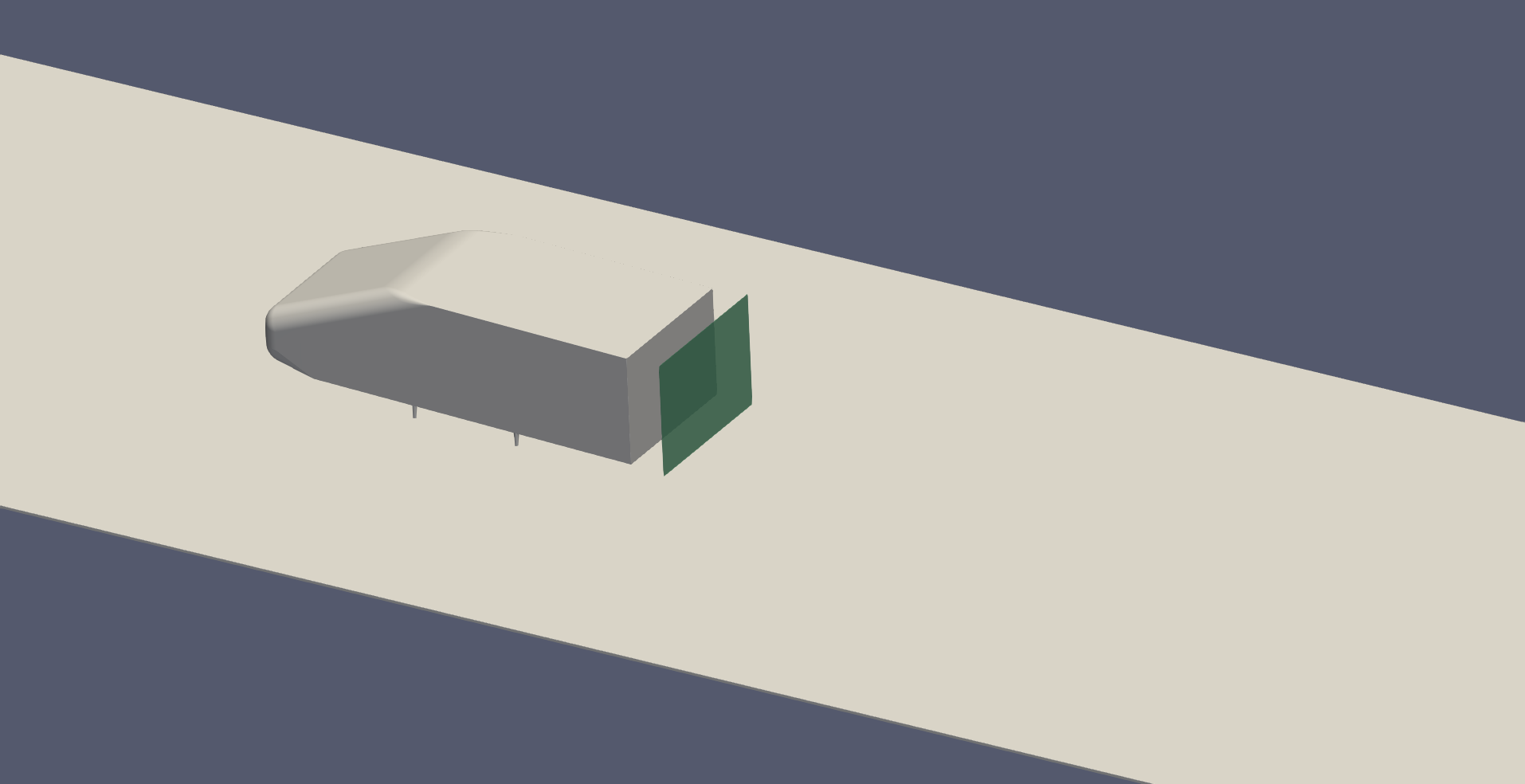}};
         \draw (5.0cm,5.25cm) node[anchor=south west,draw, inner sep=0] {\includegraphics[width=0.25\textwidth, trim={5.cm 9.cm 19.0cm 5.0cm}, clip]{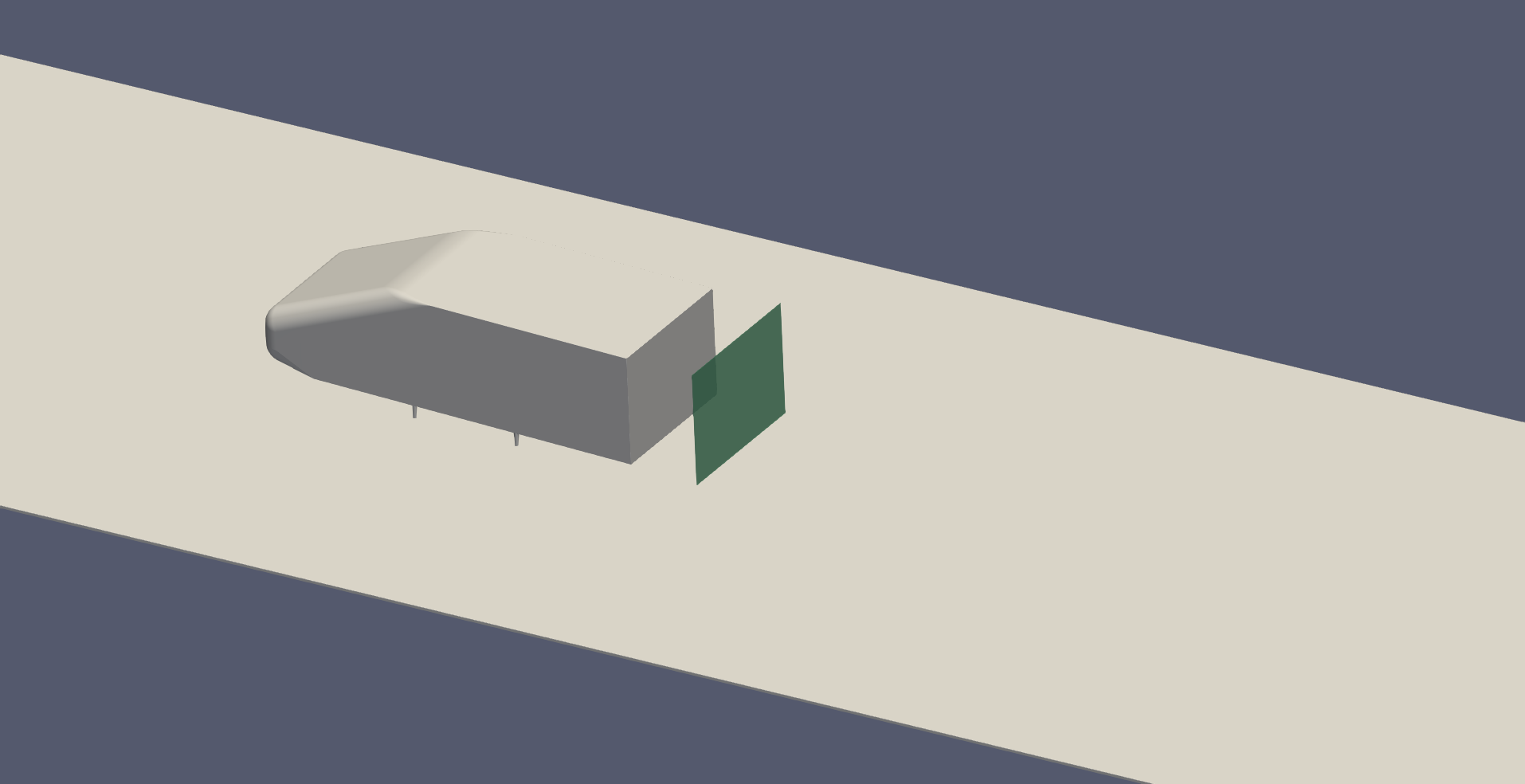}};

         \draw (3.5cm, 0.0cm) node[anchor=south west,draw, inner sep=0] {\includegraphics[width=0.50\textwidth, trim={0.cm 0.cm 0.0cm 0.0cm}, clip]{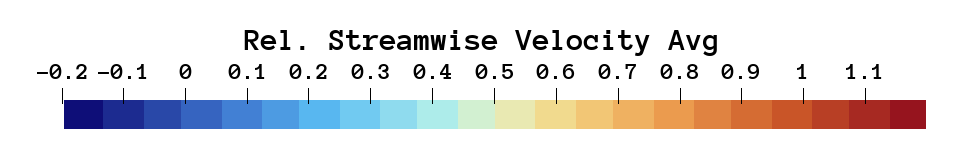}};
         
         \draw (2.0cm,  5.9cm) node[anchor=south west,draw,fill=white,rounded corners] {Averaged Solution};
         \draw (9.2cm, 5.9cm) node[anchor=south west,draw,fill=white,rounded corners] {Tomo. PIV Data};
     \end{tikzpicture}
     \caption{Time-averaged streamwise velocity at $x=\qty{0.578}{\m}$ (top) and $x=\qty{0.67048}{\m}$ (bottom).}
     \label{fig:windsor_tomopiv_1}
 \end{figure}

\begin{figure}[hbt!]
     \centering
     \begin{tikzpicture}
         \draw[use as bounding box, draw=none] (0,0) rectangle (\textwidth,13.5cm);
         
         \draw (0.2cm,6.90cm) node[anchor=south west,draw, inner sep=0] {\includegraphics[width=0.45\textwidth, trim={12.5cm 4.3cm 12.5cm 5.0cm}, clip]{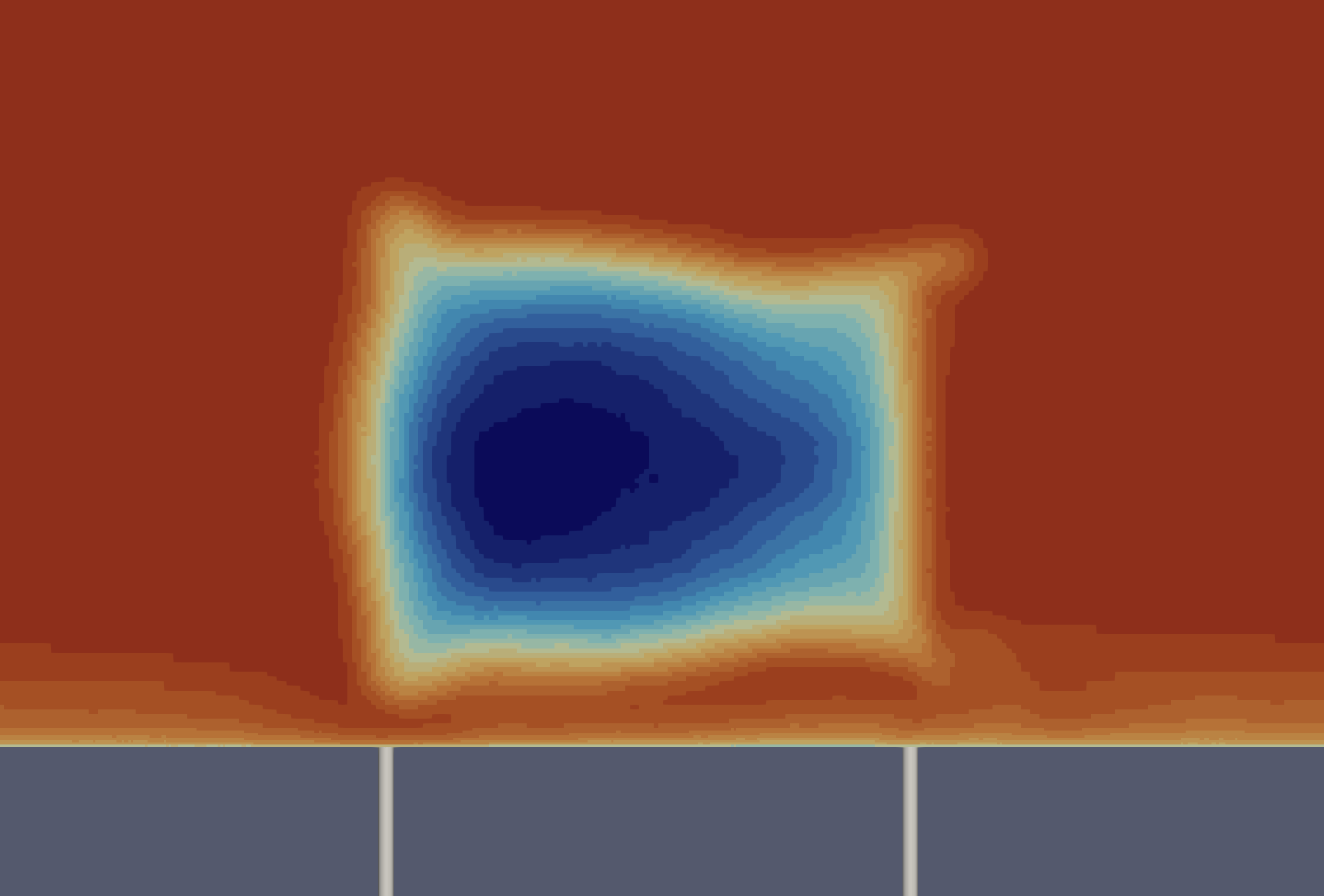}};
         \draw (7.0cm,6.90cm) node[anchor=south west,draw, inner sep=0] {\includegraphics[width=0.45\textwidth, trim={12.5cm 4.3cm 12.5cm 5.0cm}, clip]{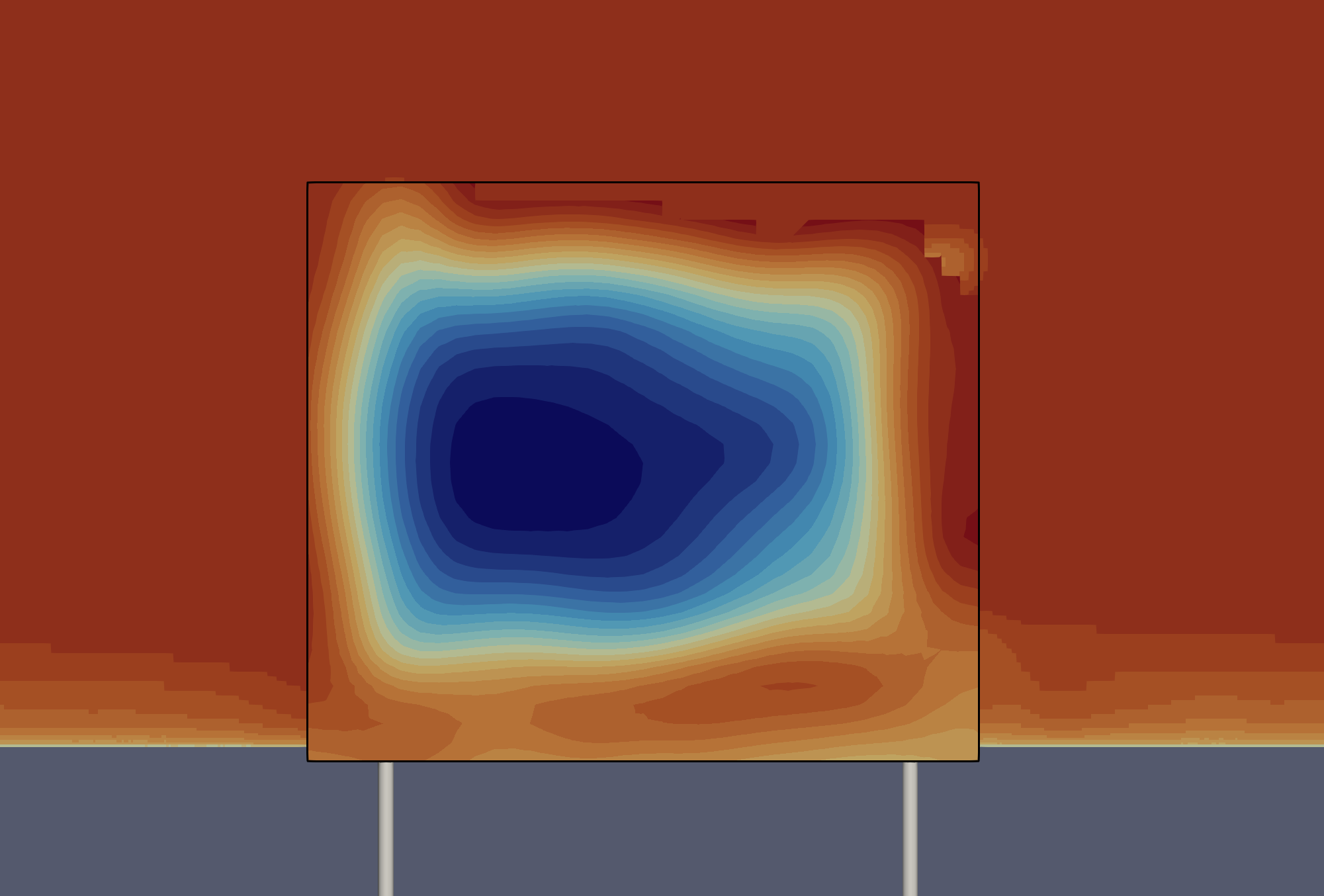}};
 
         \draw (2.0cm, 11.4cm) node[anchor=south west,draw,fill=white,rounded corners] {Averaged Solution};
         \draw (9.2cm, 11.4cm) node[anchor=south west,draw,fill=white,rounded corners] {Tomo. PIV Data};
 
         \draw (0.2cm,1.25cm) node[anchor=south west,draw, inner sep=0] {\includegraphics[width=0.45\textwidth, trim={12.5cm 4.3cm 12.5cm 5.0cm}, clip]{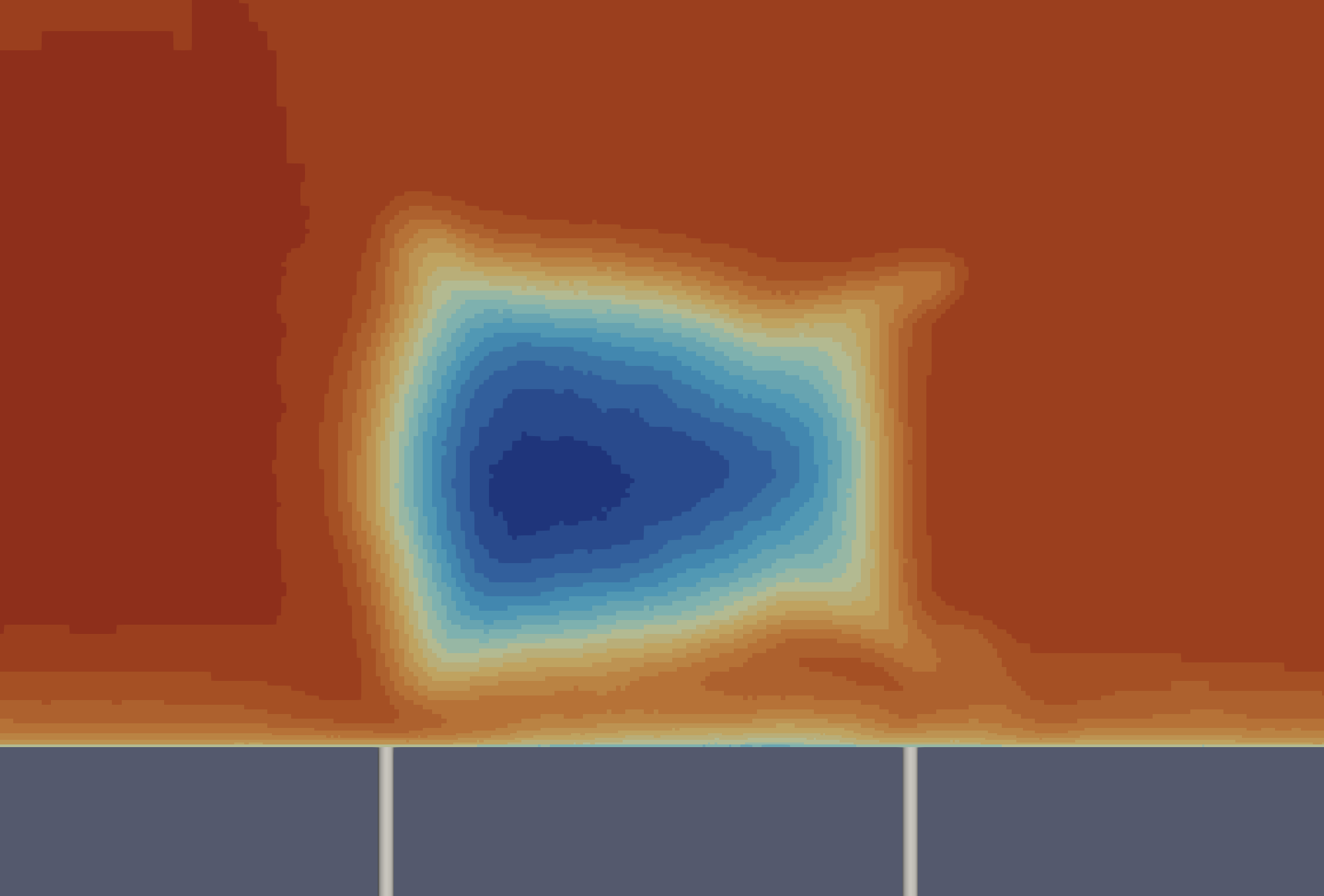}};
         \draw (7.0cm,1.25cm) node[anchor=south west,draw, inner sep=0] {\includegraphics[width=0.45\textwidth, trim={12.5cm 4.3cm 12.5cm 5.0cm}, clip]{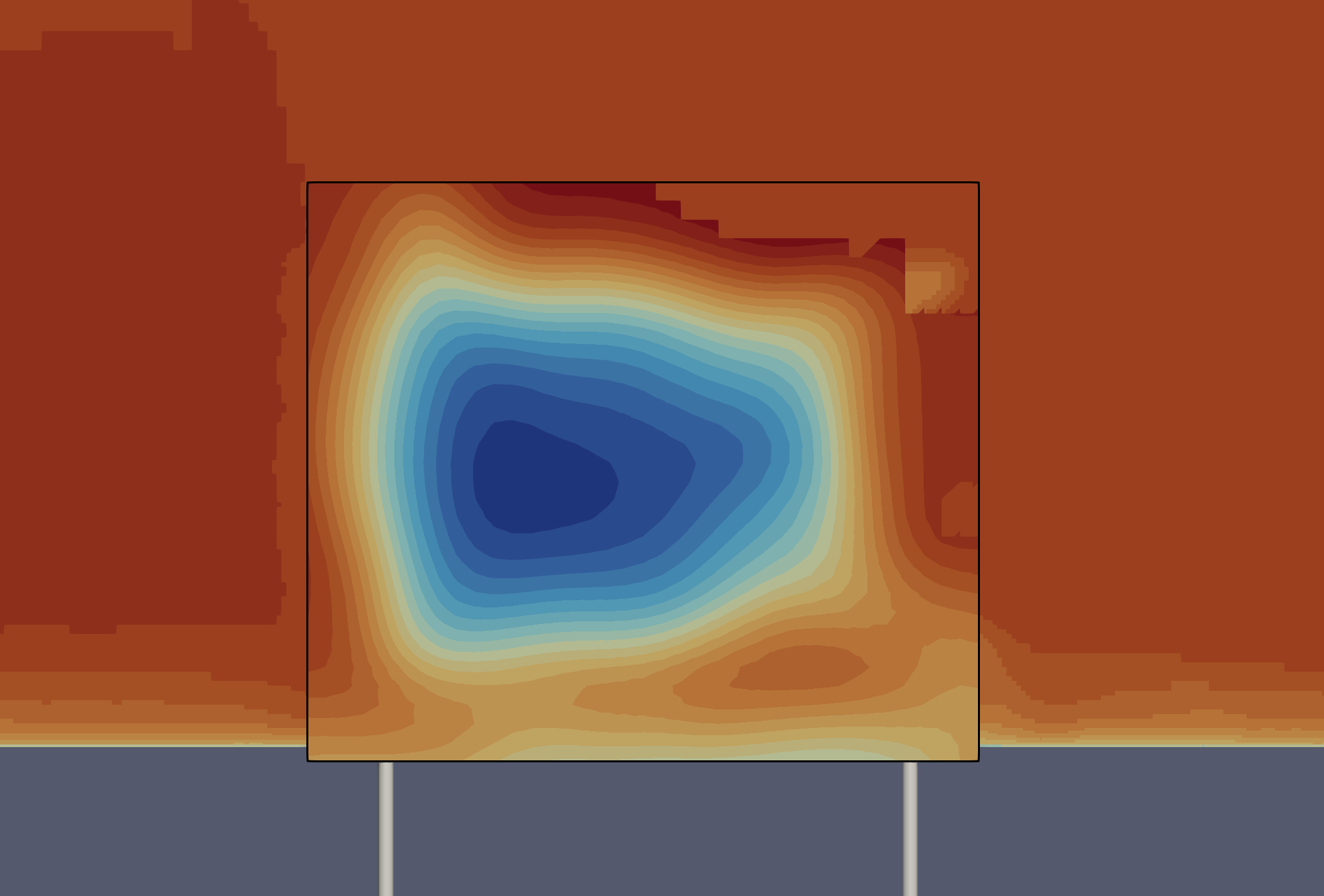}};
 
         \draw (5.0cm,11.25cm) node[anchor=south west,draw, inner sep=0] {\includegraphics[width=0.25\textwidth, trim={5.cm 9.cm 19.0cm 5.0cm}, clip]{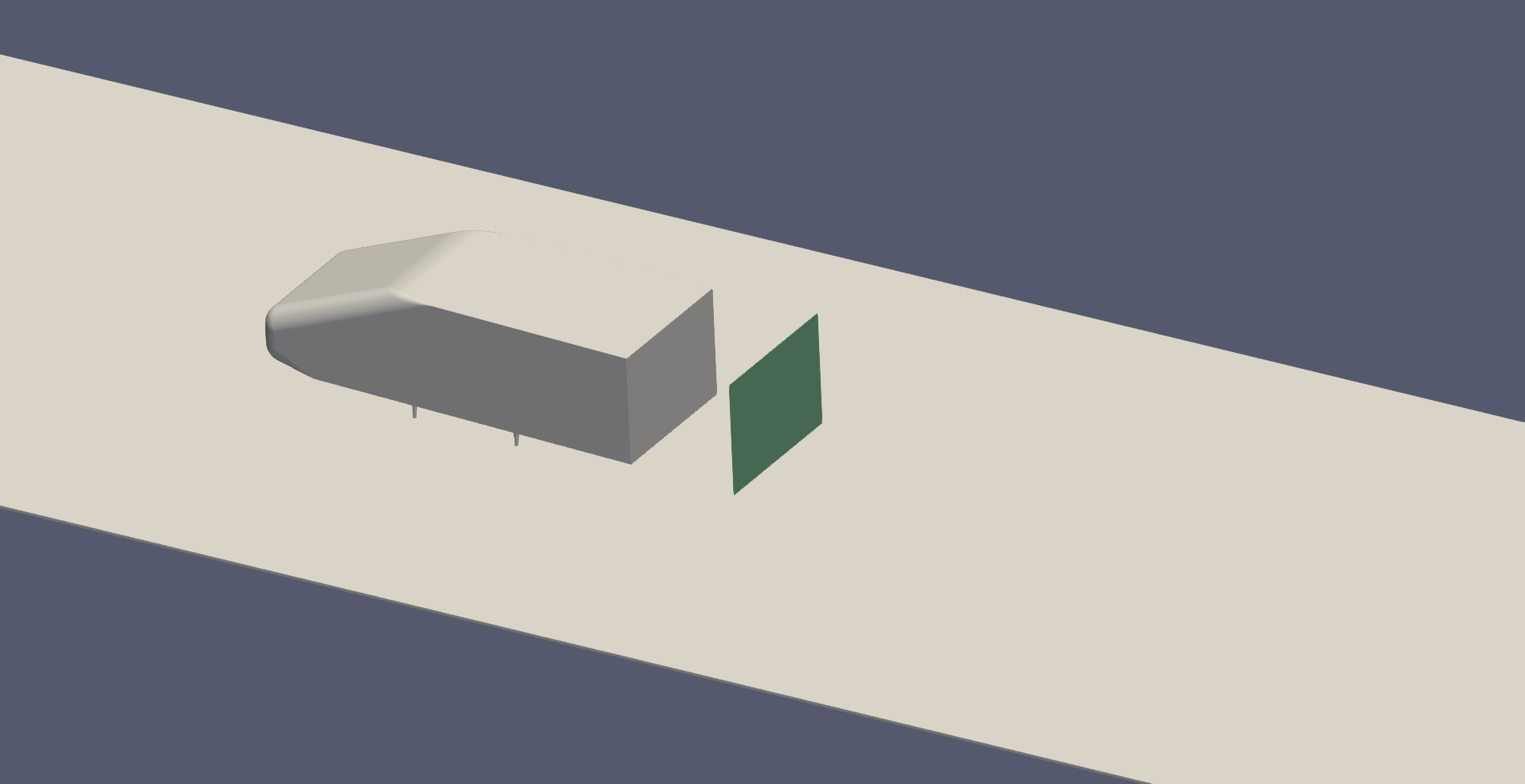}};
         \draw (5.0cm, 5.25cm) node[anchor=south west,draw, inner sep=0] {\includegraphics[width=0.25\textwidth, trim={5.cm 9.cm 19.0cm 5.0cm}, clip]{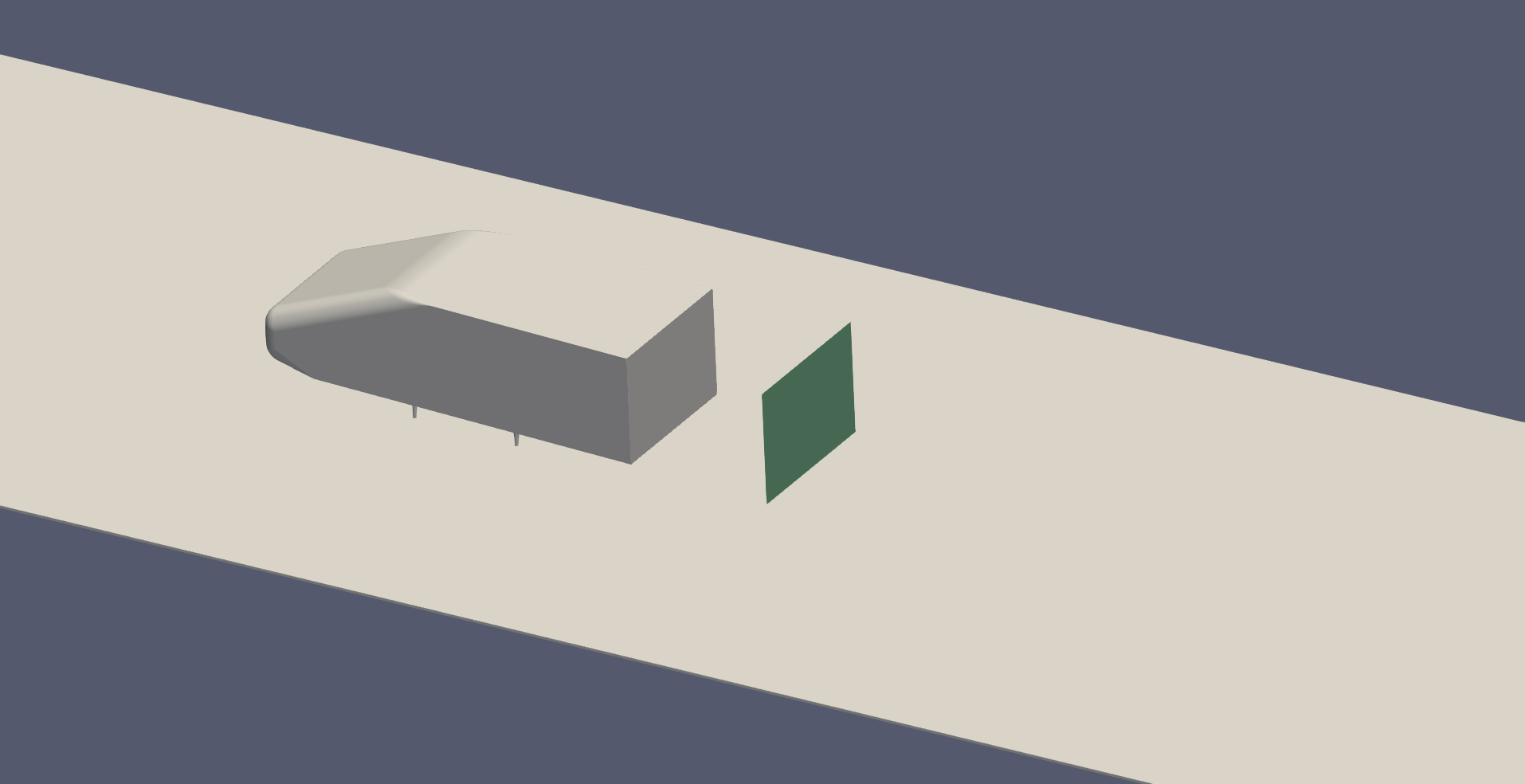}};
         
         \draw (3.5cm, 0.0cm) node[anchor=south west,draw, inner sep=0] {\includegraphics[width=0.50\textwidth, trim={0.cm 0.cm 0.0cm 0.0cm}, clip]{\images/relvel_bar.png}};
         
         \draw (2.0cm, 5.9cm) node[anchor=south west,draw,fill=white,rounded corners] {Averaged Solution};
         \draw (9.2cm, 5.9cm) node[anchor=south west,draw,fill=white,rounded corners] {Tomo. PIV Data};
     \end{tikzpicture}
     \caption{Time-averaged streamwise velocity at $x=\qty{0.77452}{\m}$ (top) and $x=\qty{0.867}{\m}$ (bottom).}
     \label{fig:windsor_tomopiv_2}
 \end{figure}

 {\bf Velocity:}
Time-averaged streamlines colored by normalized mean streamwise velocity at $y=\qty{0.195}{\m}$ are shown in
Figure~\ref{fig:windsor_piv_z0p195_lic} compared with experimental data from the thesis by Varney~\cite{Varney2020}. 
The streamlines show the large recirculation region
behind the vehicle and the correct flow topology. The data for the
$y=\qty{0.195}{\m}$ slice was taken with a freestream velocity of $\qty{40}{\m\per\s}$ which
is the same as our nominal freestream velocity but, in addition to this plane, tomographic PIV
was taken for a 3D region behind the car. This data was recorded for a flow with
a freestream velocity of $\qty{30}{\m\per\s}$. To make comparisons with the tomographic PIV
data, we normalize their velocities by their freestream velocity.  
Figures~\ref{fig:windsor_tomopiv_1} and~\ref{fig:windsor_tomopiv_2} show
progressive cuts through the wake going downstream.
Starting nearest the car, the experiment shows a thicker shear layer but
as we move downstream the predicted and experimental flow have essentially the
same shear layer thickness. The overall shape of the wake region is well
captured as well as its interaction with the floor boundary layer which can be
seen in the velocity contours.

\clearpage

%% file: sections/appendix-mlevaluation.tex
\clearpage
\section{ML evaluation}
\label{app:ml}
\subsection{Method}

We have conducted preliminary analysis on our dataset using a modified version of one of the state-of-the-art scientific machine learning (SciML) methods, MeshGraphNets \cite{pfaff2021learning} on various tasks to illustrate the practicality of the dataset for ML evaluation.  In both our method and MeshGraphNets, the node encoder and the edge encoder are 2-layer multilayer perceptrons (MLPs), and the processor consists of L message passing blocks, each containing two MLPs. There are a number of key differences between the approach taken in this work and that in the original paper of Pfaff et al.

The original justification and proof of MeshGraphNets was for transient prediction, which involves predicting dynamic quantities of the mesh at time t+1 given the current state and previous states. In contrast, our method focuses on directly predicting simulation results at the steady state  or time-averaged state. We believe steady state or time-averaged prediction is more applicable in industrial contexts, where transient simulation data is often not preserved due to the large amount of data storage required.  

This difference in task leads to distinct input node features. MeshGraphNet leverages dynamical features such as the instantaneous velocity or pressure to predict future states, but those features are not available in a steady state or time-averaged prediction. As our method predicts time-averaged results directly from meshes, the node features comprise node positions, node normals, and node defects in some use cases. Node positions represent the absolute node positions in the 3D space, as our method operates on 3D meshes. This is in contrast to the MeshGraphNets paper, which primarily focused on 2D CFD use cases. Our edge features, on the other hand, include the relative displacement vector (i.e., the difference between the positions of the two nodes forming the edge) and its norm, which closely resemble the edge features in MeshGraphNets.

Our model architecture has been adapted to address two distinct use cases: direct Key Performance Indictor (KPI) prediction and surface variable prediction. Among the two use cases, the architecture used in surface variable prediction more closely resembles that of MeshGraphNets, as both produce node level predictions. For KPI prediction, the models make graph level predictions, that is predicting one or a few values (e.g lift or drag coefficient) per graph rather than per node. To achieve this, there is a pooling layer being added between the processor and the decoder. The pooling layer converts node embeddings to graph embeddings which are then passed to the decoder.

\subsection{Results}

The entire WindsorML dataset is split into training (60\%), validation (20\%), and test (20\%) sets. For each use case, the model is trained on the training set, and the checkpoint that had the best validation error was used to obtain the inference results on the test set. 

\begin{figure}[htb]
     \centering
     \begin{subfigure}[b]{0.49\textwidth}
         \centering
         \includegraphics[width=\textwidth]{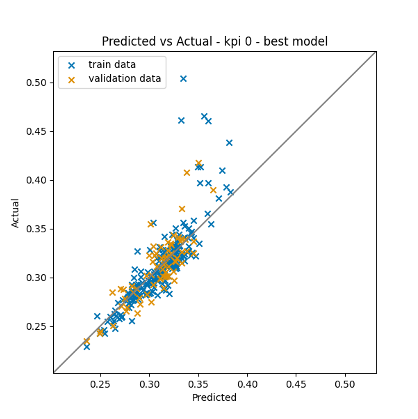}
         \caption{Predicted drag coefficient on training/validation split}
         \label{fig:mlcd1}
     \end{subfigure}
     \hfill
     \begin{subfigure}[b]{0.49\textwidth}
         \centering
         \includegraphics[width=\textwidth]{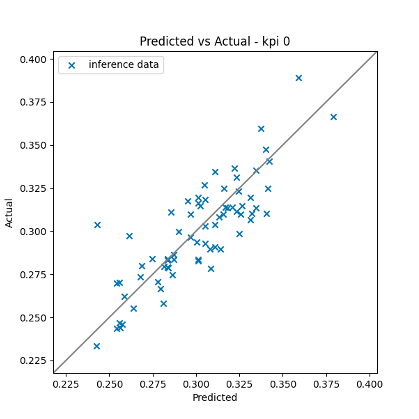}
         \caption{Predicted drag coefficient on test split}
         \label{fig:mlcd2}
     \end{subfigure}
              \caption{Prediction of the drag coefficient using the direct KPI method}
    \label{fig:mlpredictions1}
\end{figure}

\begin{figure}[htb]
     \centering
     \begin{subfigure}[b]{0.49\textwidth}
         \centering
         \includegraphics[width=\textwidth]{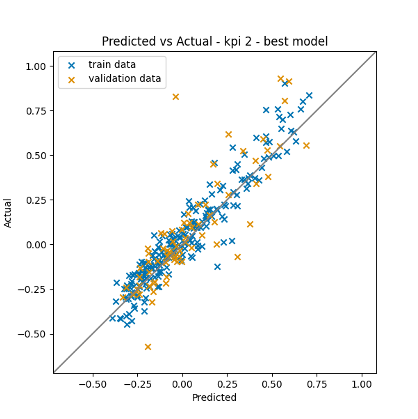}
         \caption{Predicted lift coefficient on training/validation split}
         \label{fig:mlcl1}
     \end{subfigure}
     \hfill
     \begin{subfigure}[b]{0.49\textwidth}
         \centering
         \includegraphics[width=\textwidth]{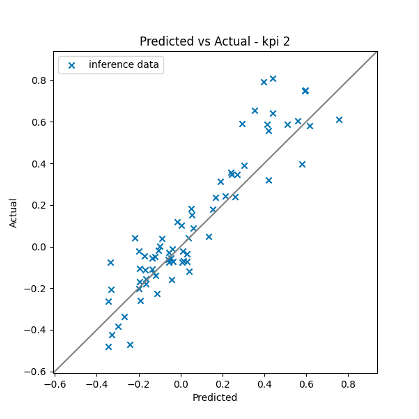}
         \caption{Predicted lift coefficient on test split}
         \label{fig:mlcl2}
     \end{subfigure}
              \caption{Prediction of the lift coefficient using the direct KPI method}
    \label{fig:mlpredictions2}
\end{figure}

Using the first method to directly predict the lift and drag coefficients we find that using a 60/20/20 split of train, validation and test data, it is possible to obtain a MSE of less than 0.00028 for the drag coefficient and a MSE less than 0.0175 for the lift coefficient. An example of the prediction accuracy is shown in Figure \ref{fig:mlpredictions1} \& \ref{fig:mlpredictions2}  for the training, validation and test data for the lift and drag coefficient. Using x8 Nvidia L40s GPUs (Amazon EC2 g6e.48xlarge instances via Amazon Web Services), the training completes in approximately 2hrs and the inference time for each new predicted geometry is 0.15 seconds. We ran this training for 1000 epochs using 5 message passing layers with a batch size of 4. The Adams optimizer was used with an initial learning rate of 0.003.

\begin{figure}[htb]
     \centering
     \begin{subfigure}[b]{0.49\textwidth}
         \centering
         \includegraphics[width=\textwidth]{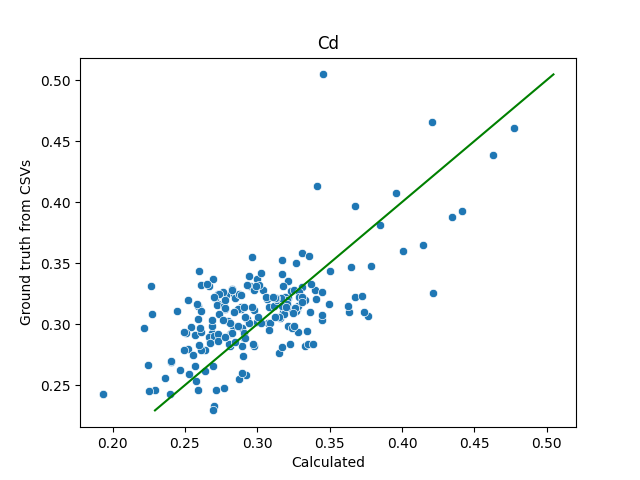}
         \caption{Predicted drag coefficient on the training split}
         \label{fig:mlcl1}
     \end{subfigure}
     \hfill
     \begin{subfigure}[b]{0.49\textwidth}
         \centering
         \includegraphics[width=\textwidth]{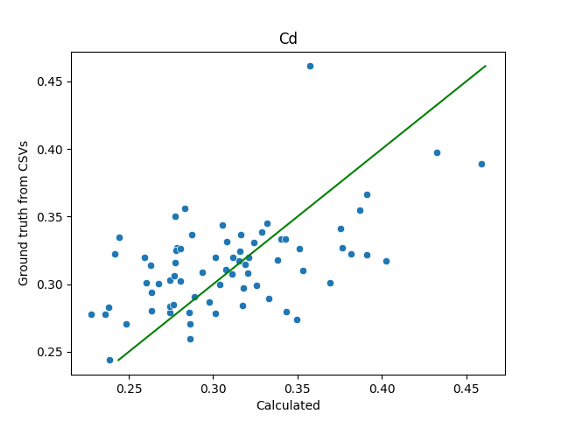}
         \caption{Predicted drag coefficient on the test split}
         \label{fig:mlcl2}
     \end{subfigure}
              \caption{Prediction of the drag coefficient obtained through integration of the wall-shear stress and pressure}
    \label{fig:mlpredictions3}
\end{figure}

\begin{figure}[htb]
     \centering
     \begin{subfigure}[b]{0.49\textwidth}
         \centering
         \includegraphics[width=\textwidth]{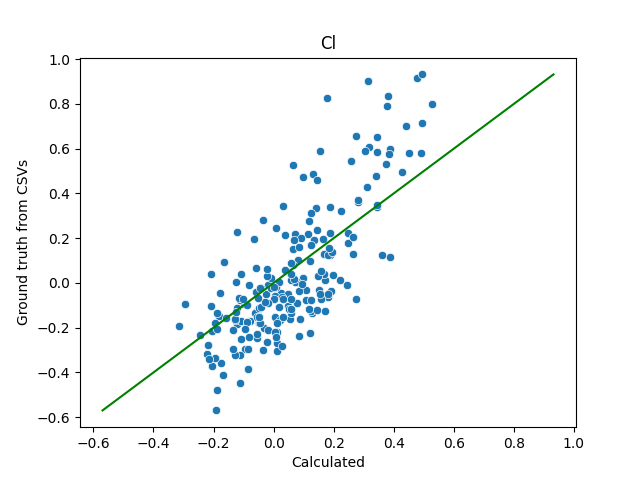}
         \caption{Predicted lift coefficient on the training split}
         \label{fig:mlcl1}
     \end{subfigure}
     \hfill
     \begin{subfigure}[b]{0.49\textwidth}
         \centering
         \includegraphics[width=\textwidth]{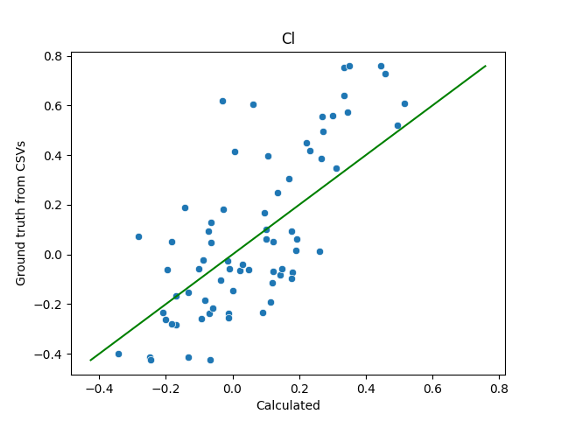}
         \caption{Predicted lift coefficient on the test split}
         \label{fig:mlcl2}
     \end{subfigure}
              \caption{Prediction of the lift coefficient obtained through integration of the wall-shear stress and pressure}
    \label{fig:mlpredictions4}
\end{figure}

For the second method, where the lift and drag coefficient are obtained through the integration of the predicted surface pressure and wall-shear stress on the 4.4M node VTP surface mesh, we obtain predictions for the drag coefficient with a mean absolute error (MAE) of 0.03269 and a weighted mean absolute percentage error (WMAPE) of 0.1040. For the lift coefficient the mean absolute error (MAE) is 0.1854 and the weighted mean absolute percentage error (WMAPE) of 0.698 (as shown in Figures \ref{fig:mlpredictions3} \& \ref{fig:mlpredictions4}). Training time is approximately 60 hours on x8 NVIDIA L40s GPUs (Amazon EC2 g6e.48xlarge instances via Amazon Web Services) and the inference time is less than a minute on the same hardware. We ran this training with BF16 for 1200 epochs using 20 message passing layers with a batch size of 1. The Adams optimizer was used with an initial learning rate of 0.001.

It is clear from these results that the WindsorML dataset represents a challenging use-case for ML methods to predict, with noticeable difference in accuracy based upon the method used. Please note that these ML evaluations are preliminary and purely serve to illustrate how this dataset can be used for ML evaluation. We hope other groups will use this dataset to do a more thorough evaluation of different ML methodologies. 

\newpage

%% file: sections/appendix-datasheet.tex
\clearpage
\section{Datasheet}
\subsection{Motivation}

\begin{itemize}

\item \textbf{For what purpose was the dataset created?} The dataset was created to address the current limitations of a lack of high-fidelity training data for the development and testing of machine learning methods for Computational Fluid Dynamics. In addition, it was created to be used as a dataset for the 4th Automotive CFD Prediction Workshop \footnote{https://autocfd.org} 

\item \textbf{Who created the dataset (e.g., which team, research group) and on behalf of which entity (e.g., company, institution, organization)?} The dataset was created by a consortium of developers and scientists from academia and industry. 

\item \textbf{Who funded the creation of the dataset?} The project was internally funded within each author organisation i.e no external grants. 
\end{itemize}

\subsection{Distribution}

\begin{itemize}

\item \textbf{Will the dataset be distributed to third parties outside of the entity (e.g., company, institution, organization) on behalf of which the dataset was created?} Yes, the dataset is open to the public

\item \textbf{How will the dataset will be distributed (e.g., tarball on website, API, GitHub)?} The dataset is free to download from Amazon S3 (without the need for an AWS account) and is fully described on the dataset website \footnote{https://caemldatasets.org}. Additional download sites and options are in the process of being created and will be shared on the website once ready. 

\item \textbf{When will the dataset be distributed?} The dataset is already available to download via Amazon S3. Additional download sites and options are in the process of being created and will be shared on the website once ready. 

\item \textbf{Will the dataset be distributed under a copyright or other intellectual property (IP) license, and/or under applicable terms of use (ToU)?} The dataset is licensed under CC-BY-SA license. 

\item \textbf{Have any third parties imposed IP-based or other restrictions on the data associated with the instances?} No

\item \textbf{Do any export controls or other regulatory restrictions apply to the dataset or to individual instances?} No

\end{itemize}

\subsection{Maintenance}

\begin{itemize}

\item \textbf{Who will be supporting/hosting/maintaining the dataset?} The dataset is being managed by the collection of authors and a public website \footnote{https://caemldatasets.org} will provide on-going updates on hosting/maintenance. 

\item \textbf{How can the owner/curator/manager of the dataset be contacted (e.g., email address)?} The owner/curator/manager of the dataset can be contacted at contact@caemldatasets.org (these are also provided in the dataset README and paper). 

\item \textbf{Is there an erratum?} No, but if we find errors we will provide updates to the dataset and note any changes in the dataset README and website.

\item \textbf{Will the dataset be updated (e.g., to correct labeling
    errors, add new instances, delete instances)?} Yes the dataset will be updated to address errors or provided extra functionality. The README of the dataset and the dataset website will be updated to reflect this. 

\item \textbf{If the dataset relates to people, are there applicable
    limits on the retention of the data associated with the instances
    (e.g., were the individuals in question told that their data would
    be retained for a fixed period of time and then deleted)?} N/A

\item \textbf{Will older versions of the dataset continue to be
    supported/hosted/maintained?} Yes, if there are substantial changes or additions, older versions will still be kept. 

\item \textbf{If others want to extend/augment/build on/contribute to
    the dataset, is there a mechanism for them to do so?} We will consider this on a case by case basis and they can contact contact@caemldatasets.org to discuss this further. 

  \end{itemize}

\subsection{Composition}

\begin{itemize}

\item \textbf{What do the instances that comprise the dataset
    represent (e.g., documents, photos, people, countries)?} Computational Fluid Dynamics simulations.

\item \textbf{How many instances are there in total (of each type, if appropriate)?} Table \ref{output-table2} details the specific outputs that are contained in the dataset for each of the 355 geometric variations of the Windsor body.

\item \textbf{Does the dataset contain all possible instances or is it
    a sample (not necessarily random) of instances from a larger set?}
 The dataset is complete collection of simulations run to date.

\item \textbf{What data does each instance consist of?} Table \ref{output-table2} details the specific outputs that are contained in the dataset for each of the 355 geometric variations of the Windsor body.

\item \textbf{Is there a label or target associated with each
    instance?} N/A

\item \textbf{Is any information missing from individual instances?}
  No, the dataset is fully described.

\item \textbf{Are relationships between individual instances made
    explicit (e.g., users' movie ratings, social network links)?} N/A

\item \textbf{Are there recommended data splits (e.g., training,
    development/validation, testing)?} 60/20/20 is the recommended split based upon initial testing. 

\item \textbf{Are there any errors, sources of noise, or redundancies
    in the dataset?} The errors associated with the Computational Fluid Dynamics method is discussed in the validation section of the main paper and SI.

\item \textbf{Is the dataset self-contained, or does it link to or
    otherwise rely on external resources (e.g., websites, tweets,
    other datasets)?} It is self-contained.

\item \textbf{Does the dataset contain data that might be considered
    confidential (e.g., data that is protected by legal privilege or
    by doctor patient confidentiality, data that includes the content
    of individuals' non-public communications)?} The data is fully open-source and not considered confidential.

\item \textbf{Does the dataset contain data that, if viewed directly,
    might be offensive, insulting, threatening, or might otherwise
    cause anxiety?} No

\end{itemize}

\subsection{Collection Process}

\begin{itemize}

\item \textbf{How was the data associated with each instance
    acquired?} The data was obtained through Computational Fluid Dynamics (CFD) simulations and then post-processed to extract only the required quantities. 

\item \textbf{What mechanisms or procedures were used to collect the
    data (e.g., hardware apparatus  or sensors, manual human
    curation, software programs, software APIs} The main paper discusses the HPC hardware used

\item \textbf{If the dataset is a sample from a larger set, what was
    the sampling strategy (e.g., deterministic, probabilistic with
    specific sampling probabilities)?} N/A

\item \textbf{Who was involved in the data collection process (e.g.,
    students, crowdworkers, contractors) and how were they compensated
    (e.g., how much were crowdworkers paid)?} N/A

\item \textbf{Over what timeframe was the data collected?} Simulation were run over the year of 2024.

\item \textbf{Were any ethical review processes conducted (e.g., by an
    institutional review board)?} N/A

\end{itemize}

\subsection{Preprocessing/cleaning/labeling}

\begin{itemize}

\item \textbf{Was any preprocessing/cleaning/labeling of the data done
    (e.g., discretization or bucketing, tokenization, part-of-speech
    tagging, SIFT feature extraction, removal of instances, processing
    of missing values)?} N/A
\end{itemize}
\subsection{Uses}

\begin{itemize}

\item \textbf{Has the dataset been used for any tasks already?} Yes, limited testing with various ML approaches has been undertaken by the author team to ensure that the data provided in the dataset is suitable for ML training and inference. 

\item \textbf{Is there a repository that links to any or all papers or systems that use the dataset?} No

\item \textbf{What (other) tasks could the dataset be used for?} The primary focus is for ML development and testing but it could also be used for the study of turbulent flows over bluff bodies.

\item \textbf{Is there anything about the composition of the dataset or the way it was collected and preprocessed/cleaned/labeled that might impact future uses?} Not to the knowledge of the authors.

\item \textbf{Are there tasks for which the dataset should not be used?} No

\end{itemize}